\documentclass[a4paper,11pt]{article}
\usepackage{jcappub} 
\usepackage{lineno}
\usepackage{orcidlink} 
\usepackage{xspace}
\usepackage[dvipsnames]{xcolor}
\usepackage{float}
\usepackage{subcaption}
\usepackage[percent]{overpic}
\usepackage[table,dvipsnames]{xcolor}
\usepackage{tabularx}
\usepackage{multirow}
\usepackage{booktabs}
\usepackage[table]{xcolor}
\usepackage{amsmath} 
\usepackage{array}
\usepackage{colortbl}
\usetikzlibrary{shapes,arrows,positioning,calc}
\usepackage{xcolor}
\usepackage{cleveref}
\usepackage{hhline}
\usepackage{nicematrix}
\usepackage{tikz}
\usepackage{pifont}
\usepackage{graphicx}
\usepackage[normalem]{ulem}
\usetikzlibrary{arrows.meta, positioning, calc, fit, backgrounds, shadows.blur}

\newcommand{\cmark}{\ding{51}}

\definecolor{merblue}{HTML}{DCEBFF}
\definecolor{merdarkblue}{HTML}{244A7A}
\definecolor{mergreen}{HTML}{E3F4E8}
\definecolor{merdarkgreen}{HTML}{2F6B3F}
\definecolor{merorange}{HTML}{FFF1D6}
\definecolor{merdarkorange}{HTML}{8A5A00}
\definecolor{mergray}{HTML}{F4F4F4}
\definecolor{merdarkgray}{HTML}{444444}

\usetikzlibrary{shapes.geometric, arrows, positioning, fit}
\definecolor{lightgray}{gray}{0.93}
\definecolor{headergray}{gray}{0.82}
\usetikzlibrary{shadows, arrows.meta}
\definecolor{deepblue}{RGB}{25,100,180}
\definecolor{deeporange}{RGB}{230,110,30}
\definecolor{lightblue}{RGB}{210,230,255}
\definecolor{lightorange}{RGB}{255,230,210}
\newcolumntype{M}[1]{>{\centering\arraybackslash}m{#1}}

\usepackage{tikz}
\usetikzlibrary{
  calc,              
  arrows.meta, 
  positioning, 
  shapes.geometric
}

\usetikzlibrary{
  matrix,          
  calc,            
  arrows.meta,     
  positioning,     
  shapes.geometric 
}

\definecolor{crisp}{RGB}{200, 200, 200}
\colorlet{Mycolor1}{green!10!orange}
\usepackage{tcolorbox}
\tcbuselibrary{theorems, skins}

\usepackage{mdframed}

\title{\boldmath Mercury (Hg): a relativistic $N$-body simulation for Horndeski gravity}

\newcommand{\hiclass}{\texttt{hi\_class}\xspace}
\newcommand{\class}{\texttt{CLASS}\xspace}
\newcommand{\camb}{\texttt{CAMB}\xspace}
\newcommand{\mochi}{\texttt{mochi\_class}\xspace}

\newcommand{\gev}{\texttt{gevolution}\xspace}
\newcommand{\mer}{\texttt{Mercury}\xspace}
\newcommand{\kev}{\texttt{$k$-evolution}\xspace}
\newcommand{\kgb}{\texttt{KGB-evolution}\xspace}

\author[a]{Ahmad Nouri-Zonoz \orcidlink{0009-0006-6164-8670},}
\affiliation[a]{Département de Physique Théorique and Centre for Astroparticle Physics,
    Université de Genève,
    24 quai Ernest-Ansermet,
    1211 Genève 4,
    Switzerland}
\author[b]{Farbod Hassani \orcidlink{0000-0003-2640-4460},}
\affiliation[b]{
    Institute of Theoretical Astrophysics,
    University of Oslo,
    Sem Sælands vei 13,
    0371 Oslo,
    Norway
}
\author[a,c,d]{Julian Adamek \orcidlink{0000-0002-0723-6740},}
\affiliation[c]{Institut f\"ur Teilchen- und Astrophysik, ETH Z\"urich, Wolfgang-Pauli-Strasse 27, 8093 Z\"urich, Switzerland}
\affiliation[d]{Institut f\"ur Astrophysik, Universit\"at Z\"urich,
Winterthurerstrasse 190, 8057 Z\"urich, Switzerland}
\author[e]{Emilio Bellini \orcidlink{0000-0003-4762-0795},}
\affiliation[e]{    Center for Astrophysics and Cosmology,
    University of Nova Gorica,
    Vipavska cesta 13,
    Rožna Dolina,
    SI-5000 Nova Gorica,
    Slovenia}
\author[a]{and Martin Kunz \orcidlink{0000-0002-3052-7394}\,}

\emailAdd{Ahmadreza.Nourizonoz@unige.ch}
\emailAdd{farbod.hassani@astro.uio.no}
\emailAdd{adamekj@ethz.ch}
\emailAdd{emilio.bellini@ung.si}
\emailAdd{martin.kunz@unige.ch}

\abstract{ We present \mer, a relativistic $N$-body code for cosmological simulations in the $\alpha$-basis formulation of Horndeski gravity. \mer extends a particle-mesh scheme by incorporating the dynamics of the additional scalar degree of freedom and the corresponding modified Einstein equations, expressed in terms of the time-dependent functions $\alpha_{\rm K}, \alpha_{\rm B}, \alpha_{\rm M}, \alpha_{\rm T}$ and the effective Planck mass $M_*^2$. This generalises our previous implementation of Horndeski subclass models, namely $k$-essence and kinetic gravity braiding (KGB), to the general linear Horndeski $\alpha$-basis. The code evolves the additional scalar degree of freedom and its effective stress--energy tensor alongside the nonlinear matter and metric fields, so that nonlinear structure formation sources the scalar-field evolution and feeds back on the gravitational potentials. 
Through its interface with \hiclass, \mer supports both phenomenological $\alpha$-function parametrisations and covariant Horndeski theories; we demonstrate the latter using nKGB, the Cubic Galileon, and Brans--Dicke models.
We validate the implementation against \hiclass in the linear regime and recover the expected nonlinear $k$-essence and KGB limits. For the luminal Horndeski models considered here, we show how a running effective Planck mass generates gravitational slip through the scalar anisotropic stress and modifies the friction terms governing the vector and tensor metric perturbations. The evolution of $M_*^2$ also modifies the normalisation and scale dependence of the effective dark-energy density contrast, leaving a corresponding imprint on its power spectrum. The present framework provides an intermediate step towards nonlinear Horndeski simulations, in which nonlinear scalar self-interactions and screening effects can be incorporated consistently within the same relativistic framework.}

\begin{document}
\maketitle
\flushbottom

\definecolor{lightblue}{RGB}{173,216,230}
\definecolor{lightgreen}{RGB}{144,238,144}
\definecolor{lightyellow}{RGB}{255,255,224}
\definecolor{lightred}{RGB}{255,182,193}

\section{Introduction}
\label{sec:intro}

Numerical tools have become an essential component in the study of cosmological perturbations and the interpretation of large-scale structure observations. Einstein--Boltzmann solvers such as \class \citep{Blas:2011rf} and \camb \cite{Lewis:2026mif} provide accurate predictions for the evolution of linear perturbations and the corresponding cosmological observables in the standard cosmological framework. Extensions such as \hiclass \citep{Zumalacarregui:2016pph,Bellini:2019syt}, \mochi \cite{Cataneo:2024uox}, and \texttt{EFTCAMB} \citep{Hu:2013twa} have generalised this framework to scalar--tensor theories by incorporating Horndeski gravity \citep{Horndeski:1974wa}, allowing the evolution of the background and linear perturbations to be computed for a broad range of dark-energy and modified-gravity models. In particular, \hiclass provides two complementary descriptions of Horndeski theories: one based on the $\alpha$ functions introduced within the effective field theory description of dark energy \citep{Bellini:2014fua} and another based on the underlying covariant Lagrangian \citep{Bellini:2019syt}. This has made it possible to confront a wide range of scalar--tensor theories with cosmological data.

While linear Einstein--Boltzmann calculations provide accurate predictions for a wide range of cosmological observables, current and forthcoming large-scale structure surveys, such as Euclid and DESI \citep{Euclid:2024yuv,Aghamousa:2016zmz}, motivate the development of accurate predictions beyond the linear regime. In modified gravity theories, this requires numerical frameworks capable of following the coupled evolution of matter, metric perturbations, and additional degrees of freedom on nonlinear scales \citep{Koyama:2015vza}. Relativistic $N$-body simulations provide such a framework by extending conventional structure formation simulations with a consistent treatment of relativistic gravitational effects. The \gev code \citep{Adamek:2015eda,Adamek:2016zes, Adamek:2026vof} incorporates relativistic metric perturbations within a weak-field formulation of general relativity and provides a relativistic description of cosmological structure formation.

A number of $N$-body frameworks have been developed to study nonlinear structure formation in modified-gravity and scalar--tensor theories, including recent implementations based on the effective field theory of dark energy and Galileon-type models \citep{Ganjoo:2026ugf,Atayde:2026dle}. Building upon the relativistic \gev framework, several extensions have incorporated additional scalar-field dynamics, including $k$-essence in \kev \citep{Hassani:2019lmy,Hassani:2019wed} and scalar-field domain formation and associated relativistic signatures of the (a)symmetron model in \texttt{Asimulation} \citep{Christiansen:2023tfy, Christiansen:2024vqv,Christiansen:2024uyr}. More recently, we extended this framework in \kgb to kinetic gravity braiding (KGB) models, incorporating the braiding interaction into the relativistic simulation framework and studying its signatures in nonlinear structure formation and relativistic observables \citep{Nouri-Zonoz:2025cul,Nouri-Zonoz:2026bwx}. The progression from \kev to \kgb established the numerical workflow that we generalise here towards a broader treatment of Horndeski gravity.

In this work, we introduce \mer, extending the \kgb code to the general linear Horndeski $\alpha$-basis. The code incorporates the Horndeski background and perturbation dynamics, with the linear perturbation sector described by the time-dependent $\alpha$ functions $\alpha_{\rm K}$, $\alpha_{\rm B}$, $\alpha_{\rm M}$ and $\alpha_{\rm T}$, together with the evolution of the effective Planck mass $M_*^2$. The corresponding scalar contribution is implemented as an effective stress--energy tensor source on the simulation lattice, while the evolution of the effective Planck mass is consistently incorporated into the gravitational sector. Following the approach adopted in \kgb, the additional scalar degree of freedom is evolved using the linear Horndeski equations while remaining consistently coupled to the nonlinear matter and metric sectors. A key feature of \mer is its direct connection to the linear Horndeski framework implemented in \hiclass \citep{Zumalacarregui:2016pph,Bellini:2019syt}. Following the same model description, \mer supports both approaches available in \hiclass: the effective-theory approach, in which the $\alpha$ functions are parametrised directly, and the covariant-theory approach, in which the model is specified through its underlying Horndeski Lagrangian and the corresponding $\alpha$ functions are determined from the covariant dynamics. We validate the implementation by recovering the known $k$-essence and KGB limits, comparing the linear regime with \hiclass predictions, and testing the framework using both effective-theory and covariant descriptions of Horndeski models.

The paper is organised as follows. In Sec.~\ref{sec:Horndeski_gravity}, we introduce the Horndeski framework and the description of the background and linear perturbation dynamics adopted in \mer. Section~\ref{sec:implementation} presents the implementation of the Horndeski contributions within the relativistic simulation framework, including the construction of the effective scalar sources in real space and the solution of the equations in Fourier space. In Sec.~\ref{sec:mercury_workflow}, we describe the numerical workflow of \mer, from the specification of Horndeski models and the computation of the background evolution to the generation of initial conditions and the nonlinear evolution in the main simulation loop. Section~\ref{sec:results} presents the numerical validation of \mer and the results for six representative Horndeski scenarios, comprising three
phenomenological EFT parametrisations and three covariant Horndeski models, comparing the background evolution and perturbation spectra with \hiclass
and examining the deviations that emerge in the nonlinear regime.  Finally, we summarise our results and discuss future perspectives in Sec.~\ref{sec:discussion}.

\section{Horndeski gravity}
\label{sec:Horndeski_gravity}

We consider the Horndeski class of scalar--tensor theories \citep{Horndeski:1974wa}, which constitutes the most general covariant theory of a metric and a single scalar field whose equations of motion contain derivatives up to second order. This second-order structure avoids the additional unstable degree of freedom that generically arises in non-degenerate higher-derivative theories, commonly associated with the Ostrogradsky instability \citep{Woodard:2015zca}. 
In the formulation used in this work, the equations are kept in the Jordan frame and organised such that the effects of the Horndeski scalar can be treated as effective dark-energy contributions to the metric equations. The scalar sector is therefore represented by an effective stress--energy tensor, while the non-minimal gravitational coupling is captured by the time-dependent effective Planck mass $M_*^2$. This representation is particularly convenient for the implementation in \gev, since the Horndeski contributions can be incorporated as additional gravitational sources while retaining the overall structure of the existing metric solver. At the same time, the minimal coupling of matter in the Jordan frame allows the particle equations of motion to retain their standard \gev form. 

\subsection{Horndeski action and $\alpha$-basis description}
\label{sec:Horndeski}

Throughout this work we assume that the Einstein equivalence principle is respected. All matter species are therefore minimally and universally coupled to the same Jordan-frame metric $g_{\mu\nu}$, with no direct coupling between the matter fields and the Horndeski scalar. Under these assumptions, the total action can be written as\footnote{
We work in natural units, $\hbar=c=1$, and follow the
convention adopted in \hiclass, in which the locally measured
Planck mass $M_{\rm Pl}^2$ is factored out of the Horndeski
Lagrangian, as in Eq.~\eqref{eq:Horndeski_action}.
In this convention, the scalar field $\phi$ is dimensionless,
while its kinetic term $X=-\nabla_\mu\phi\nabla^\mu\phi/2$
has mass dimension two. Consequently, the functions
$G_2$, $G_3$, $G_4$, and $G_5$ have mass dimensions
$2$, $0$, $0$, and $-2$, respectively, such that each
$\mathcal{L}_i$ has mass dimension two.
The factor $M_{\rm Pl}^2$ then ensures that the total
Lagrangian density has mass dimension four.
In this normalisation, general relativity corresponds to
$G_4=1/2$ and $G_2=G_3=G_5=0$.
}
\begin{equation}
S_{\rm tot}
=
\int d^4x\,\sqrt{-g}
\left[
M_{\rm Pl}^2\sum_{i=2}^{5}\mathcal{L}_i
+
\mathcal{L}_{\rm m}(g_{\mu\nu},\psi_{\rm m})
\right]\,,
\label{eq:Horndeski_action}
\end{equation}
where $\mathcal{L}_i$ are given by
\begin{align*}
\mathcal{L}_2 &= G_2(\phi, X) \, ,\\
\mathcal{L}_3 &= -G_3(\phi, X) \Box \phi \, ,\\
\mathcal{L}_4 &= G_4(\phi, X) R + G_{4X} \left[ (\Box \phi)^2 - (\nabla_\mu \nabla_\nu \phi)^2 \right] \, ,\\
\mathcal{L}_5 &= G_5(\phi, X) G_{\mu\nu} \nabla^\mu \nabla^\nu \phi - \frac{1}{6} G_{5X} \left[ (\Box \phi)^3 - 3\Box \phi (\nabla_\mu \nabla_\nu \phi)^2 + 2(\nabla_\mu \nabla_\nu \phi)^3 \right]  \, .
\end{align*}
The freedom of the theory is encoded in the functions $G_i(\phi,X)$, which are arbitrary functions of the scalar field $\phi$ and its canonical kinetic term
$X \equiv -\frac{1}{2}\nabla^\mu\phi\,\nabla_\mu\phi$.
We denote derivatives with respect to $X$ by a subscript, $G_{iX}\equiv \partial G_i/\partial X$. 

Several well-known dark energy and modified gravity models are recovered as particular choices of the functions $G_i$. For instance, minimally coupled quintessence and $k$-essence correspond to fixing $G_4=1/2$ and $G_3=G_5=0$, so that the scalar dynamics are entirely determined by $G_2(\phi,X)$. Allowing in addition a non-vanishing $G_3(\phi,X)$ while keeping $G_4=1/2$ and $G_5=0$ gives the kinetic gravity braiding (KGB) subclass, where the scalar kinetic structure mixes with the metric through the braiding interaction \citep{Deffayet:2010qz}. More general choices of $G_4$ and $G_5$ introduce non-minimal interactions between the scalar field and curvature, affecting the effective gravitational coupling and the dynamics of metric perturbations.

To describe cosmological perturbations, we work around a spatially flat Friedmann--Lema\^{\i}tre--Robertson--Walker (FLRW) background and adopt the Poisson gauge, which is the gauge used by \gev. The line element is written as 
\begin{equation} ds^2 = a^2(\tau) \left[ - e^{2\Psi} d\tau^2 - 2 B_i dx^i d\tau + \left( e^{-2\Phi}\delta_{ij} + h_{ij} \right) dx^i dx^j \right]\,, 
\label{eq:FLRW}
\end{equation} 
where $a(\tau)$ is the scale factor and $\tau$ denotes conformal time, related to cosmic time by $d\tau = dt/a(t)$. The scalar perturbations $\Psi$ and $\Phi$ correspond to the temporal and spatial metric potentials, respectively. The vector perturbation $B_i$ is transverse, $\partial^i B_i=0$, while the tensor perturbation $h_{ij}$ is transverse and traceless, $\partial^i h_{ij}=0$ and $h^i_{\ i}=0$.

The covariant metric equations and the scalar-field equation follow from the variation of the total action and form the starting point for the implementation. When expressed directly in terms of the $G_i$ functions, these equations involve many different combinations of the functions and their derivatives with respect to $\phi$ and $X$. For cosmological applications, however, the linear perturbation dynamics can be organised more efficiently in terms of the effective Planck mass $M_*^2$ and the four dimensionless $\alpha$ functions $\alpha_{\rm K}$, $\alpha_{\rm B}$, $\alpha_{\rm M}$ and $\alpha_{\rm T}$ \citep{Bellini:2014fua,Zumalacarregui:2016pph,Frusciante:2019puu}. These quantities are evaluated on the homogeneous background and can either be derived from a specified covariant Horndeski action or supplied directly through a phenomenological parametrisation.

The effective Planck mass and the four $\alpha$ functions characterise different aspects of the scalar--tensor dynamics. In the following, primes denote derivatives with respect to conformal time, $H\equiv a'/a^2$ is the physical Hubble rate, $\mathcal{H}\equiv a'/a=aH$ is the conformal Hubble rate, and the background kinetic term is $\bar X=\bar\phi'^2/(2a^2)$. All functions $G_i$ and their derivatives are understood to be evaluated on the homogeneous background. We follow the \hiclass convention in which $M_*^2$ is dimensionless and denotes the squared ratio of the cosmological effective Planck mass to the locally measured Planck mass $M_*^2 \equiv M_{\rm eff}^2/M_{\rm Pl}^2$.

\begin{itemize}

    \item $\alpha_{\rm K}$, \textit{kineticity}, characterises the kinetic contribution of the scalar degree of freedom and contributes to the kinetic coefficient governing the propagating scalar mode. In the covariant formulation, $\alpha_{\rm K}$ can receive contributions from all four Horndeski functions $G_2$, $G_3$, $G_4$ and $G_5$:
\begin{equation}
\begin{aligned}
    H^2 M_*^2 \alpha_{\rm K}
    ={}&
    2\bar X
    \left(
        G_{2X}
        +2\bar X G_{2XX}
        -2G_{3\phi}
        -2\bar X G_{3\phi X}
    \right)
    \\
    &+
    \frac{12\bar\phi'H\bar X}{a}
    \left(
        G_{3X}
        +\bar X G_{3XX}
        -3G_{4\phi X}
        -2\bar X G_{4\phi XX}
    \right)
    \\
    &+
    12H^2\bar X
    \Big[
        G_{4X}
        +8\bar X G_{4XX}
        +4\bar X^2 G_{4XXX}
    \Big]
    \\
    &-
    12H^2\bar X
    \Big[
        G_{5\phi}
        +5\bar X G_{5\phi X}
        +2\bar X^2 G_{5\phi XX}
    \Big]
    \\
    &+
    \frac{4\bar\phi'H^3\bar X}{a}
    \left(
        3G_{5X}
        +7\bar X G_{5XX}
        +2\bar X^2 G_{5XXX}
    \right)\,.
\end{aligned}
\label{eq:alphaK_def}
\end{equation}

\item $\alpha_{\rm B}$, \textit{braiding}, quantifies the kinetic mixing between the scalar and metric perturbations. A non-vanishing $\alpha_{\rm B}$ allows the scalar and metric sectors to source one another, modifying the response of the scalar degree of freedom to matter perturbations. In the covariant formulation, it receives contributions from the derivative interaction $G_3$ and from the non-minimal sectors $G_4$ and $G_5$:
    \begin{align}
        HM_*^2\alpha_{\rm B}
        =&\;
        \frac{2\bar\phi'}{a}
        \left(
            \bar X G_{3X}
            -G_{4\phi}
            -2\bar X G_{4\phi X}
        \right)
        \nonumber \\
        &+
        8H\bar X
        \left(
            G_{4X}
            +2\bar XG_{4XX}
            -G_{5\phi}
            -\bar XG_{5\phi X}
        \right)
        \nonumber \\
        &+
        \frac{2\bar\phi'H^2\bar X}{a}
        \left(
            3G_{5X}
            +2\bar XG_{5XX}
        \right)\,.
        \label{eq:alphaB_def}
    \end{align}

    \item $M_*^2$, the dimensionless \textit{effective Planck mass squared}, sets the normalisation of the gravitational sector and may evolve with the cosmological background. In the covariant formulation, it is determined by the non-minimal coupling functions $G_4$ and $G_5$:
    \begin{equation}
        M_*^2
        =
        2\left[
            G_4
            -2\bar X G_{4X}
            +\bar X G_{5\phi}
            -
            \frac{H\bar\phi'\bar X}{a}
            G_{5X}
        \right]\,.
        \label{eq:Mstar_def}
    \end{equation}

    \item $\alpha_{\rm M}$, the \textit{Planck-mass run rate}, measures the logarithmic evolution of the effective Planck mass:
    \begin{equation}
        \alpha_{\rm M}
        =
        \frac{1}{aH}
        \frac{\left(M_*^2\right)'}{M_*^2}\,.
        \label{eq:alphaM_def}
    \end{equation}

\item $\alpha_{\rm T}$, the \textit{tensor-speed excess}, measures the deviation of the propagation speed of tensor perturbations from the speed of light. In the covariant formulation, it is determined by derivative interactions in $G_4$ and $G_5$:
    \begin{equation}
        M_*^2\alpha_{\rm T}
        =
        2\bar X
        \left[
            2G_{4X}
            -2G_{5\phi}
            -
            \frac{\bar\phi''-2aH\bar\phi'}{a^2}
            G_{5X}
        \right]\,.
        \label{eq:alphaT_def}
    \end{equation}

\end{itemize}

The above relations provide the mapping between the covariant Horndeski action and the quantities used in \mer. The evolution equations can then be expressed in terms of the effective background density and pressure, $\bar\rho_\phi$ and $\bar P_\phi$, together with $M_*^2$ and the $\alpha$ functions, without explicitly retaining the individual functions $G_i$. An important feature of this formulation is that it separates the background expansion history from the independent properties of linear perturbations encoded by the $\alpha$ functions \citep{Bellini:2014fua}.

For the scalar perturbation, we follow the notation used in our previous implementations and introduce the variables
\begin{equation}
    \pi
    =
    \frac{\delta\phi}{\bar\phi'}\,,
    \qquad
    \zeta
    =
    \pi'
    +
    \mathcal{H}\pi
    -
    \Psi\,.
    \label{eq:zeta_pi}
\end{equation}

\subsection{Background equations}

The homogeneous background evolution used by \mer is obtained from \hiclass. The expansion rate is determined from the energy density of all homogeneous components through the Friedmann equation
\begin{equation}
    H^2
    =
    \sum_i \bar\rho_i\,,
    \label{eq:hiclass_friedmann}
\end{equation}
where the sum runs over all species included in the background evolution. We adopt the \hiclass convention, based on the normalisation used in \class \citep{Lesgourgues:2011re}, in which the components of the energy--momentum tensor are rescaled according to
\begin{equation}
    \bar\rho_i^{\texttt{hiclass}}
    =
    \frac{8\pi G_N}{3}\bar\rho_i^{\rm standard}
    =
    \frac{1}{3M_\textrm{Pl}^2}\bar\rho_i^{\rm standard}\,,
    \label{eq:CLASS_normalization}
\end{equation}
with $G_N$ fixed to Newton's constant as measured today. In the \hiclass convention, the effective Planck mass is normalised to the locally measured Planck mass. Consequently, $M_*^2$ is dimensionless and represents their squared ratio, with the general relativistic limit corresponding to $M_*^2=1$.

At the background level, the Horndeski scalar is described in terms of an effective dark-energy density $\bar\rho_{\phi}$ and pressure $\bar P_{\phi}$. For a covariant Horndeski model, these quantities are obtained by evaluating the action on the homogeneous background and are given by
\begin{align}
    \bar\rho_{\phi}^{\texttt{hiclass}}
    =&\;
    -\frac{1}{3}G_2
    +
    \frac{2}{3}\bar X
    \left(
        G_{2X}
        -
        G_{3\phi}
    \right)
    -
    \frac{2H^3\bar\phi'\bar X}{3a}
    \left(
        7G_{5X}
        +
        4\bar XG_{5XX}
    \right)
    \nonumber \\
    &+
    H^2
    \left[
        1
        -
        \left(
            1-\alpha_{\rm B}
        \right)
        M_*^2
        -
        4\bar X
        \left(
            G_{4X}
            -
            G_{5\phi}
        \right)
        -
        4\bar X^2
        \left(
            2G_{4XX}
            -
            G_{5\phi X}
        \right)
    \right]\,,
    \label{eq:rhoDE_hiclass}
\end{align}
and
\begin{align}
    \bar P_{\phi}^{\texttt{hiclass}}
    =&\;
    \frac{1}{3}G_2
    -
    \frac{2}{3}\bar X
    \left(
        G_{3\phi}
        -
        2G_{4\phi\phi}
    \right)
    +
    \frac{4H\bar\phi'}{3a}
    \left(
        G_{4\phi}
        -
        2\bar XG_{4\phi X}
        +
        \bar XG_{5\phi\phi}
    \right)
    \nonumber \\
    &-
    \frac{\bar\phi''-aH\bar\phi'}{3a\bar\phi'}
    HM_*^2\alpha_{\rm B}
    -
    \frac{4}{3}H^2\bar X^2G_{5\phi X}
    -
    \left(
        H^2
        +
        \frac{2H'}{3a}
    \right)
    \left(
        1
        -
        M_*^2
    \right)
    \nonumber \\
    &+
    \frac{2H^3\bar\phi'\bar X}{3a}
    G_{5X}\,.
    \label{eq:pDE_hiclass}
\end{align}
Here all functions $G_i$ and their derivatives are evaluated on the homogeneous background.

With this choice of variables, all homogeneous contributions from the Horndeski sector, including those associated with the time dependence of the effective Planck mass, are absorbed into the effective quantities $\bar\rho_{\phi}$ and $\bar P_{\phi}$. This allows the background scalar sector to be treated as an effective dark-energy fluid. Since matter is minimally coupled to the Jordan-frame metric, the matter sector remains separately conserved, while the effective dark-energy component satisfies
\begin{equation}
    \bar\rho_{\phi}'
    +
    3\mathcal{H} \bar\rho_{\phi}
    \left(
        1
        +
        w_\phi
    \right)
    =
    0\,,
    \label{eq:DE_background_conservation}
\end{equation}
where $w_\phi \equiv \bar P_\phi/\bar\rho_\phi$ is the effective dark-energy equation-of-state parameter.

The quantities $\bar\rho_\phi$ and $\bar P_\phi$ defined in this way are the background inputs used in the perturbation equations below. The practical determination of these quantities for the different \hiclass input modes is described in Sec.~\ref{sec:mercury_workflow}.

\subsection{Linear perturbation equations}

The linear perturbation equations provide the starting point for the numerical implementation. They are written in Poisson gauge and in conformal time, following the notation introduced in the previous subsections.

\subsubsection*{Einstein field equations}

The Hamiltonian constraint (Einstein $^0{}_{0}$ component) takes the form
\begin{equation}
    \nabla^2\Phi -3\mathcal{H}^2 \Psi -3 \mathcal{H}\Phi^{'}  =\frac{a^2}{2M_{*}^2}\big( \delta\rho_{\rm m} + \delta\rho_{\phi} \big)  \, ,
    \label{eq:Hamiltonian}
\end{equation}
where the scalar-field contribution to the effective energy density is 
\begin{equation}
\begin{aligned}
\delta\rho_{\phi}  &=   - \frac{M_*^2}{a^2} \bigg\{\alpha_\textrm{B} \mathcal{H}  \,\nabla ^{2}{}\pi  + 3 \alpha_\textrm{B} \mathcal{H} ^2 \Psi  - (3 \alpha_\textrm{B}  + \alpha_\textrm{K}) \mathcal{H} ^2 \zeta  + 3 \alpha_\textrm{B} \mathcal{H}  \Phi ^{\prime}
    \\&\hspace{2.2cm}+\Big[2\mathcal{H}^2 -\alpha_\textrm{B} \mathcal{H} ^2  + \alpha_\textrm{B} \mathcal{H} ^{\prime}  - 2\mathcal{H}'   -\frac{a^2}{M_*^2} (\bar{\rho}_{\rm m}  + \bar{P}_{\rm m} )\Big]3 \mathcal{H} \pi \bigg\} \, .
    \label{eq:deltarhoDE}
\end{aligned}
\end{equation}
Equation \eqref{eq:Hamiltonian} is used to determine the spatial curvature potential $\Phi$. The remaining metric perturbations are obtained from the trace-free spatial part of the Einstein equations, which is given by
\begin{equation}
\begin{aligned}
\mathcal D_{ij}\chi
+ B'_{(i,j)}
+ (2+\alpha_{\rm M})\mathcal H B_{(i,j)}
+ \frac{1}{2}h''_{ij}
+ \left(1+\frac{\alpha_{\rm M}}{2}\right)\mathcal H h'_{ij}
&- \frac{1}{2}(1+\alpha_{\rm T})\nabla^2 h_{ij}
\\=
&
\frac{a^2}{M_*^2}
\left(
\Pi^{\rm m}_{ij}
+
\Pi^\phi_{ij}
\right) \, ,
\end{aligned}
\label{eq:linear_tracefree_spatial}
\end{equation}
where $\chi \equiv \Phi-\Psi$ is the gravitational slip, $\mathcal D_{ij}\equiv \partial_i\partial_j-\frac{1}{3}\delta_{ij}\nabla^2$ is the traceless second-derivative operator and $\Pi^\phi_{ij}$ is the scalar-field anisotropic stress given by 
\begin{equation}
\Pi^\phi_{ij}
=\frac{M_*^2}{a^2}
\mathcal D_{ij}
\left[-\alpha_{\rm T}\Phi
+ (\alpha_{\rm M}-\alpha_{\rm T})\mathcal H\pi\right]\,.
\label{eq:anisotropicEQ}
\end{equation}
Thus the trace-free spatial equation contains five independent components. These determine the scalar slip $\chi$, the two vector components of $B_i$, and the two tensor components of $h_{ij}$. In the numerical implementation that we will discuss in the next section, these components are separated and solved by the corresponding spin projections in Fourier space.

The effects of the non-minimal coupling appear in both metric equations, Eqs.~\eqref{eq:Hamiltonian} and \eqref{eq:linear_tracefree_spatial}. In the Hamiltonian constraint, $M_*^2$ enters both the effective scalar-field energy density perturbation $\delta\rho_\phi$ and the prefactor multiplying the total density perturbation $\delta\rho_m+\delta\rho_\phi$. In the trace-free spatial equation, $M_*^2$ analogously enters both the effective scalar-field anisotropic stress $\Pi^{\phi}_{ij}$ and the effective gravitational coupling of the total anisotropic-stress source. The time variation of the effective Planck mass further modifies the damping of the vector and tensor perturbations through $\alpha_{\rm M}$, while the tensor-speed excess $\alpha_{\rm T}$ changes the tensor Laplacian term, corresponding to a modified propagation speed for gravitational waves.

Once the Hamiltonian constraint and the trace-free spatial equations are imposed, the remaining Einstein equations are related to them by the contracted Bianchi identities and the covariant conservation of the total stress--energy tensor. They therefore do not provide independent metric equations, but serve as consistency relations. In particular, the time-space component gives the momentum constraint

\begin{equation}
-\frac{2M_*^2 }{a^2}\Big( \frac{1}{4}\nabla^2 B_{i} + \partial_{i}\Phi'
+ \mathcal{H}\partial_{i}\Psi
\Big) = \delta T_{i}^{0 (\rm m)} + \delta T_{i}^{0 (\phi)} \, ,
\label{eq:momentum_constraint}
\end{equation}
where
\begin{equation}
 \delta T_{i}^{0 (\phi)} =\frac{M_*^2}{a^2}\bigg\{ \bigg[\frac{ a^2}{M_*^2}\Big(\bar{\rho}_{\rm m}+\bar{P}_{\rm m}\Big) -2\mathcal{H}^2+ 2\mathcal{H}'\bigg]\partial _{i}\pi + \alpha_\textrm{B}\mathcal{H}\partial_i \zeta \,\bigg\}\,.
\end{equation}
Although Eq.~\eqref{eq:momentum_constraint} is not solved independently in our implementation, its transverse projection provides an optional consistency test of the vector perturbation obtained from the spin-1 projection of Eq.~\eqref{eq:linear_tracefree_spatial}.

\subsubsection*{Scalar-field equation of motion}
The evolution of the additional scalar degree of freedom is determined by its linear equation of motion. Using the variables $\pi$ and $\zeta$ introduced above, we can write
\begin{equation}
\textcolor{BrickRed}{A_{\zeta'}} \zeta'
+ \textcolor{BrickRed}{A_{\zeta}} \zeta
+ \textcolor{BrickRed}{A_{\pi}} \pi
+ \textcolor{BrickRed}{A_{\nabla^2 \pi}} \nabla^2 \pi
+ \textcolor{BrickRed}{A_{\nabla^2 \Psi}} \nabla^2 \Psi
+ \textcolor{BrickRed}{A_{\nabla^2 \Phi}} \nabla^2 \Phi
+ \textcolor{BrickRed}{A_{\delta P_\textrm{m}}} \delta P_\textrm{m}
+ \textcolor{BrickRed}{A_{\Phi'}} \Phi'
+ \textcolor{BrickRed}{A_{\Psi}} \Psi
= 0 \, ,
\label{eq:fullEOM_factored}
\end{equation}
where
\begin{align*}
\textcolor{BrickRed}{A_{\zeta'}} &= 
- \mathcal{H}^2 \left( \frac{3}{2} \alpha_\textrm{B}^2 + \alpha_\textrm{K} \right) \, , \quad
\textcolor{BrickRed}{A_{\nabla^2 \Psi}} = 
  \alpha_\textrm{B} \mathcal{H} \, , \quad
\textcolor{BrickRed}{A_{\nabla^2 \Phi}} = 
2 \mathcal{H} \left( \alpha_\textrm{M} - \alpha_\textrm{T} \right) \, , \quad
\textcolor{BrickRed}{A_{\delta P_\textrm{m}}} = 
\frac{3 a^2 \mathcal{H}}{2 M_*^2} \, \alpha_\textrm{B} \, , \\[6pt]
\textcolor{BrickRed}{A_{\zeta}} &= 
\mathcal{H}^3 \Bigg[
3 \alpha_\textrm{B} (1 - \alpha_\textrm{B})
- \alpha_\textrm{K}
- \frac{3}{2} \alpha_\textrm{B}^2 \alpha_\textrm{M}
- \alpha_\textrm{K} \alpha_\textrm{M}
\Bigg]
- \mathcal{H}^2 \Bigg[
\frac{3}{2} \alpha_\textrm{B} \alpha_\textrm{B}' + \alpha_\textrm{K}'
\Bigg] \\[2pt]
&\quad
- \mathcal{H} \mathcal{H}' \Bigg[
3 \alpha_\textrm{B} + \frac{3}{2} \alpha_\textrm{B}^2 + 2 \alpha_\textrm{K}
\Bigg]
- \frac{3a^2 \mathcal{H}}{2 M_*^2} \, \alpha_\textrm{B} 
\left( \bar{\rho}_\textrm{m} + \bar{P}_\textrm{m} \right) \, , \\[6pt]
\textcolor{BrickRed}{A_{\pi}} &= 
3 \Bigg\{
\mathcal{H}^4 \left( \alpha_\textrm{B} - 2 \right)
- \mathcal{H}^3 \alpha_\textrm{B}'
+ \mathcal{H}^2 \mathcal{H}' \left( 4 - 2 \alpha_\textrm{B} \right)
+ \mathcal{H} \mathcal{H}' \alpha_\textrm{B}'
+ \mathcal{H}'^{2} \left( \alpha_\textrm{B} - 2 \right) \\[2pt]
&\qquad
+ \frac{a^2}{M_*^2} \Big[
\left( \bar{\rho}_\textrm{m} + \bar{P}_\textrm{m} \right) \mathcal{H}^2
- \left( \bar{\rho}_\textrm{m} + \bar{P}_\textrm{m} \right) \mathcal{H}'
- \frac{1}{2} \alpha_\textrm{B} \mathcal{H} \bar{P}_\textrm{m}'
\Big]
\Bigg\} \, .
\\
\textcolor{BrickRed}{A_{\Phi'}} &=
3 \Bigg[
\mathcal{H}^2 \left( 2 - \alpha_\textrm{B} \right)
+ \mathcal{H} \, \alpha_\textrm{B}'
+ \mathcal{H}' \left( \alpha_\textrm{B} - 2 \right)
- \frac{a^2}{M_*^2} \left( \bar{\rho}_\textrm{m} + \bar{P}_\textrm{m} \right)
\Bigg] \, , \\[6pt]
\textcolor{BrickRed}{A_{\nabla^2 \pi}} &= 
\mathcal{H}^2 \left(
2 + 2 \alpha_\textrm{M} + \alpha_\textrm{B} \alpha_\textrm{M} - 2 \alpha_\textrm{T}
\right)
+ \mathcal{H} \, \alpha_\textrm{B}'
+ \mathcal{H}' \left( \alpha_\textrm{B} - 2 \right)
- \frac{a^2}{M_*^2} \left( \bar{\rho}_\textrm{m} + \bar{P}_\textrm{m} \right) \, , \\[6pt]
\textcolor{BrickRed}{A_{\Psi}} &= 
3 \mathcal{H} \Bigg[
\mathcal{H}^2 \left( 2 - \alpha_\textrm{B} \right)
+ \mathcal{H} \, \alpha_\textrm{B}'
+ \mathcal{H}' \left( \alpha_\textrm{B} - 2 \right)
- \frac{a^2}{M_*^2} \left( \bar{\rho}_\textrm{m} + \bar{P}_\textrm{m} \right)
\Bigg] \, , \\[6pt]
\end{align*}

\section{Implementation}
\label{sec:implementation}
In this section, we describe the numerical implementation of the Horndeski sector in \mer. The implementation follows the general structure of \gev, with the scalar-field contributions constructed on the real-space lattice and the metric equations solved using Fourier-space projections.

In the following, all quantities entering the numerical evolution are expressed according to the \gev unit convention. Background quantities imported from \hiclass, including the expansion rate, densities, pressure, and their time derivatives, are converted to this convention before entering the lattice equations. In particular, \gev normalises the background energy densities such that $\rho_{c,0}=1$, while spatial coordinates and conformal time are normalised by the simulation box size. With this convention, the locally measured Planck mass is related to the gravitational coupling through $M_{\rm Pl}^2=(8\pi G_N)^{-1}=(2\,{\tt fourpiG})^{-1}$, where ${\tt fourpiG}\equiv4\pi G_N=\frac{3}{2}(H_0L/c)^2$ and $L$ denotes the simulation box size in $h^{-1}\,{\rm Mpc}$. The effective Planck mass entering the equations below is then $M_{\rm eff}^2\equiv M_{\rm Pl}^2M_*^2$.

\subsection{Real-space source construction }
We construct the scalar-field contribution as an effective stress--energy tensor on the real-space lattice. For this purpose, we define a new function, \texttt{projection\_Tmunu\_smg}, in the \texttt{gevolution.hpp} file. In the following expressions, we label lattice sites by the indices $\mathbf{i},\mathbf{j},\mathbf{k}$ and time cycles by the superscript $\mathbf{n}$. Thus, for a generic field $f$, the value stored at the lattice point $(\mathbf{i},\mathbf{j},\mathbf{k})$ during the $\mathbf{n}$-th cycle is denoted by $f^{\mathbf{n}}_{\mathbf{i},\mathbf{j},\mathbf{k}}$. The resulting components provide the lattice representation of the scalar-field effective stress--energy tensor, with $\delta T^0{}_{0(\phi)}=-\delta\rho_\phi$, together with $\delta T^0{}_{i(\phi)}$ and $\Pi^\phi_{ij}$:

\begin{align}
\delta T^{0}_{0}\left(
\tau_{\mathbf{n}},
\mathbf{x}_{\mathbf{i},\mathbf{j},\mathbf{k}}
\right)
&=
\frac{M^2_{\rm eff} }{a^2}
\bigg\{
\alpha_{\rm B}\mathcal{H}
\left(\nabla^2\pi\right)_{\mathbf{i},\mathbf{j},\mathbf{k}}^{\,\mathbf{n}}
+
3\alpha_{\rm B}\mathcal{H}^2
\Psi_{\mathbf{i},\mathbf{j},\mathbf{k}}^{\,\mathbf{n}}
-
\left(
3\alpha_{\rm B}
+
\alpha_{\rm K}
\right)
\mathcal{H}^2
\zeta_{\mathbf{i},\mathbf{j},\mathbf{k}}^{\,\mathbf{n}-\frac{1}{2}}
+
3\alpha_{\rm B}\mathcal{H}
\left(\Phi'\right)_{\mathbf{i},\mathbf{j},\mathbf{k}}^{\,\mathbf{n}} \nonumber
\label{eq:deltarho_DE}
\\ 
&\quad
+
3\mathcal{H}
\bigg[
2\mathcal{H}^2
-
\alpha_{\rm B}\mathcal{H}^2
+
\alpha_{\rm B}\mathcal{H}'
-
2\mathcal{H}'
-
\frac{a^2}{M^2_{\rm eff} }
\left(
\bar\rho_{\rm m}
+
\bar P_{\rm m}
\right)
\bigg]
\pi_{\mathbf{i},\mathbf{j},\mathbf{k}}^{\,\mathbf{n}}
\bigg\}\,,
\\[10pt]
\delta T^{0}_{\ell}\left(
\tau_{\mathbf{n}},
\mathbf{x}_{\mathbf{i},\mathbf{j},\mathbf{k}}
\right)
&=
\frac{M^2_{\rm eff} }{a^2}
\bigg\{
\bigg[
\frac{a^2}{M^2_{\rm eff} }
\left(
\bar\rho_{\rm m}
+
\bar P_{\rm m}
\right)
-
2\mathcal{H}^2
+
2\mathcal{H}'
\bigg]
\left(\partial_\ell\pi\right)_{\mathbf{i},\mathbf{j},\mathbf{k}}^{\,\mathbf{n}}
+
\alpha_{\rm B}\mathcal{H}
\left(\partial_\ell\zeta\right)_{\mathbf{i},\mathbf{j},\mathbf{k}}^{\,\mathbf{n}-\frac{1}{2}}
\bigg\}\,,
\\[10pt]
\Pi_{\ell m} \left(
\tau_{\mathbf{n}},
\mathbf{x}_{\mathbf{i},\mathbf{j},\mathbf{k}}
\right)
&=
\frac{M^2_{\rm eff}}{a^2}
\left[
\mathcal D_{\ell m}
\left(
-\alpha_{\rm T}
\Phi
+
\left(
\alpha_{\rm M}
-
\alpha_{\rm T}
\right)
\mathcal{H}
\pi
\right)
\right]_{\mathbf{i},\mathbf{j},\mathbf{k}}^{\,\mathbf{n}}\,.
\label{eq:lattice_Horndeski}
\end{align} 

Here $\ell,m=1,2,3$ denote spatial components and all background quantities are evaluated at the corresponding time cycle.

\subsection{Real-space scalar-field evolution}

The scalar-field variables are advanced in real space using a Newton--St\"ormer--Verlet leapfrog scheme \citep{Hairer_Lubich_Wanner_2003}. In this scheme, $\zeta$ is stored at half-integer time cycles, while $\pi$ and the metric variables are stored at integer cycles.\footnote{
The leapfrog scheme stores $\pi$ at integer cycles $\tau^{\mathbf{n}}$ and $\zeta$ at half-integer cycles $\tau^{\mathbf{n}+\frac12}$. Since $\zeta$ determines $\pi'$ through Eq.~\eqref{eq:zeta_pi}, it is offset by half a time step relative to $\pi$, as in the standard position--velocity leapfrog update.
}
The update for $\zeta$ is written as
\begin{equation}
    \zeta_{\mathbf{i},\mathbf{j},\mathbf{k}}^{\,\mathbf{n}+\frac{1}{2}}
    =
    \zeta_{\mathbf{i},\mathbf{j},\mathbf{k}}^{\,\mathbf{n}-\frac{1}{2}}
    +
    \zeta_{\mathbf{i},\mathbf{j},\mathbf{k}}^{\prime\,\mathbf{n}}
    \Delta\tau \,,
    \label{zetaleap}
\end{equation}
where $\zeta_{\mathbf{i},\mathbf{j},\mathbf{k}}^{\prime\,\mathbf{n}}$ is obtained from the scalar-field equation of motion,
\begin{equation}
\begin{aligned}
\zeta_{\mathbf{i},\mathbf{j},\mathbf{k}}^{\prime\,\mathbf{n}}
=
-\frac{1}{\textcolor{BrickRed}{A_{\zeta'}}}
\bigg[
&
\textcolor{BrickRed}{A_{\zeta}}
\zeta_{\mathbf{i},\mathbf{j},\mathbf{k}}^{\,\mathbf{n}}
+
\textcolor{BrickRed}{A_{\pi}}
\pi_{\mathbf{i},\mathbf{j},\mathbf{k}}^{\,\mathbf{n}}
+
\textcolor{BrickRed}{A_{\Psi}}
\Psi_{\mathbf{i},\mathbf{j},\mathbf{k}}^{\,\mathbf{n}}
+
\textcolor{BrickRed}{A_{\Phi'}}
\Phi_{\mathbf{i},\mathbf{j},\mathbf{k}}^{\prime\,\mathbf{n}}
+
\textcolor{BrickRed}{A_{\delta P_{\rm m}}}
\delta P_{\mathbf{i},\mathbf{j},\mathbf{k}}^{\,\mathbf{n}}
\\
&
+
\textcolor{BrickRed}{A_{\nabla^2\pi}}
\left(\nabla^2\pi\right)_{\mathbf{i},\mathbf{j},\mathbf{k}}^{\,\mathbf{n}}
+
\textcolor{BrickRed}{A_{\nabla^2\Phi}}
\left(\nabla^2\Phi\right)_{\mathbf{i},\mathbf{j},\mathbf{k}}^{\,\mathbf{n}}
+
\textcolor{BrickRed}{A_{\nabla^2\Psi}}
\left(\nabla^2\Psi\right)_{\mathbf{i},\mathbf{j},\mathbf{k}}^{\,\mathbf{n}}
\bigg] \,.
\label{eq:zetanprime}
\end{aligned}
\end{equation}
All coefficients appearing in Eq.~\eqref{eq:zetanprime} are evaluated at the corresponding integer time cycle $\tau^{\mathbf{n}}$.
The integer-cycle value of $\zeta$ is reconstructed from the two neighbouring half-integer cycle values as
\begin{equation}
    \zeta_{\mathbf{i},\mathbf{j},\mathbf{k}}^{\,\mathbf{n}}
    =
    \frac{
    \zeta_{\mathbf{i},\mathbf{j},\mathbf{k}}^{\,\mathbf{n}+\frac{1}{2}}
    +
    \zeta_{\mathbf{i},\mathbf{j},\mathbf{k}}^{\,\mathbf{n}-\frac{1}{2}}
    }{2}\,.
    \label{eq:zetan}
\end{equation}
Substituting Eq.~\eqref{eq:zetan} into Eq.~\eqref{eq:zetanprime}, and then inserting the result into Eq.~\eqref{zetaleap}, gives the explicit update
\begin{multline}
    \zeta_{\mathbf{i},\mathbf{j},\mathbf{k}}^{\,\mathbf{n}+\frac{1}{2}}
    =
    \frac{1}{
    1+
    \left(
    \frac{\textcolor{BrickRed}{A_{\zeta}}}
         {\textcolor{BrickRed}{A_{\zeta'}}}
    \right)
    \left(
    \frac{\Delta\tau}{2}
    \right)
    }
    \Bigg[
    \zeta_{\mathbf{i},\mathbf{j},\mathbf{k}}^{\,\mathbf{n}-\frac{1}{2}}
    -
    \frac{\Delta\tau}{
    \textcolor{BrickRed}{A_{\zeta'}}
    }
    \Bigg(
    \frac{
    \textcolor{BrickRed}{A_{\zeta}}
    \zeta_{\mathbf{i},\mathbf{j},\mathbf{k}}^{\,\mathbf{n}-\frac{1}{2}}
    }{2}
    +
    \textcolor{BrickRed}{A_{\pi}}
    \pi_{\mathbf{i},\mathbf{j},\mathbf{k}}^{\,\mathbf{n}}
    +
    \textcolor{BrickRed}{A_{\Psi}}
    \Psi_{\mathbf{i},\mathbf{j},\mathbf{k}}^{\,\mathbf{n}}
    +
    \textcolor{BrickRed}{A_{\Phi'}}
    \Phi_{\mathbf{i},\mathbf{j},\mathbf{k}}^{\prime\,\mathbf{n}}
    \\
    +
    \textcolor{BrickRed}{A_{\delta P_{\rm m}}}
    \delta P_{\mathbf{i},\mathbf{j},\mathbf{k}}^{\,\mathbf{n}}
    +
    \textcolor{BrickRed}{A_{\nabla^2\pi}}
    \left(\nabla^2\pi\right)_{\mathbf{i},\mathbf{j},\mathbf{k}}^{\,\mathbf{n}}
    +
    \textcolor{BrickRed}{A_{\nabla^2\Phi}}
    \left(\nabla^2\Phi\right)_{\mathbf{i},\mathbf{j},\mathbf{k}}^{\,\mathbf{n}}
    +
    \textcolor{BrickRed}{A_{\nabla^2\Psi}}
    \left(\nabla^2\Psi\right)_{\mathbf{i},\mathbf{j},\mathbf{k}}^{\,\mathbf{n}}
    \Bigg)
    \Bigg]\, .
    \label{eq:LFzeta}
\end{multline}
This update is implemented in the function \texttt{update\_zeta} in \texttt{gevolution.hpp}.

The update for $\pi$ follows from the definition of $\zeta$ in
Eq.~\eqref{eq:zeta_pi},
\begin{align}
    \pi^{\,\mathbf{n}+1}_{\mathbf{i},\mathbf{j},\mathbf{k}}
    &=
    \pi^{\,\mathbf{n}}_{\mathbf{i},\mathbf{j},\mathbf{k}}
    +
    \pi'^{\,\mathbf{n}+\frac{1}{2}}_{\mathbf{i},\mathbf{j},\mathbf{k}}
    \Delta\tau
    \nonumber\\
    &=
    \pi^{\,\mathbf{n}}_{\mathbf{i},\mathbf{j},\mathbf{k}}
    +
    \Bigg[
    \zeta^{\,\mathbf{n}+\frac{1}{2}}_{\mathbf{i},\mathbf{j},\mathbf{k}}
    +
    \Psi^{\,\mathbf{n}+\frac{1}{2}}_{\mathbf{i},\mathbf{j},\mathbf{k}}
    -
    \mathcal{H}^{\,\mathbf{n}+\frac{1}{2}}
    \frac{
        \pi^{\,\mathbf{n}+1}_{\mathbf{i},\mathbf{j},\mathbf{k}}
        +
        \pi^{\,\mathbf{n}}_{\mathbf{i},\mathbf{j},\mathbf{k}}
    }{2}
    \Bigg]
    \Delta\tau
    \nonumber\\
    &=
    \frac{1}
    {1+\frac{1}{2}\mathcal{H}^{\,\mathbf{n}+\frac{1}{2}}\Delta\tau}
    \Bigg[
        \pi^{\,\mathbf{n}}_{\mathbf{i},\mathbf{j},\mathbf{k}}
        +
        \Delta\tau
        \Bigg(
            \zeta^{\,\mathbf{n}+\frac{1}{2}}_{\mathbf{i},\mathbf{j},\mathbf{k}}
            +
            \Psi^{\,\mathbf{n}+\frac{1}{2}}_{\mathbf{i},\mathbf{j},\mathbf{k}}
            -
            \frac{1}{2}
            \mathcal{H}^{\,\mathbf{n}+\frac{1}{2}}
            \pi^{\,\mathbf{n}}_{\mathbf{i},\mathbf{j},\mathbf{k}}
        \Bigg)
    \Bigg]\,.
    \label{eq:LPpi}
\end{align}
Here $\pi'^{\,\mathbf{n}+\frac{1}{2}}_{\mathbf{i},\mathbf{j},\mathbf{k}}$ has been replaced using
Eq.~\eqref{eq:zeta_pi}, with the midpoint value of $\pi$ approximated as
\begin{equation}
    \pi^{\,\mathbf{n}+\frac{1}{2}}_{\mathbf{i},\mathbf{j},\mathbf{k}}
    =
    \frac{
        \pi^{\,\mathbf{n}+1}_{\mathbf{i},\mathbf{j},\mathbf{k}}
        +
        \pi^{\,\mathbf{n}}_{\mathbf{i},\mathbf{j},\mathbf{k}}
    }{2}\,.
\end{equation}
The midpoint metric potential entering Eq.~\eqref{eq:LPpi} is evaluated
directly in the numerical implementation as
\begin{equation}
    \Psi^{\,\mathbf{n}+\frac{1}{2}}_{\mathbf{i},\mathbf{j},\mathbf{k}}
    =
    \Phi^{\,\mathbf{n}}_{\mathbf{i},\mathbf{j},\mathbf{k}}
    -
    \chi^{\,\mathbf{n}}_{\mathbf{i},\mathbf{j},\mathbf{k}}
    +
    \frac{\Delta\tau}{2}
    \left(\Psi'\right)^{\mathbf{n}}_{\mathbf{i},\mathbf{j},\mathbf{k}}\,.
    \label{eq:Psi_half}
\end{equation}

Equations~\eqref{eq:LPpi} and \eqref{eq:Psi_half} are implemented in the
function \texttt{update\_pi} in \texttt{gevolution.hpp}.

\subsection{Metric solver in Fourier space}

After the real-space sources have been constructed, the metric equations are solved in Fourier space following the strategy of \gev. The solver is based on two core equations: the Hamiltonian constraint, which determines the scalar potential $\Phi$, and the trace-free part of the spatial Einstein equations, which determines the gravitational slip $\chi=\Phi-\Psi$, the transverse vector $B_i$, and the transverse-traceless tensor $h_{ij}$.

For clarity, we first summarise the corresponding \gev metric solver and then introduce the modifications required in \mer. We follow this structure for both the Hamiltonian constraint and the trace-free spatial equations, which makes the Horndeski contributions transparent relative to the baseline relativistic $N$-body scheme.

\subsubsection*{Hamiltonian constraint: \gev scheme}

We use the exponential parametrisation of the scalar metric perturbations adopted in the current \gev formulation. This convention agrees with the original metric parametrisation at linear order but differs from it at higher order. In the weak-field expansion used here, the Hamiltonian constraint is written as
\begin{equation}
    \left(
        1+2\Phi
    \right)
    \Delta\Phi
    -
    3\mathcal{H}\Phi'
    +
    3\mathcal{H}^2
    \left(
        \chi-\Phi
    \right)
    -
    \frac{1}{2}
    \delta^{ij}
    \Phi_{,i}
    \Phi_{,j}
    =
    -
    \frac{a^2}{2M_{\rm Pl}^2}
    \delta T^0{}_{0}\, .
    \label{eq:gevolution_hamiltonian_constraint}
\end{equation}
Here $\delta T^0{}_{0}\equiv T^0{}_{0}-\bar T^0{}_{0}$ denotes the perturbation of the total energy density component. Equation~\eqref{eq:gevolution_hamiltonian_constraint} differs from the purely linear Hamiltonian constraint used in Sec.~\ref{sec:Horndeski_gravity} by the weak-field terms proportional to $\Phi\Delta\Phi$ and $\delta^{ij}\Phi_{,i}\Phi_{,j}$. These terms are retained in \gev\ because the scalar potentials remain small, while their spatial derivatives can become large in the nonlinear regime.

In the numerical implementation, Eq.~\eqref{eq:gevolution_hamiltonian_constraint} is not solved directly as a nonlinear parabolic equation. Instead, following the Fourier-space solver of \gev, it is rearranged into a linear finite-difference equation for the updated potential $\Phi^{\mathbf{n}+1}_{\mathbf{i},\mathbf{j},\mathbf{k}}$. This is done by dividing by the factor $1+2\Phi$ and expanding $(1+2\Phi)^{-1}\simeq 1-2\Phi$ to the order retained in the weak-field scheme. The Laplacian and time-derivative terms contain the updated potential $\Phi^{\mathbf{n}+1}$, while the nonlinear geometric corrections and the gravitational slip are evaluated using the known fields at cycle $\mathbf{n}$\footnote{This explicit treatment assumes that the metric potentials vary only weakly over a single time step. Since $\Phi^{\mathbf{n}+1}-\Phi^{\mathbf{n}}=(\Phi')^{\mathbf{n}}\Delta\tau+\mathcal{O}(\Delta\tau^2)$, using the previous-cycle potential in these terms introduces a time lag of order $(\Phi')^{\mathbf{n}}\Delta\tau$ at the field level, with an analogous estimate for the gravitational slip $\chi$.}. The lattice version of the Hamiltonian constraint can therefore be written as
\begin{multline}
\frac{
\Phi^{\mathbf{n}+1}_{\mathbf{i}-1,\mathbf{j},\mathbf{k}}
+
\Phi^{\mathbf{n}+1}_{\mathbf{i}+1,\mathbf{j},\mathbf{k}}
+
\Phi^{\mathbf{n}+1}_{\mathbf{i},\mathbf{j}-1,\mathbf{k}}
+
\Phi^{\mathbf{n}+1}_{\mathbf{i},\mathbf{j}+1,\mathbf{k}}
+
\Phi^{\mathbf{n}+1}_{\mathbf{i},\mathbf{j},\mathbf{k}-1}
+
\Phi^{\mathbf{n}+1}_{\mathbf{i},\mathbf{j},\mathbf{k}+1}
-
6\Phi^{\mathbf{n}+1}_{\mathbf{i},\mathbf{j},\mathbf{k}}
}{
\mathrm{dx}^{2}
}
-
3\mathcal{H}
\frac{
\Phi^{\mathbf{n}+1}_{\mathbf{i},\mathbf{j},\mathbf{k}}
-
\Phi^{\mathbf{n}}_{\mathbf{i},\mathbf{j},\mathbf{k}}
}{
\mathrm{d}\tau
}\\
-
3\mathcal{H}^{2}
\left(
\Phi^{\mathbf{n}}_{\mathbf{i},\mathbf{j},\mathbf{k}}
-
\chi^{\mathbf{n}}_{\mathbf{i},\mathbf{j},\mathbf{k}}
\right)
-
\frac{1}{2}
\frac{
\left(
\Phi^{\mathbf{n}}_{\mathbf{i}+1,\mathbf{j},\mathbf{k}}
-
\Phi^{\mathbf{n}}_{\mathbf{i}-1,\mathbf{j},\mathbf{k}}
\right)^{2}
+
\left(
\Phi^{\mathbf{n}}_{\mathbf{i},\mathbf{j}+1,\mathbf{k}}
-
\Phi^{\mathbf{n}}_{\mathbf{i},\mathbf{j}-1,\mathbf{k}}
\right)^{2}
+
\left(
\Phi^{\mathbf{n}}_{\mathbf{i},\mathbf{j},\mathbf{k}+1}
-
\Phi^{\mathbf{n}}_{\mathbf{i},\mathbf{j},\mathbf{k}-1}
\right)^{2}
}{
4\mathrm{dx}^{2}
}
\\
=
-
\frac{a^{2}}{2M_{\rm Pl}^{2}}
\left(
1
-
2\Phi^{\mathbf{n}}_{\mathbf{i},\mathbf{j},\mathbf{k}}
\right)
\delta T^{0}{}_{0}
\left(
\tau_{\mathbf{n}+1},
\mathbf{x}_{\mathbf{i},\mathbf{j},\mathbf{k}}
\right)\,.
\label{eq:gevolution_hamiltonian_lattice}
\end{multline}

To solve Eq.~\eqref{eq:gevolution_hamiltonian_lattice} numerically, \gev\ rewrites the constraint as an equation for the new grid field $\Phi^{\mathbf{n}+1}$. The terms involving $\Phi^{\mathbf{n}+1}$ are kept as the left-hand side of the update equation. The remaining metric terms are evaluated explicitly using the fields at cycle $\mathbf{n}$, while the total stress--energy perturbation is constructed at $\tau_{\mathbf{n}+1}$. Moving these known terms to the right-hand side, the source is defined as
\begin{multline}
s_{\mathbf{i},\mathbf{j},\mathbf{k}}
\equiv
3\mathcal{H}^{2}
\left(
\Phi^{\mathbf{n}}_{\mathbf{i},\mathbf{j},\mathbf{k}}
-
\chi^{\mathbf{n}}_{\mathbf{i},\mathbf{j},\mathbf{k}}
\right)
+
\frac{1}{8\mathrm{dx}^{2}}
\left(
\Phi^{\mathbf{n}}_{\mathbf{i}+1,\mathbf{j},\mathbf{k}}
-
\Phi^{\mathbf{n}}_{\mathbf{i}-1,\mathbf{j},\mathbf{k}}
\right)^{2}
+
\frac{1}{8\mathrm{dx}^{2}}
\left(
\Phi^{\mathbf{n}}_{\mathbf{i},\mathbf{j}+1,\mathbf{k}}
-
\Phi^{\mathbf{n}}_{\mathbf{i},\mathbf{j}-1,\mathbf{k}}
\right)^{2}
\\
+
\frac{1}{8\mathrm{dx}^{2}}
\left(
\Phi^{\mathbf{n}}_{\mathbf{i},\mathbf{j},\mathbf{k}+1}
-
\Phi^{\mathbf{n}}_{\mathbf{i},\mathbf{j},\mathbf{k}-1}
\right)^{2}
-
\frac{3\mathcal{H}}{\mathrm{d}\tau}
\Phi^{\mathbf{n}}_{\mathbf{i},\mathbf{j},\mathbf{k}}
-
\frac{a^{2}}{2M_{\rm Pl}^{2}}
\left(
1
-
2\Phi^{\mathbf{n}}_{\mathbf{i},\mathbf{j},\mathbf{k}}
\right)
\delta T^{0}{}_{0}
\left(
\tau_{\mathbf{n}+1},
\mathbf{x}_{\mathbf{i},\mathbf{j},\mathbf{k}}
\right)\,.
\label{eq:source_phi_lattice}
\end{multline}

The Hamiltonian constraint is therefore reduced to the linear implicit equation for $\Phi^{\mathbf{n}+1}$
\begin{equation}
\frac{
\Phi^{\mathbf{n}+1}_{\mathbf{i}-1,\mathbf{j},\mathbf{k}}
+
\Phi^{\mathbf{n}+1}_{\mathbf{i}+1,\mathbf{j},\mathbf{k}}
+
\Phi^{\mathbf{n}+1}_{\mathbf{i},\mathbf{j}-1,\mathbf{k}}
+
\Phi^{\mathbf{n}+1}_{\mathbf{i},\mathbf{j}+1,\mathbf{k}}
+
\Phi^{\mathbf{n}+1}_{\mathbf{i},\mathbf{j},\mathbf{k}-1}
+
\Phi^{\mathbf{n}+1}_{\mathbf{i},\mathbf{j},\mathbf{k}+1}
-
6\Phi^{\mathbf{n}+1}_{\mathbf{i},\mathbf{j},\mathbf{k}}
}{
\mathrm{dx}^{2}
}
-
\frac{3\mathcal{H}}{\mathrm{d}\tau}
\Phi^{\mathbf{n}+1}_{\mathbf{i},\mathbf{j},\mathbf{k}}
=
s_{\mathbf{i},\mathbf{j},\mathbf{k}} \,.
\label{eq:phi_linear_lattice}
\end{equation}

Because the grid is periodic, the Fourier transform is the natural way to solve this equation. It rewrites the grid field as a sum of Fourier modes, and each mode can then be solved separately for $\Phi^{\mathbf{n}+1}$. The source field $s_{\mathbf{i},\mathbf{j},\mathbf{k}}$ is therefore first transformed to Fourier space. For a grid with $N$ points in each spatial direction, the transform is defined as
\begin{equation}
\tilde{s}_{\mathbf{u},\mathbf{v},\mathbf{w}}
\equiv
\sum_{\mathbf{i},\mathbf{j},\mathbf{k}}
s_{\mathbf{i},\mathbf{j},\mathbf{k}}
\,e^{
-2\pi \mathrm{i}
\left(
\mathbf{i}\mathbf{u}
+
\mathbf{j}\mathbf{v}
+
\mathbf{k}\mathbf{w}
\right)/N
}\, .
\label{eq:source_phi_fourier_transform}
\end{equation}
Here $\mathbf{i},\mathbf{j},\mathbf{k}$ label the real-space lattice point, while $\mathbf{u},\mathbf{v},\mathbf{w}$ label the Fourier mode. The full lattice contains $N^3$ grid points.

Applying the same Fourier transform to Eq.~\eqref{eq:phi_linear_lattice}, the finite-difference equation becomes an algebraic equation for each Fourier mode,
\begin{equation}
-
\left[
\frac{4}{\mathrm{dx}^{2}}
\sin^{2}
\left(
\frac{\pi\mathbf{u}}{N}
\right)
+
\frac{4}{\mathrm{dx}^{2}}
\sin^{2}
\left(
\frac{\pi\mathbf{v}}{N}
\right)
+
\frac{4}{\mathrm{dx}^{2}}
\sin^{2}
\left(
\frac{\pi\mathbf{w}}{N}
\right)
+
\frac{3\mathcal{H}}{\mathrm{d}\tau}
\right]
\tilde{\Phi}^{\mathbf{n}+1}_{\mathbf{u},\mathbf{v},\mathbf{w}}
=
\tilde{s}_{\mathbf{u},\mathbf{v},\mathbf{w}} \,.
\label{eq:phi_fourier_hamiltonian}
\end{equation}
This equation contains only one unknown quantity for each mode, namely $\tilde{\Phi}^{\mathbf{n}+1}_{\mathbf{u},\mathbf{v},\mathbf{w}}$. The updated potential is therefore obtained by dividing the transformed source $\tilde{s}_{\mathbf{u},\mathbf{v},\mathbf{w}}$ by the factor multiplying $\tilde{\Phi}^{\mathbf{n}+1}_{\mathbf{u},\mathbf{v},\mathbf{w}}$.

\subsubsection*{Hamiltonian constraint: \mer modifications}

In \mer, the numerical rearrangement of the Hamiltonian constraint and the Fourier-space inversion remain unchanged. The modification enters through the source term. In addition to the matter contribution already present in \gev, the scalar-field contribution constructed on the real-space lattice, Eq.~\eqref{eq:deltarho_DE}, is added to the source. At the same time, the constant Planck mass $M_{\rm Pl}^2$ appearing in the \gev source is replaced by the time-dependent cosmological Planck mass $M^2_{\rm eff}$. This gives
\begin{equation}
-\frac{a^{2}}{2M_{\rm Pl}^{2}}
\left(
1
-
2\Phi^{\mathbf{n}}_{\mathbf{i},\mathbf{j},\mathbf{k}}
\right)
\delta T^{0}_{0}
\;\longrightarrow\;
-\frac{a^{2}}{2M^2_{\rm eff} }
\left(
1
-
2\Phi^{\mathbf{n}}_{\mathbf{i},\mathbf{j},\mathbf{k}}
\right)
\left[
\delta T^{0(\rm m)}_{0}
+
\delta T^{0(\phi)}_{0}
\right]\, .
\label{eq:source_replacement_mercury}
\end{equation}

\subsubsection*{Trace-free equation: \gev scheme}

After updating $\Phi$, the remaining metric perturbations are obtained from the trace-free part of the spatial Einstein equations. In the \gev implementation used here, these equations take the form 
\begin{align}
\left(
\delta^i_k\delta^j_l
-
\frac{1}{3}
\delta^{ij}\delta_{kl}
\right)
\bigg[
\frac{1}{2}h''_{ij}
+
\mathcal{H}h'_{ij}
-
\frac{1}{2}\nabla^2 h_{ij}
&+
B'_{(i,j)}
+
2\mathcal{H}B_{(i,j)}
+
\chi_{,ij}
-
2\Phi_{,i}\Phi_{,j}
\bigg]
\nonumber\\
&=
\frac{a^2}{M_{\rm Pl}^2}
\left(
\delta^i_k\delta^j_l
-
\frac{1}{3}
\delta^{ij}\delta_{kl}
\right)
\delta_{mi}T^m{}_j
=
\frac{a^2}{M_{\rm Pl}^2}
\Pi_{kl}\,.
\label{eq:gevolution_traceless_ij}
\end{align}

Equation~\eqref{eq:gevolution_traceless_ij} already incorporates the approximation adopted in the solver. A more complete weak-field expression, retaining all second-order terms in the scalar metric potentials, is given in \cite{Adamek:2026vof}. When written in terms of $\chi=\Phi-\Psi$, this expression contains additional nonlinear terms involving $\chi$, which are neglected here, while the linear contribution $\chi_{,ij}$ and the quadratic term in $\Phi$ are retained. These omitted terms are second order in the metric perturbations and are subdominant when the anisotropic stress provides the leading source of the trace-free metric perturbations \cite{Adamek:2026vof}.

Following the same strategy as for the Hamiltonian constraint, the nonlinear term built from the already known scalar potential is moved to the source. Equation~\eqref{eq:gevolution_traceless_ij} can therefore be rearranged as
\begin{align}
\left(
\delta^i_k\delta^j_l
-
\frac{1}{3}
\delta^{ij}\delta_{kl}
\right)
\bigg[
\frac{1}{2}h''_{ij}
+
\mathcal{H}h'_{ij}
-
\frac{1}{2}\nabla^2 h_{ij}
+
B'_{(i,j)}
+
2\mathcal{H}B_{(i,j)}
+
\chi_{,ij}
\bigg]\nonumber
&\\=
\left(
\delta^i_k\delta^j_l
-
\frac{1}{3}
\delta^{ij}\delta_{kl}
\right)
\bigg[
\frac{a^2}{M_{\rm Pl}^2}
\delta_{mi}T^m{}_{j}
+
2\Phi_{,i}\Phi_{,j}
\bigg]\, .
\label{eq:gevolution_traceless_rearranged}
\end{align}
The quantity inside the second set of brackets is then identified as the source tensor,
\begin{equation}
S_{ij}
\equiv
\frac{a^2}{M_{\rm Pl}^2}
\delta_{mi}T^m{}_{j}
+
2\Phi_{,i}\Phi_{,j}\,.
\label{eq:Sij_continuum_gevolution}
\end{equation}

The source tensor $S_{ij}$ must be constructed at the correct positions on the lattice. Scalar quantities such as $\Phi$ are stored at the vertices of the grid, and the diagonal components of a spatial tensor are also evaluated at the vertices. The components of vector fields are associated with the edges, while the off-diagonal components of a spatial tensor are evaluated on the faces of the lattice cells. The finite-difference operators used in $S_{ij}$ are therefore chosen so that all terms contributing to a given component are evaluated at the same lattice position.

For a diagonal component, all terms are evaluated at the vertex $\mathbf{x}_{\mathbf{i},\mathbf{j},\mathbf{k}}$. For example,
\begin{align}
S_{11}
\left(
\mathbf{x}_{\mathbf{i},\mathbf{j},\mathbf{k}}
\right)
\equiv\;&
\frac{a^2}{M_{\rm Pl}^2}
T^{1}{}_{1}
\left(
\tau_{\mathbf{n}+1},
\mathbf{x}_{\mathbf{i},\mathbf{j},\mathbf{k}}
\right)
+
\frac{
\left(
\Phi^{\mathbf{n}+1}_{\mathbf{i}+1,\mathbf{j},\mathbf{k}}
-
\Phi^{\mathbf{n}+1}_{\mathbf{i}-1,\mathbf{j},\mathbf{k}}
\right)^2
}{
2\mathrm{dx}^{2}
}\,.
\label{eq:S11_lattice_gevolution}
\end{align}
The second term is the lattice representation of $2\Phi_{,1}\Phi_{,1}$ at the same vertex.

For an off-diagonal component, the source is evaluated at the corresponding face of the lattice cell. For example, $S_{12}$ is evaluated at $\mathbf{x}_{\mathbf{i}+\frac{1}{2},\mathbf{j}+\frac{1}{2},\mathbf{k}}$, and is written as
\begin{align}
&S_{12}
\left(
\mathbf{x}_{\mathbf{i}+\frac{1}{2},\mathbf{j}+\frac{1}{2},\mathbf{k}}
\right)
\equiv
\frac{a^2}{M_{\rm Pl}^2}
T^{1}{}_{2}
\left(
\tau_{\mathbf{n}+1},
\mathbf{x}_{\mathbf{i}+\frac{1}{2},\mathbf{j}+\frac{1}{2},\mathbf{k}}
\right)
\nonumber\\
&\quad
+
\frac{1}{2}
\frac{
\Phi^{\mathbf{n}+1}_{\mathbf{i}+1,\mathbf{j}+1,\mathbf{k}}
-
\Phi^{\mathbf{n}+1}_{\mathbf{i},\mathbf{j}+1,\mathbf{k}}
+
\Phi^{\mathbf{n}+1}_{\mathbf{i}+1,\mathbf{j},\mathbf{k}}
-
\Phi^{\mathbf{n}+1}_{\mathbf{i},\mathbf{j},\mathbf{k}}
}{
\mathrm{dx}
}
\frac{
\Phi^{\mathbf{n}+1}_{\mathbf{i}+1,\mathbf{j}+1,\mathbf{k}}
-
\Phi^{\mathbf{n}+1}_{\mathbf{i}+1,\mathbf{j},\mathbf{k}}
+
\Phi^{\mathbf{n}+1}_{\mathbf{i},\mathbf{j}+1,\mathbf{k}}
-
\Phi^{\mathbf{n}+1}_{\mathbf{i},\mathbf{j},\mathbf{k}}
}{
\mathrm{dx}
}\,.
\label{eq:S12_lattice_gevolution}
\end{align}
The second term in Eq.~\eqref{eq:S12_lattice_gevolution} represents $2\Phi_{,1}\Phi_{,2}$, with both gradients evaluated at the same face. The other components of $S_{ij}$ are constructed in the same way.

With this definition, the trace-free equation becomes
\begin{equation}
\begin{aligned}
\frac{1}{2}h''_{ij}
+
\mathcal{H}h'_{ij}
-
\frac{1}{2}\nabla^2 h_{ij}
+
B'_{(i,j)}
+
2\mathcal{H}B_{(i,j)}
+
\chi_{,ij}
-
\frac{1}{3}\delta_{ij}\nabla^2\chi
=
S_{ij}
-
\frac{1}{3}
\delta_{ij}
\delta^{kl}S_{kl} \,.
\end{aligned}
\label{eq:gevolution_tracefree_spatial_source}
\end{equation}
The trace-free spatial equation is solved in Fourier space by decomposing the source tensor into its scalar, vector and tensor parts. We first write the projection formulae in continuum Fourier notation. Defining the transverse projector
\begin{equation}
P^i{}_{j}
\equiv
\delta^i{}_{j}-\frac{k^i k_j}{k^2} \,,
\label{eq:fourier_transverse_projector}
\end{equation}
the scalar projection gives the gravitational slip,
\begin{equation}
\tilde\chi(\mathbf{k})
=
\frac{1}{2k^2}
\left(
\delta^{ij}
-
3
\frac{k^i k^j}{k^2}
\right)
\tilde S_{ij}(\mathbf{k})\, .
\label{eq:chi_fourier_projection_gevolution}
\end{equation}
The vector projection gives the transverse frame-dragging equation,
\begin{equation}
\tilde B'_i(\mathbf{k})
+
2\mathcal{H}
\tilde B_i(\mathbf{k})
=
\frac{1}{a^2}
\left(
a^2
\tilde B_i(\mathbf{k})
\right)'
=
-
\frac{2\mathrm{i}}{k^2}
P_i{}^j
k^l
\tilde S_{jl}(\mathbf{k})\, .
\label{eq:Bi_fourier_projection_gevolution}
\end{equation}
Finally, the tensor projection gives the transverse-traceless tensor equation,
\begin{equation}
\tilde h''_{ij}(\mathbf{k})
+
2\mathcal{H}
\tilde h'_{ij}(\mathbf{k})
+
k^2
\tilde h_{ij}(\mathbf{k})
=
2
\left(
P_i{}^l P_j{}^m
-
\frac{1}{2}
P_{ij}P^{lm}
\right)
\tilde S_{lm}(\mathbf{k}) \,.
\label{eq:hij_fourier_projection_gevolution}
\end{equation}

On the lattice these equations are implemented using discrete Fourier modes. For diagonal components, which are evaluated at the vertices, the Fourier transform is
\begin{equation}
\tilde S_{11}
\left(
\mathbf{u},
\mathbf{v},
\mathbf{w}
\right)
\equiv
\sum_{\mathbf{i},\mathbf{j},\mathbf{k}}
S_{11}
\left(
\mathbf{x}_{\mathbf{i},\mathbf{j},\mathbf{k}}
\right)
e^{
-2\pi\mathrm{i}
\left(
\mathbf{i}\mathbf{u}
+
\mathbf{j}\mathbf{v}
+
\mathbf{k}\mathbf{w}
\right)/N
} \,.
\label{eq:S11_fourier_gevolution}
\end{equation}
For off-diagonal components, which are evaluated on the faces of the lattice cells, an additional phase factor accounts for the half-cell displacement. For example,
\begin{equation}
\tilde S_{12}
\left(
\mathbf{u},
\mathbf{v},
\mathbf{w}
\right)
\equiv
e^{-\pi\mathrm{i}\mathbf{u}/N}
e^{-\pi\mathrm{i}\mathbf{v}/N}
\sum_{\mathbf{i},\mathbf{j},\mathbf{k}}
S_{12}
\left(
\mathbf{x}_{\mathbf{i}+\frac{1}{2},\mathbf{j}+\frac{1}{2},\mathbf{k}}
\right)
e^{
-2\pi\mathrm{i}
\left(
\mathbf{i}\mathbf{u}
+
\mathbf{j}\mathbf{v}
+
\mathbf{k}\mathbf{w}
\right)/N
} \,.
\label{eq:S12_fourier_gevolution}
\end{equation}
The remaining components are transformed analogously.

Following \gev, the lattice momentum associated with the mode
$\left(\mathbf{u},\mathbf{v},\mathbf{w}\right)$ is defined as
\begin{equation}
\left(
k^1_{\mathbf{u},\mathbf{v},\mathbf{w}},
k^2_{\mathbf{u},\mathbf{v},\mathbf{w}},
k^3_{\mathbf{u},\mathbf{v},\mathbf{w}}
\right)
\equiv
\left(
\frac{2}{\mathrm{dx}}
\sin
\frac{\pi\mathbf{u}}{N},
\frac{2}{\mathrm{dx}}
\sin
\frac{\pi\mathbf{v}}{N},
\frac{2}{\mathrm{dx}}
\sin
\frac{\pi\mathbf{w}}{N}
\right) \,.
\label{eq:lattice_momentum_gevolution}
\end{equation}
The discrete version of Eqs.~\eqref{eq:chi_fourier_projection_gevolution}--\eqref{eq:hij_fourier_projection_gevolution} is then obtained by making the replacements
\begin{equation}
\tilde S_{ij}(\mathbf{k})
\rightarrow
\tilde S_{ij}
\left(\mathbf{u},\mathbf{v},\mathbf{w}\right)\,,
\qquad
k^i\rightarrow k^i_{\mathbf{u},\mathbf{v},\mathbf{w}}\,,
\qquad
k^2
\rightarrow
\delta_{ij}
k^i_{\mathbf{u},\mathbf{v},\mathbf{w}}
k^j_{\mathbf{u},\mathbf{v},\mathbf{w}}\, .
\label{eq:gevolution_lattice_replacement_rule}
\end{equation}
Thus the same Fourier-space source tensor
$\tilde S_{ij}\left(\mathbf{u},\mathbf{v},\mathbf{w}\right)$
is used to obtain $\chi$, $B_i$ and $h_{ij}$ through the scalar, vector and tensor projections.

\subsubsection*{Trace-free equation: \mer modifications}

In \mer, the scalar, vector and tensor projectors retain the same form as in \gev. The modifications enter through the source tensor and through the time-dependent coefficients governing the vector and tensor evolution. The source tensor becomes
\begin{equation}
S_{ij}
\;\longrightarrow\;
S^{\mer}_{ij}
\equiv
\frac{a^2}{M^2_{\rm eff} }
\left[
\delta_{mi}T^{m(\rm m)}_{j}
+
\Pi^{(\phi)}_{ij}
\right]
+
2\Phi_{,i}\Phi_{,j}\,.
\label{eq:Sij_mercury}
\end{equation}

The Horndeski anisotropic stress $\Pi^{(\phi)}_{ij}$ is trace-free and has purely scalar form, being proportional to $\mathcal D_{ij}\equiv\partial_i\partial_j-\frac{1}{3}\delta_{ij}\nabla^2$ acting on a scalar quantity. It therefore contributes directly to the scalar projection determining the gravitational slip, while its vector and tensor projections vanish at linear order. The scalar projection consequently keeps the same algebraic form as in \gev,
\begin{equation}
\tilde\chi(\mathbf{k})
=
\frac{1}{2k^2}
\left(
\delta^{ij}
-
3
\frac{k^i k^j}{k^2}
\right)
\tilde S^{\mer}_{ij}(\mathbf{k})\, .
\label{eq:chi_fourier_projection_mercury}
\end{equation}

For the vector mode, the running of the effective Planck mass modifies the damping term. Using $\alpha_{\rm M}=(M_*^2)'/(\mathcal H M_*^2)$, the vector equation becomes
\begin{equation}
\frac{1}{a^2M_*^2}
\left(
a^2M_*^2
\tilde B_i(\mathbf{k})
\right)'
=
\tilde B'_i(\mathbf{k})
+
\left(
2+\alpha_{\rm M}
\right)
\mathcal{H}
\tilde B_i(\mathbf{k})
=
-
\frac{2\mathrm{i}}{k^2}
P_i{}^j
k^l
\tilde S^{\mer}_{jl}(\mathbf{k})\, .
\label{eq:Bi_fourier_projection_mercury}
\end{equation}

Similarly, the tensor equation is modified by both the running of the effective Planck mass and the tensor-speed excess,
\begin{equation}
\frac{1}{a^2M_*^2}
\left(
a^2M_*^2
\tilde h'_{ij}(\mathbf{k})
\right)'
+
\left(
1+\alpha_{\rm T}
\right)
k^2
\tilde h_{ij}(\mathbf{k})
=
2
\left(
P_i{}^l P_j{}^m
-
\frac{1}{2}
P_{ij}P^{lm}
\right)
\tilde S^{\mer}_{lm}(\mathbf{k})\, .
\label{eq:hij_fourier_projection_mercury}
\end{equation}
Equivalently, its left-hand side is
\begin{equation}
\tilde h''_{ij}(\mathbf{k})
+
\left(
2+\alpha_{\rm M}
\right)
\mathcal{H}
\tilde h'_{ij}(\mathbf{k})
+
\left(
1+\alpha_{\rm T}
\right)
k^2
\tilde h_{ij}(\mathbf{k}) \, .
\end{equation}

The discrete implementation follows the same Fourier transforms, lattice momenta and projection operators as in \gev, with $\tilde S_{ij}$ replaced by $\tilde S^{\mer}_{ij}$.

\newpage
\section{Numerical workflow of \mer}
\label{sec:mercury_workflow}

\begin{figure*}[h]
\centering
\resizebox{\textwidth}{!}{%
\begin{tikzpicture}[
    >=Latex,
    font=\large,
    arrow/.style={
        ->,
        thick,
        draw=merdarkgray
    },
    dashedarrow/.style={
        ->,
        thick,
        dashed,
        draw=merdarkgray
    },
    modebox/.style={
        draw=merdarkblue,
        fill=merblue,
        rounded corners=3pt,
        thick,
        align=center,
        text width=5.2cm,
        minimum height=1.9cm,
        drop shadow={
            shadow xshift=4.5pt,
            shadow yshift=-4.5pt,
            opacity=0.3
        }
    },
    processbox/.style={
        draw=merdarkblue,
        fill=merblue,
        rounded corners=3pt,
        thick,
        align=center,
        text width=6cm,
        minimum height=1.05cm,
        drop shadow={
            shadow xshift=4.5pt,
            shadow yshift=-4.5pt,
            opacity=0.3
        }
    },
    initbox/.style={
        draw=merdarkorange,
        fill=merorange,
        rounded corners=3pt,
        thick,
        align=center,
        text width=8.3cm,
        minimum height=1.05cm,
        drop shadow={
            shadow xshift=4.5pt,
            shadow yshift=-4.5pt,
            opacity=0.3
        }
    },
    stepbox/.style={
        draw=merdarkgray,
        fill=white,
        rounded corners=3pt,
        align=left,
        text width=13cm,
        minimum height=0.75cm,
        inner xsep=8pt,
        inner ysep=4pt,
        drop shadow={
            shadow xshift=4.5pt,
            shadow yshift=-4.5pt,
            opacity=0.3
        }
    },
    number/.style={
        circle,
        draw=teal!70!black,
        fill=teal!20,
        thick,
        inner sep=1.3pt,
        minimum size=0.55cm,
        font=\scriptsize\bfseries,
        drop shadow={
            shadow xshift=4.5pt,
            shadow yshift=-4.5pt,
            opacity=0.3
        }
    },
    loopframe/.style={
        draw=teal!70!black,
        fill=teal!10,
        rounded corners=5pt,
        thick,
        inner sep=10pt,
        drop shadow={
            shadow xshift=4.5pt,
            shadow yshift=-4.5pt,
            opacity=0.3
        }
    },
    outputbox/.style={
        draw=teal!70!black,
        fill=teal!18,
        rounded corners=3pt,
        thick,
        align=center,
        text width=9cm,
        minimum height=1.05cm,
        drop shadow={
            shadow xshift=4.5pt,
            shadow yshift=-4.5pt,
            opacity=0.3
        }
    },
    stage1label/.style={
        font=\bfseries\large,
        text=merdarkblue,
        align=center
    },
    stage2label/.style={
        font=\bfseries\large,
        text=merdarkorange,
        align=center
    },
    stage3label/.style={
        font=\bfseries\large,
        text=teal!70!black,
        align=center
    },
    stage1dash/.style={
        dashed,
        thick,
        draw=merdarkblue
    },
    stage2dash/.style={
        dashed,
        thick,
        draw=merdarkorange
    },
    stage3dash/.style={
        dashed,
        thick,
        draw=teal!70!black
    }
]
%
%
\def\loopgap{0.5cm}
\def\initialloopgap{2.0cm}
%
%
\node[modebox] (mode) at (0,0) {
\ \textbf{Input mode}\\[2pt]
$\left\{
\begin{array}{l}
\text{EFT / }\alpha\text{ parametrisation} \\
\text{Covariant action}
\end{array}
\right.$
};
\node[processbox] (hiclass) at (7.0,0) {
\ \textbf{\hiclass\ calculation}\\
background evolution, $M_*^2$, $\alpha_i$, and transfer functions
};
\node[processbox] (tables) at (14.3,0) {
\ \textbf{\mer\ interface}\\
$H(a)$, $\rho_\phi(a)$, $P_\phi(a)$, $M_*^2(a)$,
$\alpha_i(a)$, transfer functions
};
%
%
\node[initbox] (ic) at (7.0,-3.4) {
\ \textbf{Initial conditions at $z_{\rm ini}$}\\
construct initial metric and scalar-field perturbations,
generate particle realisation
};
%
%
\node[stepbox, below=\initialloopgap of ic] (s1) {
\textbf{Project matter sources:}
construct $T^{0\,(m)}_{0}$, $T^{0\,(m)}_{i}$,
and $T^{i\,(m)}_{j}$ on the lattice.
};
\node[number, left=0.25cm of s1] {1};
\node[stepbox, below=\loopgap of s1] (s2) {
\textbf{Project Horndeski sources:}
construct $T^{0\,(\phi)}_{0}$, $T^{0\,(\phi)}_{i}$,
and $\Pi^{(\phi)}_{ij}$ on the lattice.
};
\node[number, left=0.25cm of s2] {2};
\node[stepbox, below=\loopgap of s2] (s3) {
\textbf{Solve the Hamiltonian constraint:}
update $\Phi$ in Fourier space.
};
\node[number, left=0.25cm of s3] {3};
\node[stepbox, below=\loopgap of s3] (s4) {
\textbf{Solve the trace-free spatial equations:}
construct $S_{ij}$ and determine $\chi$, $B_i$, and $h_{ij}$
through scalar, vector, and tensor projections.
};
\node[number, left=0.25cm of s4] {4};
\node[stepbox, below=\loopgap of s4] (s5) {
\textbf{Update metric derivatives:}
compute $\Phi'$ and $\Psi'$ from the metric fields.
};
\node[number, left=0.25cm of s5] {5};
\node[stepbox, below=\loopgap of s5] (s6) {
\textbf{Write outputs when required:}
save snapshots, power spectra, lightcones, and diagnostic quantities.
};
\node[number, left=0.25cm of s6] {6};
\node[stepbox, below=\loopgap of s6] (s7) {
\textbf{Update scalar-field variables:}
advance $\zeta$ and $\pi$ with the leapfrog scheme.
};
\node[number, left=0.25cm of s7] {7};
\node[stepbox, below=\loopgap of s7] (s8) {
\textbf{Advance particles and background:}
update particle momenta and positions, advance the background,
and step to the next cycle.
};
\node[number, left=0.25cm of s8] {8};
%
%
\node[outputbox, below=1.0cm of s8] (out) {
\cmark\ \textbf{Simulation complete!}
};
%
%
\begin{scope}[on background layer]
\node[
    loopframe,
    fit=(s1)(s2)(s3)(s4)(s5)(s6)(s7)(s8)
] (loopframe) {};
\node[
    fit=(mode)(hiclass)(tables),
    inner sep=6pt
] (stageIgroup) {};
\node[
    fit=(ic),
    inner sep=6pt
] (stageIIgroup) {};
\end{scope}
%
%
\node[
    stage1label,
    anchor=north
] (stagelabel1)
at ($(mode.south)+(0,-1cm)$)
{$\bullet$ Stage I};
\draw[stage1dash]
    (mode.south)
    --
    (stagelabel1.north);
\node[
    stage2label,
    anchor=east
] (stagelabel2)
at ($(stageIIgroup.west)+(-1.4cm,0)$)
{$\bullet$ Stage II};
\draw[stage2dash]
    (stagelabel2.east)
    --
    (stageIIgroup.west);
\coordinate (stageIIIanchor)
    at ($(loopframe.north west)+(0.2cm,0)$);
\node[
    stage3label,
    anchor=south
] (stagelabel3)
at ($(stageIIIanchor)+(0,1cm)$)
{$\bullet$ Stage III};
\draw[stage3dash]
    (stagelabel3.south)
    --
    (stageIIIanchor);
%
%
\coordinate (tablesdrop)
    at ($(tables.south)+(0,-0.6cm)$);
\coordinate (icentry)
    at ($(ic.north)+(0,0.85cm)$);
%
%
\draw[arrow]
    (mode.east)
    --
    (hiclass.west);
\draw[arrow]
    (hiclass.east)
    --
    (tables.west);
\draw[arrow]
    (tables.south)
    --
    (tablesdrop)
    -|
    (icentry)
    --
    (ic.north);
\draw[arrow]
    (ic.south)
    --
    (loopframe.north);
\draw[arrow] (s1.south) -- (s2.north);
\draw[arrow] (s2.south) -- (s3.north);
\draw[arrow] (s3.south) -- (s4.north);
\draw[arrow] (s4.south) -- (s5.north);
\draw[arrow] (s5.south) -- (s6.north);
\draw[arrow] (s6.south) -- (s7.north);
\draw[arrow] (s7.south) -- (s8.north);
\draw[arrow]
    (loopframe.south)
    --
    (out.north);
\draw[dashedarrow]
    (s8.east)
    --
    ++(1.0cm,0)
    |-
    node[
        pos=0.25,
        right,
        font=\large,
        align=center
    ]
    {next time\\step}
    (s1.east);
\end{tikzpicture}%
}
\caption{Schematic overview of the \mer\ workflow.}
\label{fig:mercury_flowchart}
\end{figure*}
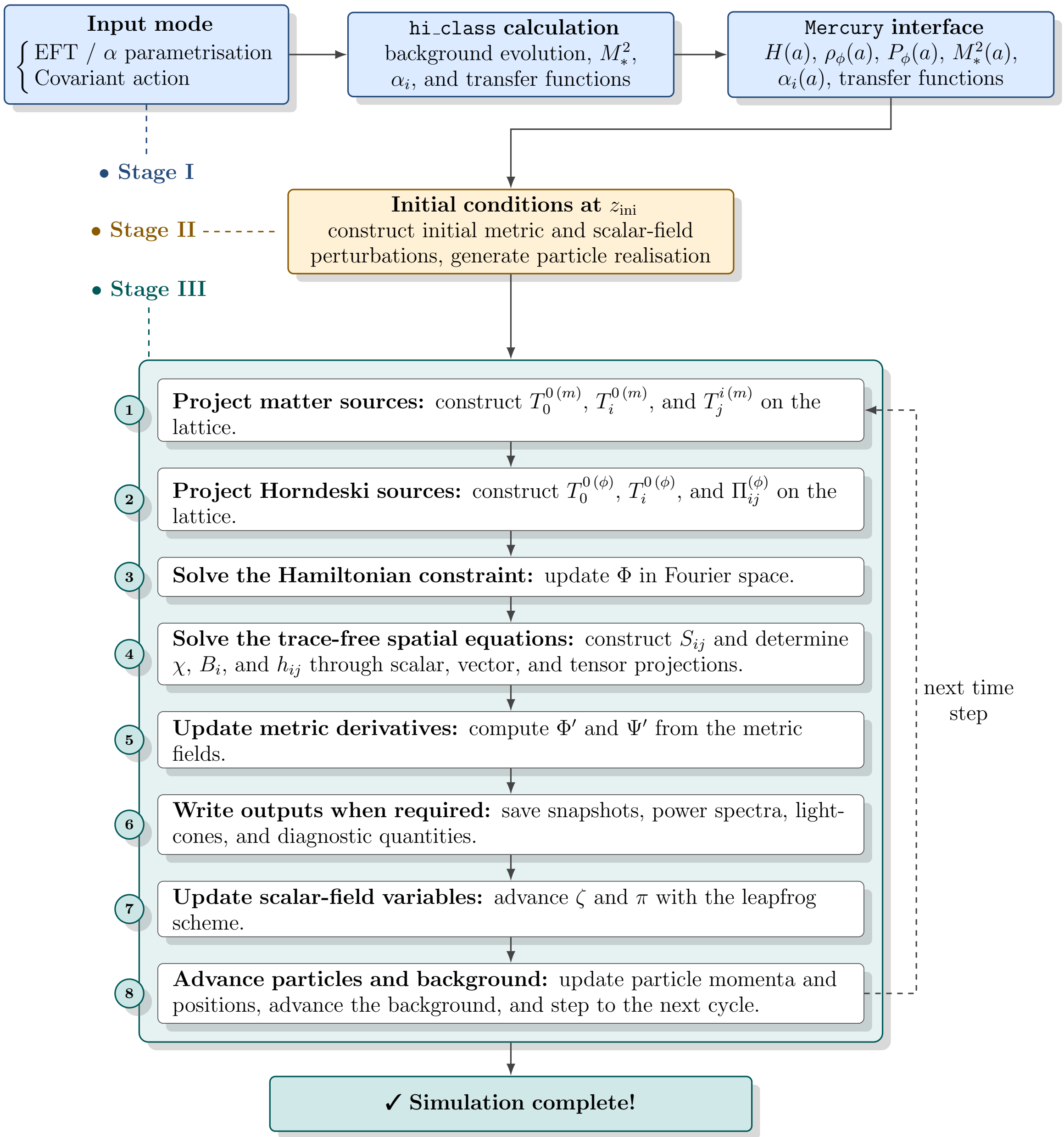

This section describes the numerical workflow of \mer, from the specification of the Horndeski model and the construction of the initial conditions to the main simulation loop. We organise the procedure into three stages, summarised schematically in Fig.~\ref{fig:mercury_flowchart} and discussed in the following subsections.

\subsection{Stage I: Model specification and background calculation}

The implementation in \mer\ separates the relativistic $N$-body evolution from the specification of the underlying Horndeski model. The metric and scalar-field equations evolved by \mer\ are written in terms of the $\alpha$ functions $\alpha_{\rm K}$, $\alpha_{\rm B}$, $\alpha_{\rm M}$ and $\alpha_{\rm T}$, together with the effective Planck mass $M_*^2$ and the corresponding background quantities. These are computed by \hiclass,  which provides the interface between the model specified by the user and the background and perturbation quantities required by the simulation. As a consequence, any model implemented in \hiclass can be run natively in \mer, and the implementation of new models requires modifications only to \hiclass.\footnote{See \url{https://hiclass-code.net/} for the implementation procedure.}

Two input workflows are supported. In the \textit{effective-theory approach}, the background expansion and the time dependence of the $\alpha$ functions are specified phenomenologically. In the \textit{covariant-theory approach}, the model is instead defined through a Horndeski action, from which \hiclass\ derives the background evolution and the corresponding $\alpha$ functions. Once these quantities have been computed and passed to \mer, the subsequent simulation evolution is independent of how the model was specified. The two input workflows are described separately below.

A common part of both workflows is the treatment of the present-day fractional density of the scalar sector, controlled in \hiclass\ by the parameter \texttt{Omega\_smg}. Three cases are possible:
\begin{itemize}
    \item \texttt{Omega\_smg} unspecified or set to zero:
    the scalar-field equations are not activated and \hiclass reduces to the standard \class\ evolution.
    \item $0<\texttt{Omega\_smg}<1$:
    the scalar sector is active and its present-day fractional density is specified directly. Either $\Omega_{\Lambda,0}$ or $\Omega_{\rm fld,0}$ must be left unspecified, and the remaining component is determined from the closure relation\footnote{Since the Horndeski implementation in \hiclass assumes spatial flatness, we set $\Omega_{k,0}=0$, and the sum in Eq.~\eqref{eq:hiclass_closure} runs over the energy-density components only.}
    \begin{equation}
        \sum_i\Omega_{i,0}=1 \,.
        \label{eq:hiclass_closure}
    \end{equation}
    \item \texttt{Omega\_smg}$<0$:
    the scalar sector remains active, but its present-day fractional density is determined internally from the closure relation. In this case, both $\Omega_{\Lambda,0}$ and $\Omega_{\rm fld,0}$ must be specified.
\end{itemize}

For the simulations presented in this work, we set
$\texttt{Omega\_smg}=-1$ together with
$\Omega_{\Lambda,0}=\Omega_{\rm fld,0}=0$.
The present-day scalar-sector density is therefore determined by \hiclass\ from Eq.~\eqref{eq:hiclass_closure}.

\subsubsection{Effective theory approach}

In this mode, the background expansion and the linear Horndeski perturbation sector are specified independently. The background history is selected through \texttt{expansion\_model}, while the time dependence of the Horndeski $\alpha$ functions is selected separately through \texttt{gravity\_model}. Consequently, different choices of $\alpha_i(a)$ may be combined with the same expansion history, and conversely the same $\alpha$-function parametrisation may be evaluated for different background histories. For the examples considered in this paper, the relevant \hiclass input parameters are summarised in Table~\ref{tab:hiclass_parametrized}. The input \texttt{Omega\_smg} fixes the target present-day scalar-sector density, either explicitly or through the closure relation. To reproduce this target, \hiclass adjusts the entry of \texttt{expansion\_smg} selected by \texttt{tuning\_index\_smg}. If $\Omega_{\rm smg,0}$ is selected for tuning, its supplied value is overwritten.

\begin{table}[H]
  \centering
  \caption{Relevant \hiclass\ input options for the parametrised
  Horndeski models used here.}
  \label{tab:hiclass_parametrized}

  \renewcommand{\arraystretch}{1.3}
  \renewcommand{\tabularxcolumn}[1]{m{#1}}

  \begin{tabularx}{\textwidth}{
    |>{\centering\arraybackslash}m{3.1cm}
    |>{\centering\arraybackslash}m{3.9cm}
    |>{\raggedright\arraybackslash}X|
  }
    \hline

    \rowcolor{gray!30}
    \textbf{Parameter}
    &
    \textbf{Value / Option}
    &
    \centering\arraybackslash\textbf{Description}
    \\
    \hline\hline \rowcolor{white} \texttt{Omega\_smg} & $0$, $(0,1)$, or $<0$ & Controls the scalar-sector density: zero disables the scalar sector, a positive value specifies its present-day density, and a negative value determines it from the closure relation. \\ \hline

    \rowcolor{white}
    \texttt{expansion\_model}
    &
    \texttt{wowa}
    &
    Parametrises the expansion history through a dark-energy
    equation of state
    $w(a)=w_0+w_a(1-a)$. 
    \\
    \hline

    \rowcolor{gray!10}
    \texttt{expansion\_smg}
    &
    $(\Omega_{\rm smg,0},w_0,w_a)$
    &
    A vector specifying the present-day scalar-field density and the
    two dark-energy equation-of-state parameters. \\ \hline \rowcolor{white} \texttt{tuning\_index\_smg} & integer index & Selects the entry of \texttt{expansion\_smg} varied to reproduce the target present-day scalar-sector density.
    \\
    \hline

    \rowcolor{gray!10}
    &
    \texttt{propto\_omega}
    &
    $\alpha_i=\hat{\alpha}_i\,\Omega_{\rm smg}(a)$
    \\
    \hhline{|>{\arrayrulecolor{gray!10}}->{\arrayrulecolor{black}}|-|-|}

    \rowcolor{gray!10}
    &
    \texttt{propto\_scale}
    &
    $\alpha_i=\hat{\alpha}_i\,a(\tau)$
    \\
    \hhline{|>{\arrayrulecolor{gray!10}}->{\arrayrulecolor{black}}|-|-|}

    \rowcolor{gray!10}
    \multirow[c]{-3}{3.1cm}{\centering\texttt{gravity\_model}}
    &
    \texttt{constant\_alphas}
    &
    $\alpha_i=\hat{\alpha}_i$
    \\
    \hline\rowcolor{white}
    \texttt{parameters\_smg}
    &
    $
    \left(
      \hat{\alpha}_{\rm K},
      \hat{\alpha}_{\rm B},
      \hat{\alpha}_{\rm M},
      \hat{\alpha}_{\rm T},
      M^2_{*,{\rm ini}}
    \right)
    $
    &
    A vector containing the coefficients that determine the chosen
    $\alpha$-function parametrisation, together with the initial value
    of $M_*^2$.
    \\
    \hline

  \end{tabularx}
\end{table}

\noindent\textbf{Background workflow:}

Once the present-day abundance has been fixed, the background evolution is constructed according to the selected \texttt{expansion\_model}. For the simulations shown in this work, we use the \texttt{wowa} option, adopting the Chevallier--Polarski--Linder (CPL) parametrisation of the scalar-sector equation of state \citep{Chevallier:2000qy,Linder:2002et},
\begin{equation}
    w_\phi(a)
    =
    w_0+w_a(1-a)\,.
    \label{eq:CPL_background}
\end{equation}
The parameters $\Omega_{\rm smg,0}$, $w_0$, and $w_a$ are specified through \texttt{expansion\_smg}. The present-day abundance fixes the normalisation of the scalar-sector density, while $w_0$ and $w_a$ determine its evolution. Integrating the corresponding conservation equation gives
\begin{equation}
    \bar\rho_\phi(a)
    =
    \bar\rho_{\phi,0}\,
    a^{-3(1+w_0+w_a)}
    \exp\!\left[
        3w_a(a-1)
    \right]\,.
    \label{eq:CPL_density_evolution}
\end{equation}
The corresponding pressure is $\bar P_\phi=w_\phi\bar\rho_\phi$, while $\bar\rho_\phi'$ follows the conservation equation. No explicit evolution of the homogeneous scalar field $\bar\phi$ is required in this mode, since the background scalar sector is specified directly through its effective density, pressure, and equation of state.

The $\alpha$ functions are specified independently through \texttt{gravity\_model}. For the \texttt{propto\_omega} parametrisation used in this work, they follow the fractional dark-energy density,
\begin{equation}
    \alpha_i(a)
    =
    \hat{\alpha}_i\,\Omega_{\rm smg}(a)\,,
    \qquad
    i\in
    \left\{
        {\rm K},
        {\rm B},
        {\rm M},
        {\rm T}
    \right\}\,.
\end{equation}
The coefficients $\hat{\alpha}_{\rm K}$, $\hat{\alpha}_{\rm B}$, $\hat{\alpha}_{\rm M}$, and $\hat{\alpha}_{\rm T}$, together with the initial value $M_{*,{\rm ini}}^2$, are specified through \texttt{parameters\_smg}. The alternatives listed in Table~\ref{tab:hiclass_parametrized} include \texttt{propto\_scale}, for which the $\alpha$ functions are proportional to the scale factor, and \texttt{constant\_alphas}, for which they remain constant in time.

In this mode, the covariant relations connecting the Horndeski functions $G_i$ to $M_*^2$, the $\alpha$ functions, and the effective scalar background quantities are not evaluated. Instead, the background evolution is fixed directly through \texttt{expansion\_model}, while $M_*^2$ and the $\alpha$ functions are constructed from the chosen phenomenological parametrisation. The resulting quantities are then supplied to \mer\ in the same form as in the covariant workflow.

\subsubsection{Covariant-theory approach}

In the covariant mode, the background expansion and the Horndeski perturbation sector are determined self-consistently from the underlying action. The user selects a theory through \texttt{gravity\_model} and specifies the corresponding model parameters and scalar-field initial conditions through \texttt{parameters\_smg}. In contrast to the effective-theory approach, no independent \texttt{expansion\_model} is required. Instead, \hiclass\ evolves the homogeneous scalar field together with the background expansion and evaluates $\bar\rho_\phi$, $\bar P_\phi$, $M_*^2$, and the $\alpha$ functions from the resulting background solution \cite{Bellini:2019syt}. The covariant models considered in this work are summarised in Table~\ref{tab:hiclass_covariant} and briefly introduced below.

For the covariant models considered below, we follow the normalisation of Eq.~\eqref{eq:Horndeski_action}, with $M_{\rm Pl}^2$ factored out of the Horndeski Lagrangian. The scalar fields are expressed as dimensionless variables, so that $X$ has dimensions of inverse length squared.

\begin{table}[t]
  \centering
  \caption{Relevant \hiclass input options for the covariant Horndeski models.}
  \label{tab:hiclass_covariant}

  \renewcommand{\arraystretch}{1.3}
  \renewcommand{\tabularxcolumn}[1]{m{#1}}

  \begin{tabularx}{\textwidth}{
    |>{\centering\arraybackslash}m{3.1cm}
    |>{\centering\arraybackslash}m{3.9cm}
    |>{\raggedright\arraybackslash}X|
  }
    \hline

    \rowcolor{gray!30}
    \textbf{Parameter}
    &
    \textbf{Value / Option}
    &
    \centering\arraybackslash\textbf{Description}
    \\
    \hline\hline

    \rowcolor{white}
    \texttt{Omega\_smg}
    &
    $0$, $(0,1)$, or $<0$
    &
    Controls the scalar-sector density: zero disables the scalar sector, a positive value specifies its present-day density, and a negative value determines it from the closure relation.
    \\
    \hline

    \cellcolor{gray!10}
    &
    \cellcolor{gray!10}
    \texttt{galileon}
    &
    \cellcolor{gray!10}
    Selects the covariant Galileon model.
    \\
    \hhline{|>{\arrayrulecolor{gray!10}}->{\arrayrulecolor{black}}|-|-|}

    \cellcolor{gray!10}
    &
    \cellcolor{gray!10}
    \texttt{nkgb}
    &
    \cellcolor{gray!10}
    Selects the $n$KGB model.
    \\
    \hhline{|>{\arrayrulecolor{gray!10}}->{\arrayrulecolor{black}}|-|-|}

    \cellcolor{gray!10}
    \multirow[c]{-3}{3.1cm}{\centering\texttt{gravity\_model}\footnotemark}
    &
    \cellcolor{gray!10}
    \texttt{brans dicke}
    &
    \cellcolor{gray!10}
    Selects the Brans--Dicke model.
    \\
    \hline

    \rowcolor{white}
    \multirow[c]{3}{3.1cm}[-4.7ex]{\centering\texttt{parameters\_smg}}
    &
    Cubic Galileon: none
    &
    The model parameters are fixed by $\Omega_{\rm smg,0}$ and the tracker condition.
    \\
    \hhline{|~|-|-|}

    &
    $n$KGB:
    $(g,n,R_{\rm shift,0})$
    &
    Specifies the braiding coupling, the power-law index, and the present departure from the shift-current attractor.
    \\
    \hhline{|~|-|-|}

    &
    Brans--Dicke:
    $(V_0,\omega_{\rm BD},\phi_{\rm ini},\phi'_{\rm ini})$
    &
    Specifies the dimensionless constant-potential parameter, the Brans--Dicke coupling, and the initial scalar-field value and velocity.
    \\
    \hline

    \rowcolor{gray!10}
    \texttt{tuning\_index\_smg}
    &
    integer index
    &
    Selects the model parameter varied to reproduce the target value of $\Omega_{\rm smg,0}$.
    \\
    \hline

  \end{tabularx}
\end{table}\footnotetext{Some models admit additional options, such as \texttt{gravity\_submodel}, which selects a particular subclass. See the \hiclass documentation at \url{https://hiclass-code.net/} for details.}

\medskip
\noindent\textbf{Cubic covariant Galileon.}
The Cubic covariant Galileon is the lowest-order member of the covariant Galileon family and contains a derivative self-interaction of the scalar field. It is defined by
\begin{equation}
    G_2=-X\,,
    \qquad
    G_3=-\frac{2c_3X}{H_0^2}\,,
    \qquad
    G_4=\frac{1}{2}\,,
    \qquad
    G_5=0 \,.
    \label{eq:cubic_galileon_covariant_functions}
\end{equation}
In the \hiclass implementation used here, the model is selected
through \texttt{gravity\_model = galileon} and
\texttt{gravity\_submodel = cubic}. While in principle this model has three free parameters, i.e.\ $(\phi_{\rm ini},\,\phi^\prime_{\rm ini},\,c_3)$, $\phi_{\rm ini}$ is irrelevant since the theory is shift-symmetric, $c_3$ is fixed to satisfy the closure relation and $\phi^\prime_{\rm ini}$ is chosen such that the solution sits on the attractor (in \hiclass, the system can also be evolved away from the attractor). Thus, no additional
\texttt{parameters\_smg} vector is required.
 Since $G_4$ is constant and $G_5=0$, the effective Planck mass is constant, $M_*^2=1$, and the tensor propagation speed is unmodified, giving $\alpha_{\rm M}=\alpha_{\rm T}=0$, while the cubic interaction produces a non-vanishing braiding function $\alpha_{\rm B}$.

\medskip
\noindent\textbf{$n$KGB.}
The $n$KGB model is a shift-symmetric kinetic gravity braiding theory that generalises the Cubic Galileon through a power-law dependence on the kinetic term. It is defined by
\begin{equation}
    G_2=-X\,,
    \qquad
    G_3=
    \frac{g^{\,n-\frac{1}{2}}}{n}
    \left(
        \frac{X}{H_0^2}
    \right)^n\,,
    \qquad
    G_4=\frac{1}{2}\,,
    \qquad
    G_5=0\,,
    \label{eq:nkgb_covariant_functions}
\end{equation}
The model is specified by the dimensionless coupling $g$,
which controls the strength of the kinetic braiding interaction,
the power-law index $n$, which determines its dependence on the
kinetic term $X$, and the quantity $R_{\rm shift,0}$, which
measures the present-day departure from the shift-current attractor
$J=0$, where $J$ denotes the scalar shift current.
The latter quantity is defined as

\begin{equation}
    R_{\rm shift,0}
    =
    \left.
    \frac{\sqrt{2X}\,J}
         {\sqrt{2X}\,J+X}
    \right|_{a=1}\,,
    \qquad
    0\leq R_{\rm shift,0}<1\,.
    \label{eq:nkgb_rshift}
\end{equation}
The attractor solution corresponds to $R_{\rm shift,0}=0$. Since $G_4$ is constant and $G_5=0$, the effective Planck mass is constant, $M_*^2=1$, and the tensor propagation speed is unmodified, giving $\alpha_{\rm M}=\alpha_{\rm T}=0$. The $n$KGB input file therefore uses $\texttt{parameters\_smg}=(g,n,R_{\rm shift,0})$.

\medskip
\noindent\textbf{Brans--Dicke gravity.}
Brans--Dicke gravity introduces a non-minimal coupling between the scalar field and the Ricci scalar. In the implementation considered here, the covariant functions are
\begin{equation}
    G_2=
    \frac{\omega_{\rm BD}}{\phi}X-3H_0^2V_0\,,
    \qquad
    G_3=0\,,
    \qquad
    G_4=\frac{\phi}{2}\,,
    \qquad
    G_5=0\,.
    \label{eq:brans_dicke_covariant_functions}
\end{equation}
Here, $V_0$ is the dimensionless constant-potential parameter, while $\omega_{\rm BD}$ controls the scalar kinetic term and the coupling strength relative to the general-relativistic limit. The GR limit is approached as $\omega_{\rm BD}\rightarrow\infty$, for which the scalar becomes increasingly difficult to excite. In this convention, the dimensionless effective Planck mass squared is $M_*^2=2G_4=\phi$.
The parameters $\phi_{\rm ini}$ and $\phi'_{\rm ini}$ specify the initial conditions for the homogeneous scalar field.
The corresponding input vector is
$\texttt{parameters\_smg}=
    \left(
        V_0,
        \omega_{\rm BD},
        \phi_{\rm ini},
        \phi'_{\rm ini}
    \right).
$
The optional parameter \texttt{M\_pl\_today\_smg} can be used to fix the present-day effective Planck mass by adjusting the scalar-field initial value. When \texttt{normalize\_G\_NR} is also activated, \hiclass\ instead normalises the effective gravitational coupling felt by nonrelativistic matter.
\\

\noindent\textbf{Background workflow:}

Once the present-day fractional scalar-sector density has been fixed through \texttt{Omega\_smg}, \hiclass\ treats $\Omega_{\rm smg,0}$ as the target abundance that the covariant background solution must reproduce. When \texttt{Omega\_smg}$<0$, this target is first inferred from the cosmological closure relation discussed above. In the normalisation used by \hiclass,
\begin{equation}
    \bar\rho_{\phi,0}^{\rm target}
    =
    H_0^2\Omega_{\rm smg,0}^{\rm target}\,.
    \label{eq:covariant_target_density}
\end{equation}
The selected \texttt{gravity\_model}, together with the corresponding entries of \texttt{parameters\_smg}, determines the covariant functions $G_i(\phi,X)$ and the scalar-field initial conditions. Starting from a trial set of model parameters, \hiclass\ constructs an initial configuration satisfying the Friedmann constraint and evolves the homogeneous scalar field together with the cosmological background using its constraint-preserving integration scheme \cite{Bellini:2019syt}. The effective scalar density and pressure are evaluated along this trajectory from the covariant expressions introduced in Section~\ref{sec:Horndeski}.

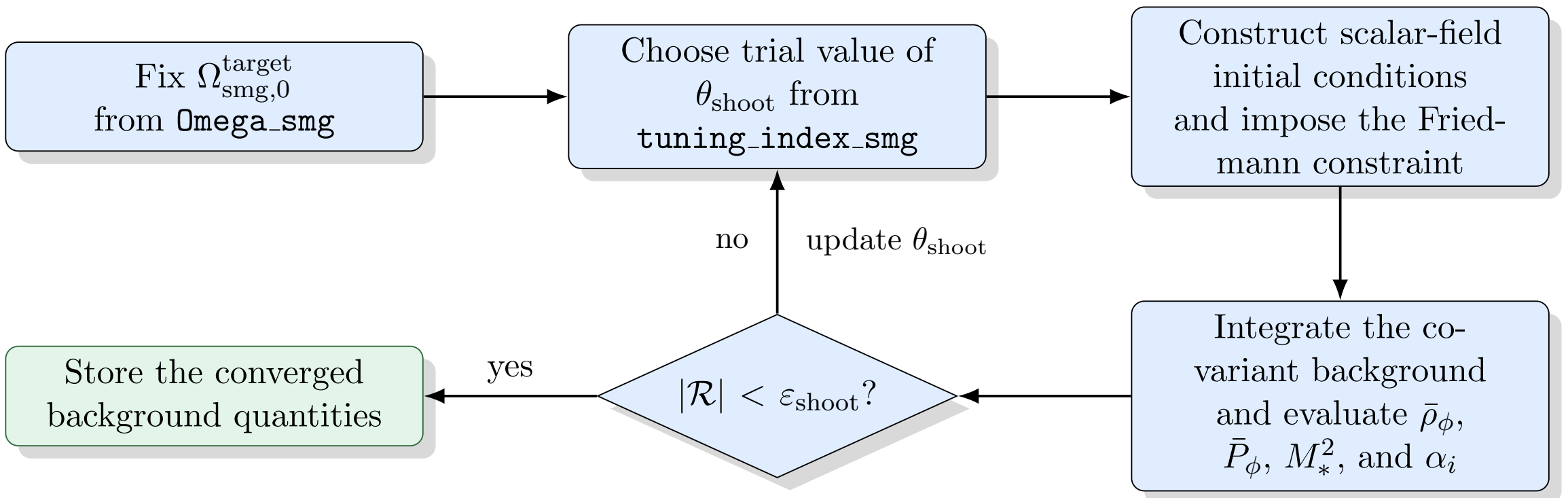
\begin{figure}[H]
\centering
\resizebox{0.98\textwidth}{!}{%
\begin{tikzpicture}[
    node distance=1.5cm and 1.6cm,
    box/.style={
        draw,
        fill=merblue!90,
        rounded corners,
        align=center,
        text width=4.2cm,
        minimum height=1.0cm,
        inner xsep=6pt,
        inner ysep=4pt,
        drop shadow={
            shadow xshift=4pt,
            shadow yshift=-4pt,
            opacity=0.3
        }
    },
    outputbox/.style={
        draw=merdarkgreen,
        fill=mergreen,
        rounded corners,
        align=center,
        text width=4.2cm,
        minimum height=1.0cm,
        inner xsep=6pt,
        inner ysep=4pt,
        drop shadow={
            shadow xshift=4pt,
            shadow yshift=-4pt,
            opacity=0.3
        }
    },
    decision/.style={
        draw,
        fill=merblue!90,
        diamond,
        aspect=2.2,
        align=center,
        text width=2.8cm,
        inner sep=1.5pt,
        drop shadow={
            shadow xshift=4pt,
            shadow yshift=-4pt,
            opacity=0.3
        }
    },
    arrow/.style={
        ->,
        thick,
        >=Latex
    }
]

\node[box] (target)
{
Fix $\Omega_{\rm smg,0}^{\rm target}$\\
from \texttt{Omega\_smg}
};

\node[box, right=of target] (trial)
{
Choose trial value of\\
$\theta_{\rm shoot}$ from
\texttt{tuning\_index\_smg}
};

\node[box, right=of trial] (initial)
{
Construct scalar-field initial conditions\\
and impose the Friedmann constraint
};

\node[decision, below=1.6cm of trial] (check)
{
$|\mathcal{R}|<\varepsilon_{\rm shoot}$?
};

\node[box] (integrate) at (initial |- check)
{
Integrate the covariant background\\
and evaluate
$\bar\rho_\phi$, $\bar P_\phi$, $M_*^2$, and $\alpha_i$
};

\node[outputbox] (output) at (target |- check)
{
Store the converged background quantities
};

\draw[arrow] (target.east) -- (trial.west);
\draw[arrow] (trial.east) -- (initial.west);
\draw[arrow] (initial.south) -- (integrate.north);
\draw[arrow] (integrate.west) -- (check.east);

\draw[arrow]
(check.west)
--
node[above]{yes}
(output.east);

\draw[arrow]
(check.north)
--
(trial.south);

\coordinate (loopmid) at ($(check.north)!0.5!(trial.south)$);

\node[
    font=\small,
    anchor=east
]
at ($(loopmid)+(-0.18cm,0)$)
{no};

\node[
    font=\small,
    anchor=west
]
at ($(loopmid)+(0.18cm,0)$)
{update $\theta_{\rm shoot}$};

\end{tikzpicture}
}
\caption{Schematic representation of the shooting procedure used to construct the covariant Horndeski background.}
\label{fig:covariant_background_shooting}
\end{figure}

For models in which a parameter must be adjusted to reproduce the desired present-day abundance, \texttt{tuning\_index\_smg} selects the corresponding shooting parameter, which we denote by $\theta_{\rm shoot}$. After each trial background evolution, the resulting value $\Omega_{\rm smg,0}^{\rm trial}$ is compared with the target through
\begin{equation}
    \mathcal{R}(\theta_{\rm shoot})
    =
    \Omega_{\rm smg,0}^{\rm trial}
    -
    \Omega_{\rm smg,0}^{\rm target}\,.
    \label{eq:covariant_shooting_residual}
\end{equation}
If $|\mathcal{R}|$ exceeds the required tolerance $\varepsilon_{\rm shoot}$, the shooting parameter is updated and the background evolution is repeated. Once the target abundance is reproduced, the converged solution is retained and the quantities required by \mer, including $H(a)$, $\bar\rho_\phi(a)$, $\bar P_\phi(a)$, $M_*^2(a)$, and the $\alpha$ functions, are stored along the background trajectory. The procedure is summarised in Fig.~\ref{fig:covariant_background_shooting}.

\subsection{Stage II: Construction of the initial conditions}
\label{sec:stageII_initial_conditions}

Once the background evolution and linear transfer functions have been computed by \hiclass, they are used to construct the initial conditions at a redshift $z_{\rm ini}$ chosen by the user. In the simulations presented in this work, we set $z_{\rm ini}=100$.

The perturbations evolved by \hiclass are expressed in synchronous gauge, whereas \mer evolves the metric and scalar-field perturbations in Poisson gauge. The transfer functions required for the initialisation are therefore transformed consistently to Poisson gauge. The corresponding gauge-transformation relations are given in Appendix~C of \cite{Nouri-Zonoz:2025cul}.

The initial perturbations are generated as correlated Gaussian realisations by combining the linear transfer functions with the primordial curvature spectrum. The same random seed and Fourier-mode ordering are used for the different perturbation variables, ensuring that the matter, metric, and Horndeski scalar fields correspond to the same primordial realisation.

\subsubsection{Realisation of the perturbation fields}
To initialise the scalar-field perturbation $\pi$ and its evolution variable $\zeta$ on the real-space lattice, we first transform their Fourier-space perturbations from \hiclass into Poisson gauge and then apply an inverse fast Fourier transform (FFT).
These fields provide the initial conditions for the subsequent scalar-field evolution and for the construction of the Horndeski stress--energy tensor.

The initial spatial metric potential $\Phi$ is constructed from the total density and velocity-divergence transfer functions using the linearised Hamiltonian and momentum constraints, Eqs.~\eqref{eq:Hamiltonian} and \eqref{eq:momentum_constraint}.
The gravitational slip $\chi$ is obtained from the trace-free spatial metric equation, with the linear \hiclass slip used at early times when the optional linear-$\chi$ treatment is active. The temporal potential is then reconstructed as
\begin{equation}
    \Psi=\Phi-\chi\,,
\end{equation}
and is therefore not stored as an independent lattice field. The initial vector potential $B_i$ is obtained from the corresponding vector projection of the matter momentum source.

\subsubsection{Initialisation of the particle phase space}

The particle phase space is initialised following the procedure implemented in \gev\ and described in Appendix~A of \cite{Adamek:2016zes}. A homogeneous particle template is displaced using the linear matter transfer functions, while the corresponding velocity transfer functions determine the initial particle momenta. The required relativistic gauge corrections are included consistently in the construction of these fields.

In the blended treatment adopted in this work, CDM and baryons are represented by a single particle ensemble. Their density and velocity transfer functions are combined using the corresponding background-density weights before the displacement and momentum fields are generated. The vector potential $B_i$ is computed from the initial particle momentum density, whereas $h_{ij}$ is obtained from the tensor source at the first evolution cycle and is therefore not included among the fields initialised at this stage. At the end of this stage, the initialised lattice fields are 
\begin{equation}
    \left\{
        \Phi,\chi,B_i,\pi,\zeta
    \right\}\,,
\end{equation}
while the particle sector is specified by
\begin{equation}
    \left\{
        \mathbf{x}_p,\mathbf{q}_p
    \right\}\,,
\end{equation}
where $\mathbf{x}_p$ and $\mathbf{q}_p$ denote the position
and canonical momentum of particle $p$, respectively. Together with the background and Horndeski quantities obtained in Stage~I, these variables define the initial state passed to the main evolution loop in Stage~III.

\subsection{Stage III: Main evolution loop}
\label{sec:stageIII_evolution}

After the initial conditions have been constructed, the simulation enters the main evolution loop. At each cycle, the matter and Horndeski sources are projected onto the lattice, the metric fields are updated, the Horndeski scalar perturbations $\pi$ and $\zeta$ are advanced, and finally the particle phase-space coordinates and background scale factor are evolved.

The first step is the construction of the matter stress--energy tensor on the simulation lattice. The particle positions and canonical momenta are used to evaluate the components $T^{0\,(m)}_{0}$, $T^{0\,(m)}_{i}$, and $T^{i\,(m)}_{j}$, which are deposited onto the lattice using the particle-to-mesh projection scheme of \gev. The explicit expressions for the particle stress--energy tensor and the corresponding projection procedure are given in \cite{Adamek:2016zes}.

The deposited matter energy-density component is separated into its homogeneous and perturbed parts, with $\delta T^0{}_0{}^{(m)}$ contributing to the scalar source $s_{\mathbf{i},\mathbf{j},\mathbf{k}}$ defined in Eq.~\eqref{eq:source_phi_lattice}, while the momentum density and spatial stress contribute to the vector and trace-free spatial sources.

The Horndeski contribution to the stress--energy tensor is then evaluated from the current scalar and metric fields. The relevant components are $\delta T^{0\,(\phi)}_{0}$, $\delta T^{0\,(\phi)}_{\ell}$, and $\Pi^{(\phi)}_{\ell m}$, whose lattice expressions are given in Eqs.~\eqref{eq:deltarho_DE}--\eqref{eq:lattice_Horndeski}. These contributions are added to the corresponding gravitational sources before the metric equations are solved.

The Hamiltonian constraint is solved first to update the spatial metric potential $\Phi$. The scalar source is transformed to Fourier space, where the discretised constraint in Eq.~\eqref{eq:phi_fourier_hamiltonian} is solved mode by mode, and the resulting $\Phi$ field is transformed back to the real-space lattice.

The trace-free spatial equations are treated next. The source $S_{ij}$ is transformed to Fourier space and decomposed into its scalar, vector, and tensor parts. The scalar projection determines the gravitational slip $\chi$ through Eq.~\eqref{eq:chi_fourier_projection_mercury}. At early times, when the linear-$\chi$ treatment is active, the full linear slip supplied by \hiclass is used instead. The vector sector determines the frame-dragging potential $B_i$ through either the transverse projection of the momentum constraint in Eq.~\eqref{eq:momentum_constraint} or the evolution equation in Eq.~\eqref{eq:Bi_fourier_projection_mercury}, depending on the selected treatment. The tensor perturbation $h_{ij}$ can either be evolved dynamically through Eq.~\eqref{eq:hij_fourier_projection_mercury} or obtained from its elliptic, quasistatic limit by neglecting the time-derivative terms. The tensor results presented in this work use the latter treatment.

Once $\Phi$ and $\chi$ have been updated, their conformal-time derivatives are reconstructed from the current and previous time levels. Since $\Psi=\Phi-\chi$, this provides both $\Phi'$ and $\Psi'$, which enter the subsequent evolution of the Horndeski scalar perturbations. Snapshots, power spectra, and lightcones are written whenever the corresponding output redshifts are reached.

The scalar variables $\zeta$ and $\pi$ are advanced using the staggered leapfrog scheme described in Section~\ref{sec:implementation}. The variable $\zeta$ is evaluated at half-integer time levels and $\pi$ at integer time levels. When required by the scalar stability condition, the main simulation step is subdivided into smaller scalar steps, with the background evaluated consistently throughout the subcycling.

Finally, the particle phase-space coordinates are advanced using the updated metric fields. The canonical momenta are updated first, the background scale factor is advanced by half a time step, and the particle positions are then evolved using the updated momenta. The remaining half-step of the background evolution is subsequently completed. This preserves the time staggering of the particle leapfrog scheme. The conformal time and time step are then updated, and the procedure is repeated for the next simulation cycle.

\section{Numerical results and validation}
\label{sec:results}
In this section, we present the numerical results obtained with \mer and assess the implementation against the corresponding linear predictions from \hiclass.

\subsection{Simulation setup}
\label{sec:simulation_setup}

We consider six representative Horndeski models, divided into two
groups according to the input approach used in \hiclass. The first
group consists of three phenomenological models specified directly in
the effective-theory basis. We refer to these simulations as EFT-$k$ess, EFT-KGB, and EFT-LH, corresponding respectively to the $k$-essence subclass, the kinetic gravity braiding subclass, and the luminal Horndeski subclass. For EFT-LH, we adopt the best-fit values obtained from
the CMB+DESI+ISWL+PantheonPlus data combination in
\citep{Chudaykin:2025gdn}. We further choose a large kineticity,
$\hat{\alpha}_{\rm K}=3\times10^{3}$, which leads to a small
speed of sound. The EFT-KGB and EFT-$k$ess models are then obtained as
successive limits of EFT-LH by setting
$\hat{\alpha}_{\rm M}=0$ and subsequently
$\hat{\alpha}_{\rm B}=0$, respectively, while keeping
$\hat{\alpha}_{\rm T}=0$ in all three cases. For EFT-LH and EFT-KGB, the phantom divide can be crossed without generating a singularity in the perturbation evolution. We therefore retain the best-fit background values
$(w_0,w_a)=(-0.88,-0.40)$ obtained from
\citep{Chudaykin:2025gdn}. By contrast, the minimally coupled
$k$-essence model considered here does not allow a regular crossing of
$w=-1$. For EFT-$k$ess, we therefore set $w_a=0$ while keeping
$w_0=-0.88$. Moreover, the time dependence of the effective theory functions follows the
\texttt{propto\_omega} parametrisation,
\begin{equation}
    \alpha_i(a)
    =
    \hat{\alpha}_i\,\Omega_{\rm smg}(a)\,.
    \label{eq:proptoomega}
\end{equation}

The second group consists of three theories specified through a covariant Horndeski action: the nKGB model, the Cubic Galileon, and the Brans--Dicke model. For these simulations, the background evolution and the effective functions $M_*^2$ and $\alpha_i$ are computed self-consistently from the covariant model by \hiclass.

All simulations are initialised at $z_{\rm ini}=100$ using the same
baseline cosmological parameters, summarised in
Table~\ref{table:cosmoparams}, and adopt a common numerical setup with
$N_{\rm grid} = N_{\rm pcl}=3072^3$ and $L_{\rm box}=4000\,h^{-1}{\rm Mpc}$.
In all cases, the present-day dark-energy density is determined from
the spatial-flatness closure relation,
\begin{equation}
    \Omega_{{\rm smg},0}
    =
    1-\Omega_{{\rm m},0}-\Omega_{{\rm r},0}\,.
\end{equation}

\begin{table}[h]
\caption{Baseline cosmological parameters used in all simulations.}
\centering
\renewcommand{\arraystretch}{1.2}
\begin{tabular}{|*{6}{c|}}
\hline
\rowcolor{crisp}
$\Omega_{{\rm cdm},0}$
&
$\Omega_{{\rm b},0}$
&
$\Omega_{{\rm r},0}$
&
$h$
&
$A_{\rm s}$
&
$n_{\rm s}$
\\
\hline\hline
0.26081
&
0.04817
&
$9.15 \times10^{-5}$
&
0.6763
&
$2.02\times10^{-9}$
&
0.9619
\\
\hline
\end{tabular}
\label{table:cosmoparams}
\end{table}

The model-dependent parameters are summarised in Table~\ref{tab:simulation_models}.

\begin{table}[h]
\centering

\caption{Model parameters used in the simulations. All simulations use $3072^3$ grid cells and particles, and a box size of $L_{\rm box}=4000\,h^{-1}{\rm Mpc}$. For the EFT models, $\alpha_i(a)=\hat{\alpha}_i\Omega_{\rm smg}(a)$ and $M_{\ast,{\rm ini}}^2=1$.}

\label{tab:simulation_models}

\renewcommand{\arraystretch}{1.2}
\setlength{\tabcolsep}{10pt}

\begin{NiceTabular}{|c|lll|}[cell-space-limits=2pt]
\hline
Group & Model & Parameter & Value \\
\hline


\Block[fill=merblue!15]{6-1}{
\rotatebox[origin=c]{90}{
\shortstack{\textbf{Group 1}\\\textbf{EFT}}
}
}
&
EFT-$k$ess
&
$(\hat{\alpha}_{\rm K},
  \hat{\alpha}_{\rm B},
  \hat{\alpha}_{\rm M},
  \hat{\alpha}_{\rm T})$
&
$(3.0\times10^{3},\,0,\,0,\,0)$
\\
&
&
$(w_0,w_a)$
&
$(-0.88,\,0)$
\\

\cline{2-4}

&
EFT-KGB
&
$(\hat{\alpha}_{\rm K},
  \hat{\alpha}_{\rm B},
  \hat{\alpha}_{\rm M},
  \hat{\alpha}_{\rm T})$
&
$(3.0\times10^{3},\,0.49,\,0,\,0)$
\\
&
&
$(w_0,w_a)$
&
$(-0.88,\,-0.40)$
\\

\cline{2-4}

&
EFT-LH
&
$(\hat{\alpha}_{\rm K},
  \hat{\alpha}_{\rm B},
  \hat{\alpha}_{\rm M},
  \hat{\alpha}_{\rm T})$
&
$(3.0\times10^{3},\,0.49,\,1.28,\,0)$
\\
&
&
$(w_0,w_a)$
&
$(-0.88,\,-0.40)$
\\

\hline


\Block[fill=merorange!15]{6-1}{
\rotatebox[origin=c]{90}{
\shortstack{\textbf{Group 2}\\\textbf{Covariant}}
}
}
&
\Block{2-1}{nKGB}
&
\Block{2-1}{$(g,n,R_{\rm shift,0})$}
&
\Block{2-1}{$(4\times10^{-5},\,2,\,0)$}
\\
&
&
&
\\

\cline{2-4}

&
\Block{2-1}{Cubic Galileon}
&
\Block{2-1}{Background branch}
&
\Block{2-1}{Tracker solution}
\\
&
&
&
\\

\cline{2-4}

&
\Block{2-1}{Brans--Dicke}
&
\Block{2-1}{$(V_0,\omega_{\rm BD},\phi_{\rm ini},\phi'_{\rm ini})$}
&
\Block{2-1}{$(0.7,\,1000,\,1,\,10^{-4})$}
\\
&
&
&
\\

\hline
\end{NiceTabular}
\end{table}

\subsection{Background evolution}

\begin{figure*}
    \centering
    \includegraphics[width=\textwidth]{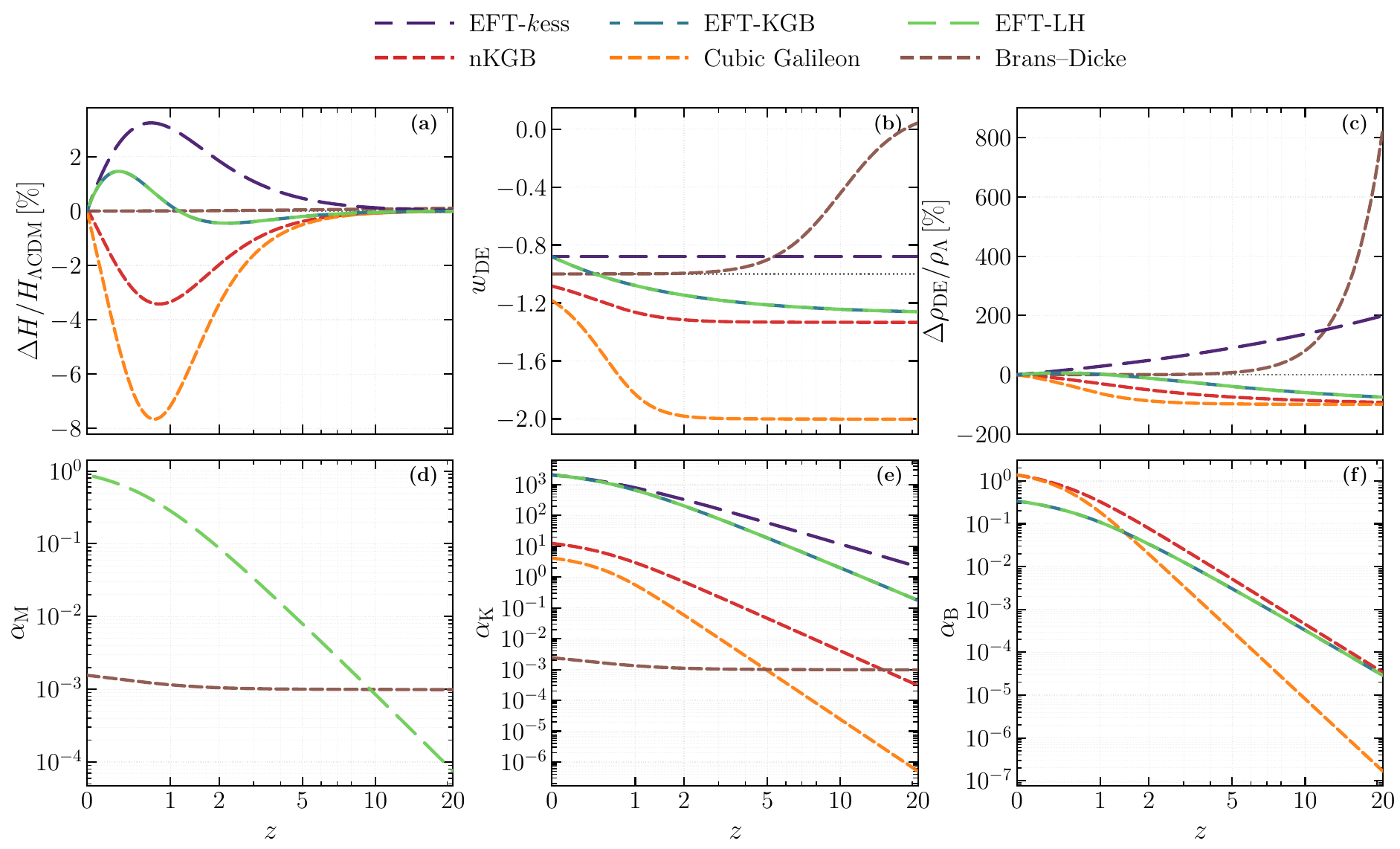}
    \caption{Background evolution of the six models considered in this work. The upper panels show the fractional deviation of the Hubble parameter from the reference $\Lambda$CDM cosmology, the effective dark-energy equation of state $w_{\rm DE}$, and the fractional difference in the physical dark-energy density, $\Delta\rho_{\rm DE}/\rho_{\Lambda}$. The lower panels show the evolution of $\alpha_{\rm M}$, $\alpha_{\rm K}$, and $\alpha_{\rm B}$. The function $\alpha_{\rm M}$ vanishes for all models except EFT-LH and Brans--Dicke. For Brans--Dicke, $\alpha_{\rm B}=-\alpha_{\rm M}$; its negative values are therefore not displayed on the logarithmic scale used in the $\alpha_{\rm B}$ panel.}
    \label{fig:background_evolution}
\end{figure*}

Before turning to the perturbation spectra, we first compare the background evolution of the six models introduced above. Figure~\ref{fig:background_evolution} presents the redshift evolution of the relevant background quantities for the six models. The upper row shows the expansion history and the effective dark-energy evolution, while the lower row presents the time-dependent $\alpha$ functions entering the perturbation equations.

The differences in the expansion rate shown in panel~(a) can be interpreted in terms of the evolution of $w_{\rm DE}$ and $\rho_{\rm DE}$ in panels~(b) and~(c). In panel~(c), we plot the fractional deviation of the dark-energy density from its $\Lambda$CDM value,
\begin{equation}
    \frac{\Delta\rho_{\rm DE}}{\rho_{\Lambda}}
    \equiv
    \frac{\rho_{\rm DE}(z)-\rho_{\Lambda}}{\rho_{\Lambda}}\,,
\end{equation}
where $\rho_{\Lambda}$ denotes the dark-energy density of the reference $\Lambda$CDM cosmology and is constant. Since all models share the same present-day Hubble parameter and dark-energy density, the deviations in panels~(a) and~(c) vanish at $z=0$. Quintessence-like behaviour, $w_{\rm DE}>-1$, generally implies a larger dark-energy density in the past than for a cosmological constant, whereas phantom behaviour, $w_{\rm DE}<-1$, implies a smaller one. These differences modify the total background energy density and, through the Friedmann equation, the evolution of $H(z)$.

The EFT-$k$ess model provides the clearest example of quintessence-like behaviour. Its equation of state is constant, with $w_{\rm DE}=-0.88$, so its physical dark-energy density increases towards the past relative to the constant $\rho_{\Lambda}$. This is reflected by the positive deviation in panel~(c) and leads to an enhanced Hubble rate at low and intermediate redshifts, reaching a few per cent around $z\sim1$. EFT-KGB and EFT-LH share the same prescribed background evolution and therefore overlap throughout the upper row. Their equation of state is quintessence-like at late times but crosses into the phantom regime at higher redshift. Accordingly, their dark-energy density is initially larger than $\rho_{\Lambda}$ when moving away from $z=0$, but later falls below it. The Hubble-rate deviation follows the same transition, becoming positive at low redshift, reaching a maximum below $z\sim1$, and turning slightly negative at higher redshift.

The covariant KGB models exhibit phantom behaviour over the redshift interval in which their dark-energy sector contributes appreciably to the expansion. Their physical dark-energy density is therefore lower than $\rho_{\Lambda}$ over this range, producing the negative deviations in panel~(c) and a corresponding suppression of the Hubble rate. This effect is relatively mild for nKGB, whose equation of state remains closer to $-1$, but is considerably stronger for the Cubic Galileon, for which $w_{\rm DE}$ approaches $-2$. The Cubic Galileon consequently shows the largest departure in panel~(a), with $\Delta H/H_{\Lambda{\rm CDM}}$ reaching about $-8\%$ around $z\sim1$.

Brans--Dicke remains close to the $\Lambda$CDM expansion history at low redshift despite the pronounced evolution of its effective equation of state.
The effective equation of state remains close to $-1$ at low redshift, with only small variations around the phantom divide. At higher redshift, the effective dark-energy density grows rapidly relative to $\rho_{\Lambda}$, producing the pronounced rise in panel~(c). Although this fractional difference becomes large, its impact on $H(z)$ remains modest because the background is increasingly dominated by matter at these redshifts. Consequently, the Brans--Dicke expansion history stays close to that of $\Lambda$CDM.

The lower panels of Fig.~\ref{fig:background_evolution} show the evolution of the $\alpha$ functions. For the EFT models, their time dependence follows the parametrisation introduced in Eq.~\eqref{eq:proptoomega}, such that each non-vanishing $\alpha_i$ scales with $\Omega_{\rm smg}(a)$. As the dark-energy contribution becomes increasingly important towards low redshift, the corresponding $\alpha$ functions grow and approach their largest values at the present epoch. An instructive example is $\alpha_{\rm K}$ in panel~(e). Although the same value of $\hat{\alpha}_{\rm K}$ is adopted for all three EFT models, EFT-KGB and EFT-LH follow the same evolution, whereas EFT-$k$ess departs from them at higher redshift. This difference originates entirely from the background evolution entering the parametrisation: EFT-KGB and EFT-LH share the same dynamical dark-energy equation of state and therefore the same $\Omega_{\rm smg}(a)$, while EFT-$k$ess has $w_a=0$ and consequently a different evolution of $\Omega_{\rm smg}$. The three models nevertheless approach the same value of $\alpha_{\rm K}$ at $z=0$, where their present-day dark-energy density fractions coincide.

For the covariant models, by contrast, the evolution of the $\alpha$ functions is determined by the underlying scalar--tensor theory rather than by the phenomenological parametrisation adopted for the EFT models. This results in  different redshift dependences. In Brans--Dicke, both $\alpha_{\rm M}$ and $\alpha_{\rm K}$ vary only mildly over the redshift range shown and remain approximately constant compared with the stronger evolution found in the other models. Although $G_3=0$, the non-minimal coupling through $G_4(\phi)$ generates a non-vanishing braiding function, which for this model satisfies $\alpha_{\rm B}=-\alpha_{\rm M}$. The nKGB and Cubic Galileon models, by contrast, show a strong increase in both $\alpha_{\rm K}$ and $\alpha_{\rm B}$ towards low redshift. Although their evolution differs, both functions decrease rapidly at higher redshifts, with the Cubic Galileon showing a steeper decline.

\subsection{Power-spectrum analysis}
\begin{figure}[!t]
    \centering
    \includegraphics[width=\textwidth]{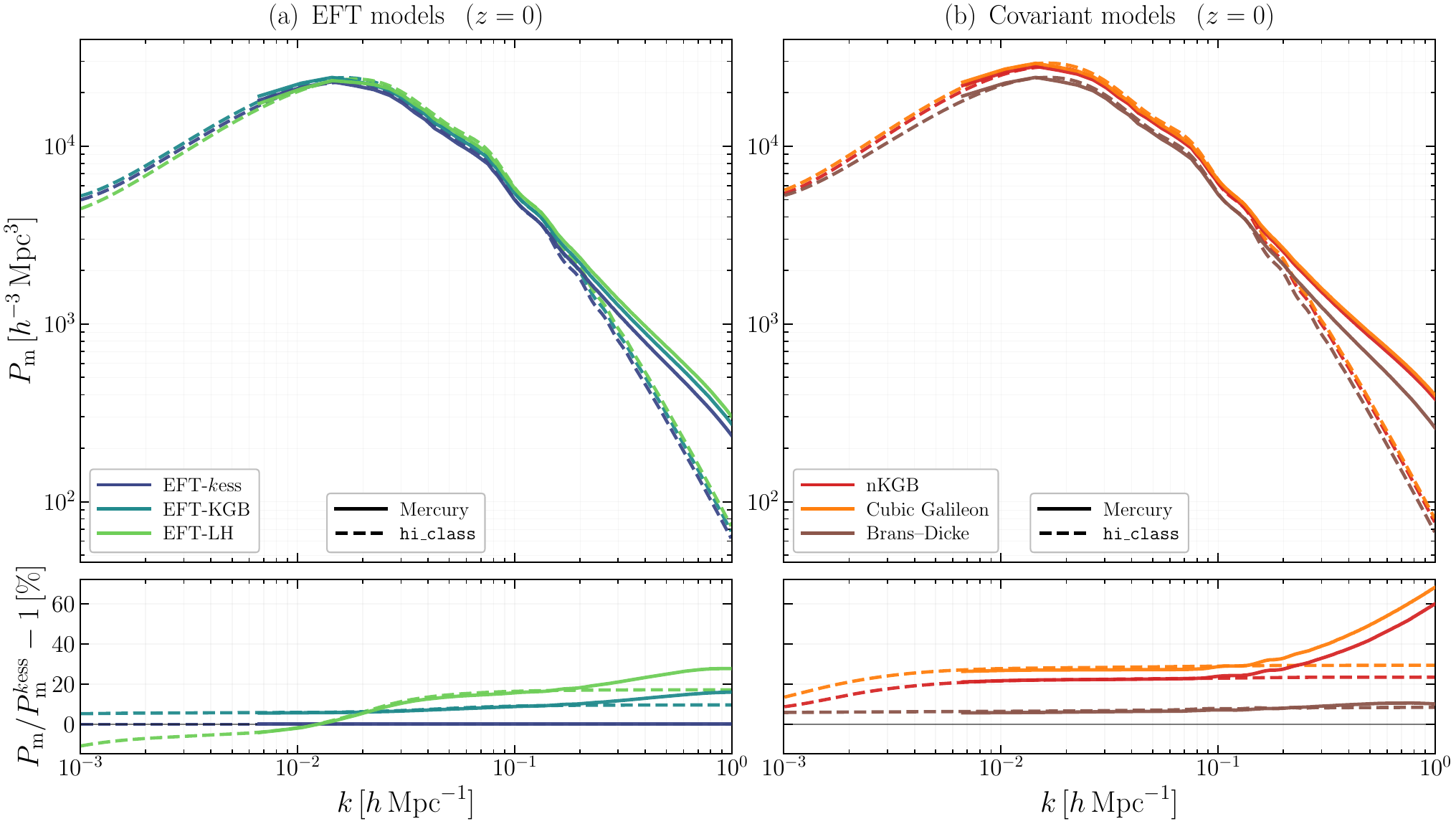}
    \caption{Matter power spectra at $z=0$ for the EFT models (left) and the covariant models (right). Solid lines show the results from \mer, while dashed lines show the corresponding linear predictions from \hiclass. The lower panels show the relative differences with respect to EFT-$k$ess, expressed as percentages.}
    \label{fig:MatterPower}
\end{figure}

We next examine the power spectra of the perturbations at $z=0$ and compare the results from \mer with the corresponding linear predictions from \hiclass. On large scales, the two closely match, as expected when the evolution remains in the linear regime. Towards smaller scales, the \mer spectra depart from the linear predictions as nonlinear structure formation becomes important. These nonlinear effects propagate through the coupled matter, scalar-field, and metric sectors, modifying the Horndeski perturbations and subsequently feeding back onto the metric perturbations. 

The convergence tests presented in Appendix~\ref{app:convergence} indicate that, for the simulations with $N_{\rm grid}=N_{\rm pcl}=3072^3$ discussed in this section, resolution effects at the per cent level become relevant in the individual power spectra at $k\gtrsim0.3$--$0.4\,h\,{\rm Mpc}^{-1}$. These effects largely cancel in power-spectrum ratios, with the tested ratios showing a resolution dependence within approximately $2\%$ over the full plotted range. This supports a reliable interpretation of the relative model signatures down to the smallest scales shown.

Figure~\ref{fig:MatterPower} shows the matter power spectra at $z=0$ for the EFT and covariant models. On large scales, the \mer results match the corresponding linear predictions from \hiclass, as can also be seen in the lower panels showing the relative differences with respect to EFT-$k$ess. The differences between the models in this regime reflect their different growth histories, which accumulate from the initial conditions to the present epoch. Towards smaller scales, nonlinear evolution increases the power relative to the linear prediction while preserving the ordering of the models compared to the linear case.

Within the EFT family, EFT-KGB and EFT-LH are particularly useful to compare, since they share the same prescribed background evolution as well as the same $\alpha_{\rm K}$ and $\alpha_{\rm B}$, but differ through the non-vanishing $\alpha_{\rm M}$ and the associated evolution of $M_*^2$ in EFT-LH. The resulting difference in the matter power spectrum therefore isolates the effect of the additional non-minimal coupling within this EFT parametrisation. This difference is already present in the linear evolution and persists into the nonlinear regime, where the overall separation between the spectra becomes more pronounced. EFT-$k$ess, which also differs in its background evolution and has vanishing braiding, shows the lowest matter power among the three models at small scales.

For the covariant models, the Cubic Galileon exhibits the largest matter power, followed by nKGB and Brans--Dicke. The same ordering is already apparent in the linear spectra and is maintained after nonlinear evolution. Unlike the EFT-KGB and EFT-LH comparison, however, the covariant models differ simultaneously in their background expansion, effective Planck mass, and the evolution of the $\alpha$ functions. Their relative amplitudes therefore reflect the combined effect of these differences on the growth of matter perturbations and cannot be associated with a single Horndeski function. The matter spectra thus provide a useful reference for the following discussion, since the scale-dependent departure from the linear predictions marks the regime in which nonlinear matter evolution can begin to source corresponding scalar-field and metric perturbations.

\begin{figure}[t]
    \centering
    \includegraphics[width=\textwidth]{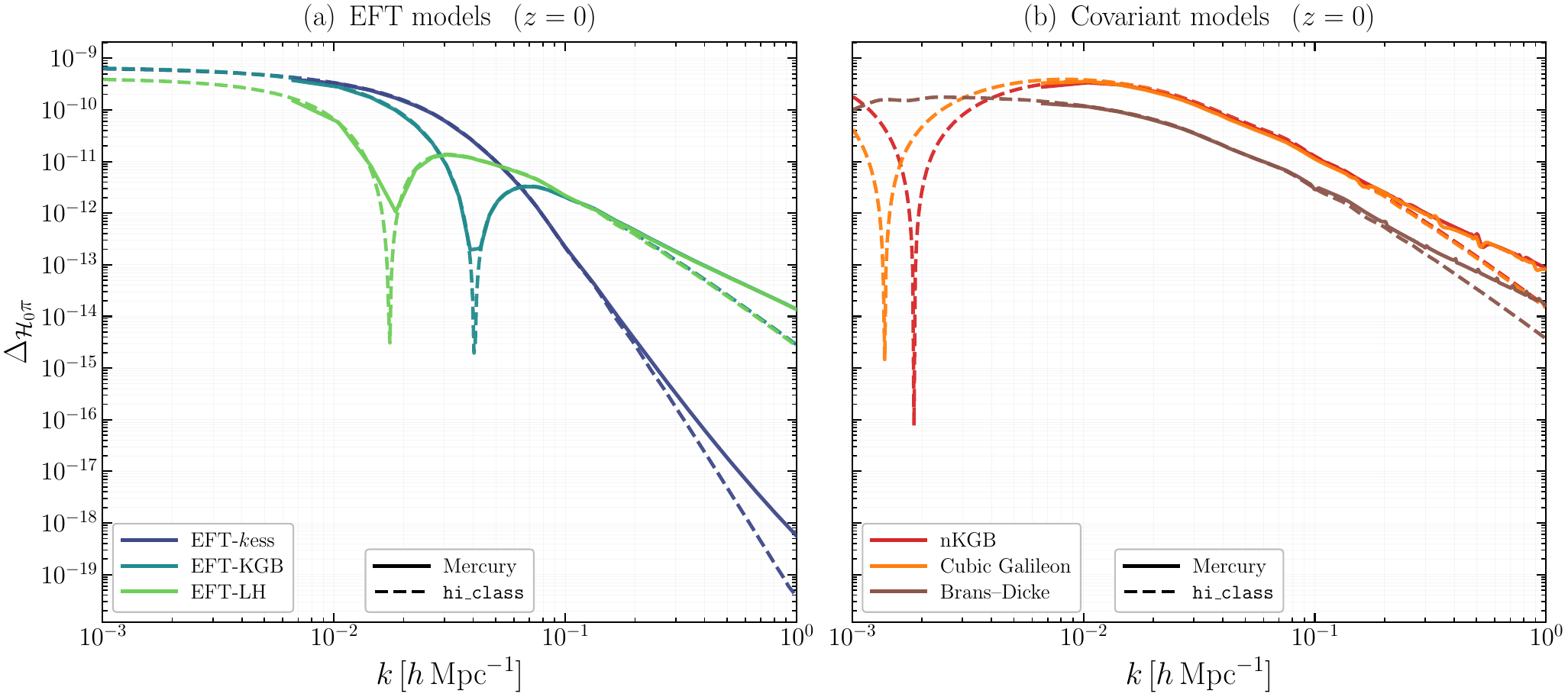}
    \caption{Dimensionless power spectra of the scalar-field perturbation $\mathcal H_0\pi$ at $z=0$ for the EFT models (left) and the covariant models (right). Solid lines show the results from \mer, while dashed lines show the corresponding linear predictions from \hiclass.}
    \label{fig:PiPower}
\end{figure}

Figure~\ref{fig:PiPower} shows the power spectrum of the scalar-field perturbation $\pi$. A characteristic feature of the braided models is the pronounced minimum associated with a zero crossing and sign change of $\pi$, previously discussed in detail for KGB models in \citep{Nouri-Zonoz:2025cul}. This behaviour is clearly visible for EFT-KGB and EFT-LH, both of which have a non-vanishing $\alpha_{\rm B}$, while it is absent in the EFT-$k$ess limit. Similar zero crossings are also present in the covariant nKGB and Cubic Galileon models at lower wavenumbers, around $k\sim10^{-3}\,h\,\mathrm{Mpc}^{-1}$. Their different scalar dynamics, including the evolution of the sound speed, shift the characteristic sign change of $\pi$ towards larger scales.

\begin{figure}[t]
    \centering
    \includegraphics[width=\textwidth]{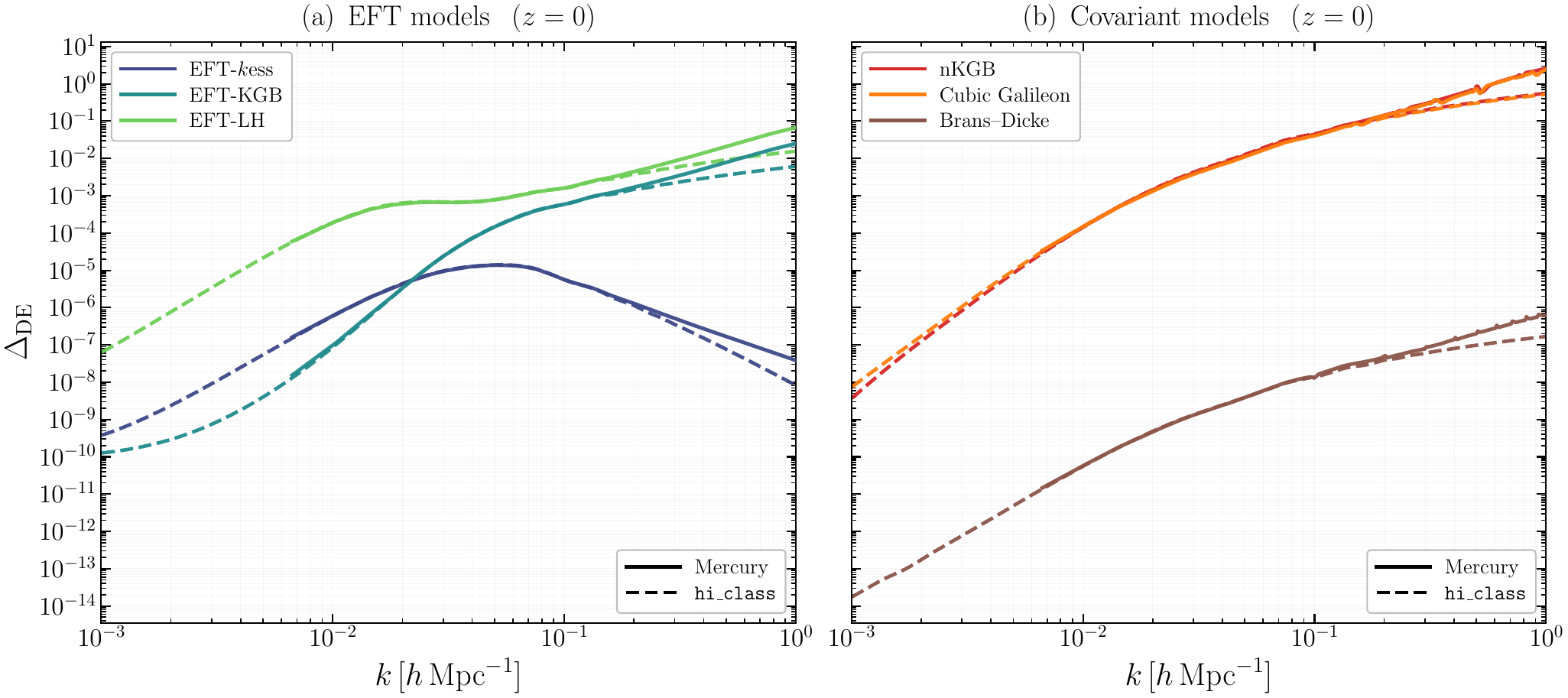}
    \caption{Dimensionless power spectra of the effective dark-energy density contrast $\delta_{\rm DE}$ at $z=0$ for the EFT models (left) and the covariant models (right). Solid lines show the results from \mer, while dashed lines show the corresponding linear predictions from \hiclass.}
    \label{fig:DEPower}
\end{figure}

Figure~\ref{fig:DEPower} shows the dimensionless power spectrum of the effective dark-energy density contrast, $\delta_{\rm DE}=\delta\rho_\phi/\bar\rho_\phi$. Its behaviour can be understood directly from the structure of the effective density perturbation in Eq.~\eqref{eq:deltarhoDE}. In particular, $\delta\rho_\phi$ receives contributions from $\pi$, $\zeta$, and the scalar metric perturbations, with their relative importance controlled by the $\alpha$ functions. Of particular relevance is the braiding contribution $\alpha_{\rm B}\mathcal H\nabla^2\pi$, which in Fourier space is proportional to $\alpha_{\rm B}k^2\pi$. This gradient enhancement explains why the dark-energy density perturbations can continue to grow towards small scales even though the scalar-field spectrum shown in Fig.~\ref{fig:PiPower} decreases in this regime.

The distinction is particularly clear among the EFT models. In the EFT-$k$ess limit, $\alpha_{\rm B}=0$, and all terms in Eq.~\eqref{eq:deltarhoDE} proportional to the braiding function vanish. The density perturbation is then mainly determined by the terms involving $\alpha_{\rm K}\zeta$ and $\pi$, resulting in the turnover and subsequent decrease of $\Delta_{\rm DE}$ towards small scales. Once braiding is introduced in EFT-KGB, the $k^2\pi$ contribution becomes increasingly important and produces a strong enhancement of the dark-energy clustering at high wavenumbers, consistent with the behaviour previously found in \citep{Nouri-Zonoz:2025cul}. EFT-LH has the same background evolution and the same $\alpha_{\rm K}$ and $\alpha_{\rm B}$ as EFT-KGB, but additionally includes a non-vanishing $\alpha_{\rm M}$ and an evolving effective Planck mass. Note that although $\alpha_{\rm M}$ does not appear explicitly in Eq.~\eqref{eq:deltarhoDE}, it affects $\delta \rho_\phi$ both through the evolution of $M_*^2$ and indirectly through the modified evolution of the scalar and metric perturbations. The effective Planck mass changes the normalisation of the effective density perturbation through the overall prefactor and also enters the coefficient of the term proportional to $\pi$. Its evolution can therefore modify not only the amplitude of $\delta_{\rm DE}$ but also its scale dependence, as shown in Fig.~\ref{fig:DEPower}.

The covariant nKGB and Cubic Galileon models show an even stronger growth of the dark-energy density perturbations towards small scales. Both models possess a non-vanishing braiding interaction, while their $\alpha_{\rm K}$ and $\alpha_{\rm B}$ functions grow strongly towards low redshift, as shown in Fig.~\ref{fig:background_evolution}. In particular, at $z=0$ their braiding functions are of order unity, substantially larger than the value $\alpha_{\rm B}\simeq0.34$ reached by EFT-KGB and EFT-LH. The larger braiding amplitude enhances the gradient term proportional to $\alpha_{\rm B}k^2\pi$ in Eq.~\eqref{eq:deltarhoDE}, which helps to produce the stronger dark-energy clustering seen in the covariant KGB models. This contribution therefore becomes increasingly important at late times and on small scales. The similar amplitudes of the nKGB and Cubic Galileon dark-energy power spectra at $z=0$ are consistent with their comparable values of $\alpha_{\rm K}$ and $\alpha_{\rm B}$, although the resulting density perturbations also depend on the evolution of other background quantities and the scalar-field perturbations.

Brans--Dicke behaves very differently. As discussed above, its Horndeski functions remain small and evolve only mildly over the relevant redshift range. In particular, $\alpha_{\rm K}$ stays at the level of $\sim10^{-3}$, several orders of magnitude below the values reached by the EFT models and substantially smaller than in nKGB and the Cubic Galileon, while $\alpha_{\rm B}$ and $\alpha_{\rm M}$ likewise remain small and vary only mildly over the same redshift range. Together with the large value $\omega_{\rm BD}=1000$, this keeps the model relatively close to its GR limit and leads to much weaker scalar perturbations, resulting in a dark-energy density power spectrum several orders of magnitude below those of nKGB and the Cubic Galileon.

\begin{figure}[t]
    \centering
    \includegraphics[width=\textwidth]{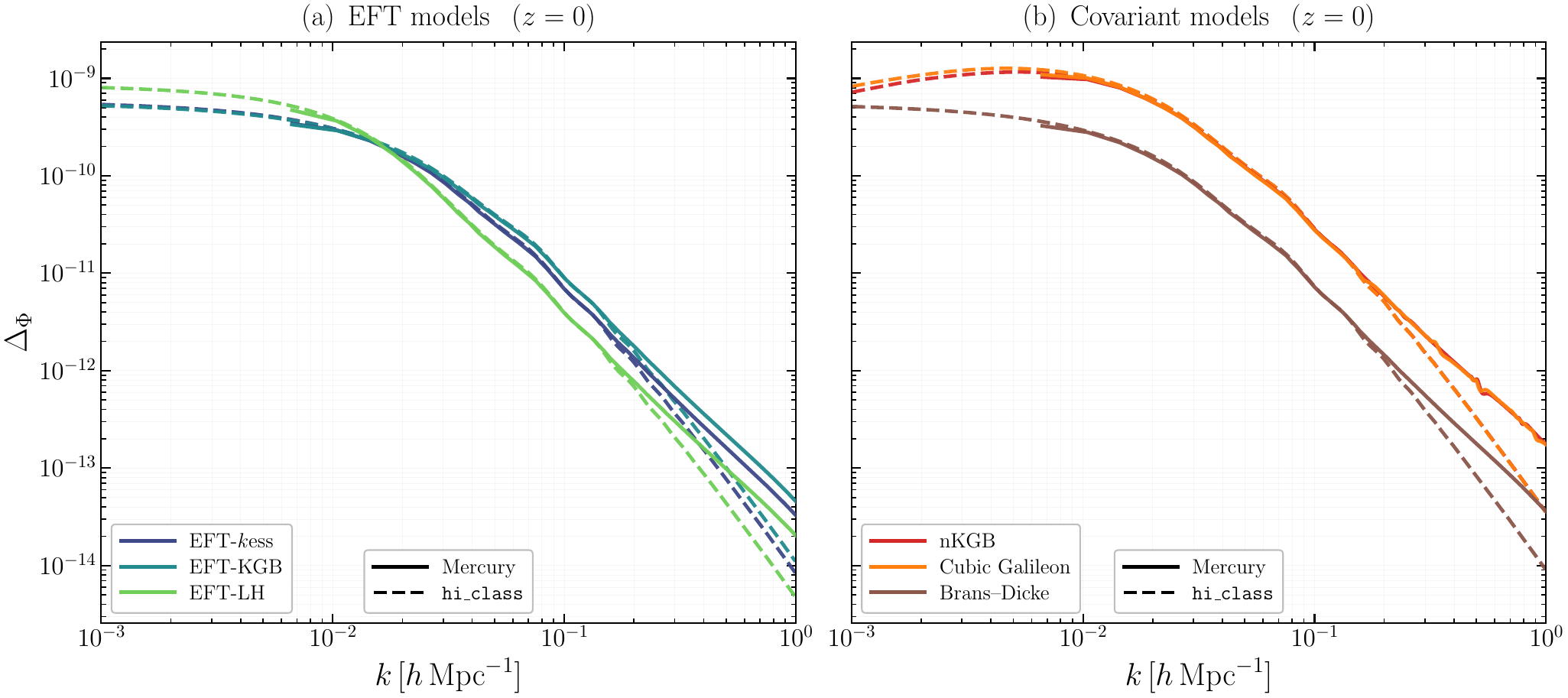}
    \caption{Dimensionless power spectra of the gravitational potential $\Phi$ at $z=0$ for the EFT models (left) and the covariant models (right). Solid lines show the results from \mer, while dashed lines show the corresponding linear predictions from \hiclass.}
    \label{fig:PhiPower}
\end{figure}

Figure~\ref{fig:PhiPower} shows the dimensionless power spectrum of the gravitational potential $\Phi$. Its behaviour can be understood from the Hamiltonian constraint in Eq.~\eqref{eq:Hamiltonian}, where $\Phi$ is sourced by the combined matter and scalar-field density perturbations, with the gravitational response normalised by $M_*^2$. On sub-horizon scales, this approximately gives $k^2\Phi\propto(\delta\rho_{\rm m}+\delta\rho_\phi)/M_*^2$, so that the decrease of the gravitational potential spectra towards large wavenumbers reflects the increasing $k^2$ suppression, modulated by the scale dependence of the matter and dark-energy sources.

Within the EFT models, EFT-KGB retains slightly more potential power towards smaller scales than EFT-$k$ess, consistent with the enhanced scalar clustering produced by braiding. EFT-LH instead shows an earlier and steeper decrease. This behaviour can be understood from the running of the effective Planck mass, characterised by $\alpha_{\rm M}>0$. Since $M_{*,\mathrm{ini}}^2=1$, the effective Planck mass increases towards the present epoch, suppressing the gravitational response to matter density perturbations through the factor $1/M_*^2$ in the Hamiltonian constraint. Furthermore, the effective dark-energy density perturbation in  Eq.~\eqref{eq:deltarhoDE} contains an overall prefactor $M_*^2$, which cancels against the factor $1/M_*^2$ in the Hamiltonian constraint, although additional dependence on $M_*^2$ remains within the expression. Consequently, its enhanced power does not necessarily translate into a larger contribution to the gravitational potential. In addition, the non-vanishing $\alpha_{\rm M}$ generates gravitational slip, allowing the temporal potential $\Psi$, which governs matter dynamics, to evolve differently from the spatial potential $\Phi$. These effects provide a qualitative explanation for why EFT-LH exhibits less potential power despite having enhanced matter and effective dark-energy density power spectra compared to EFT-KGB and EFT-$k$ess.
For the covariant models, nKGB and the Cubic Galileon exhibit the largest potential power and a mild enhancement before the subsequent decline, consistent with their strongly clustered scalar sectors seen in Fig.~\ref{fig:DEPower}. In both models $M_*^2$ is constant, so this behaviour is consistent with their strong braiding and the resulting enhanced scalar contribution to the total density source. Brans--Dicke remains lower in amplitude, in agreement with the much weaker scalar clustering found in the preceding spectra.

\begin{figure}[t]
    \centering
    \includegraphics[width=\textwidth]{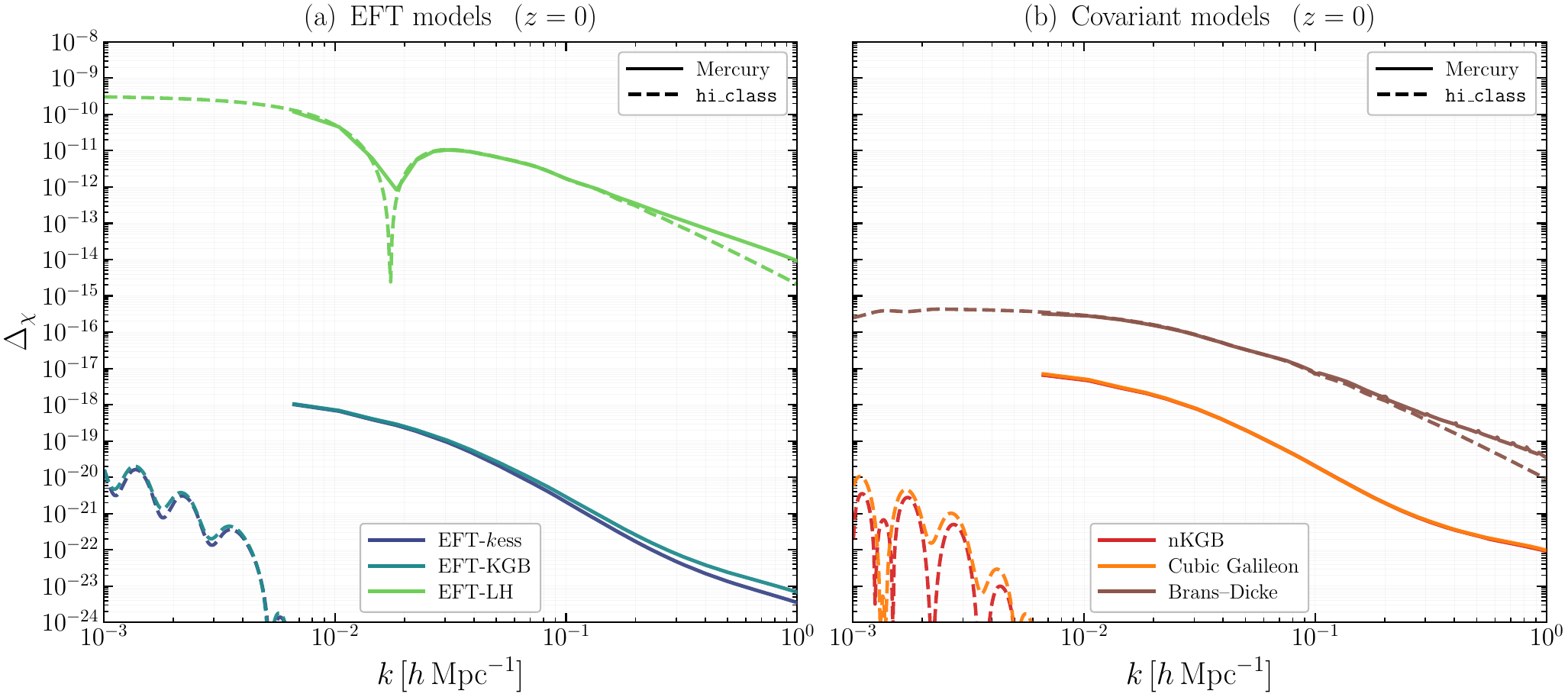}
    \caption{Dimensionless power spectra of the gravitational slip $\chi=\Phi-\Psi$ at $z=0$ for the EFT models (left) and the covariant models (right). Solid lines show the results from \mer, while dashed lines show the corresponding linear predictions from \hiclass.}
    \label{fig:chiPower}
\end{figure}

Figure~\ref{fig:chiPower} shows the dimensionless power spectrum of the gravitational slip, $\chi=\Phi-\Psi$. For EFT-$k$ess, EFT-KGB, nKGB, and the Cubic Galileon, both $\alpha_{\rm M}$ and $\alpha_{\rm T}$ vanish, so the scalar-field anisotropic stress is zero at linear order. The corresponding \hiclass prediction for $\chi$ is not strictly zero, owing to the anisotropic stress of relativistic species, including photons and free-streaming massless neutrinos. In the \mer simulations, the nonlinear evolution of dark matter generates an additional contribution to the gravitational slip through its anisotropic stress. Deep inside the horizon, this second-order effect is larger than the linear contribution of relativistic species. Consequently, the power spectrum of $\chi$ in these models exceeds the linear \hiclass prediction, particularly on smaller scales.

The situation is different for EFT-LH and Brans--Dicke, where $\alpha_{\rm M}\neq0$ while $\alpha_{\rm T}=0$. In this case, the Horndeski scalar generates a non-zero anisotropic stress already at linear order through the term proportional to $\alpha_{\rm M}\mathcal{H}\pi$ in Eq.~\eqref{eq:anisotropicEQ}, which is much larger than the slip generated by nonlinear matter evolution in the models with $\alpha_{\rm M}=0$. The \mer and \hiclass spectra consequently agree closely on large scales, while towards smaller scales the \mer results develop additional power as nonlinear matter and metric sources become increasingly important.

\begin{figure}[t]
    \centering
    \includegraphics[width=\textwidth]{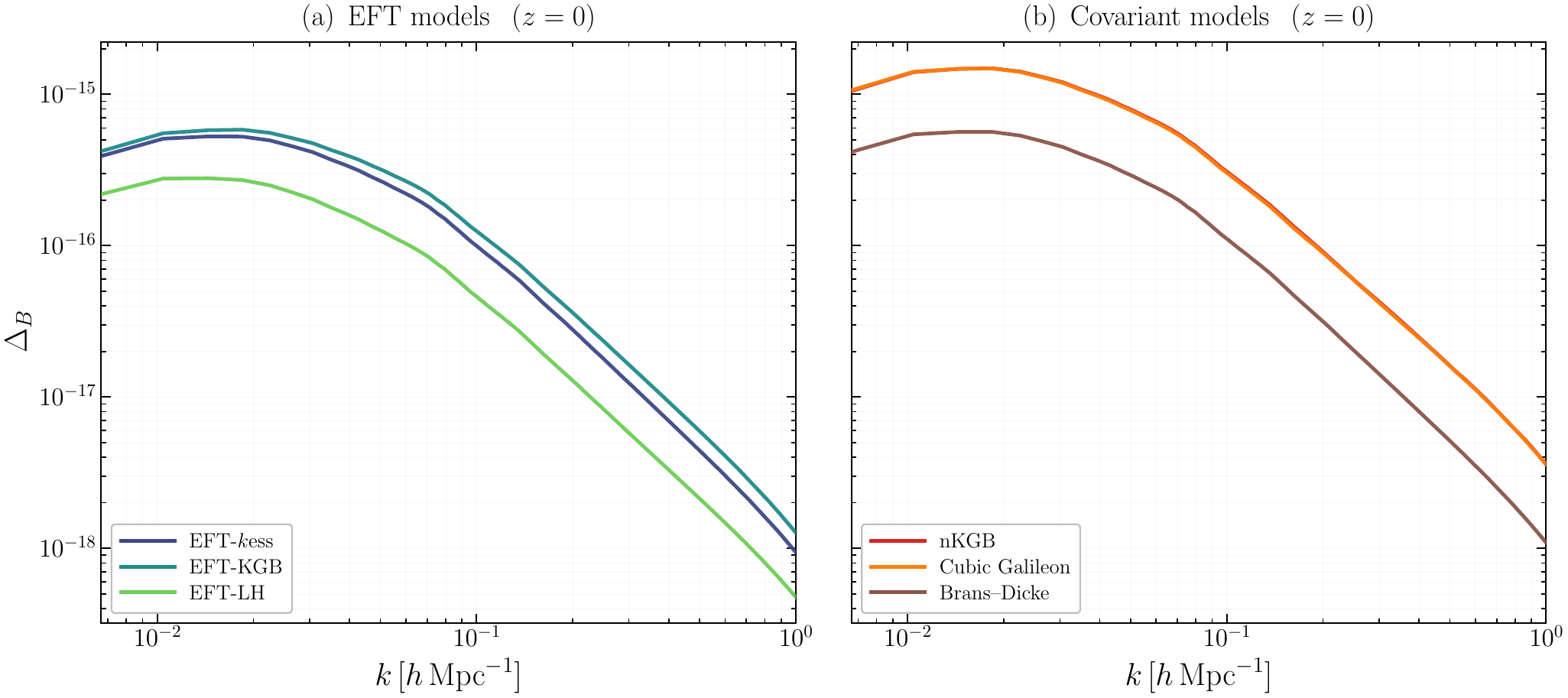}
    \caption{Dimensionless power spectra of the vector metric perturbation $B_i$ at $z=0$ for the EFT models (left) and the covariant models (right). Solid lines show the results from \mer.}
    \label{fig:BPower}
\end{figure}

\begin{figure}[t]
    \centering
    \includegraphics[width=\textwidth]{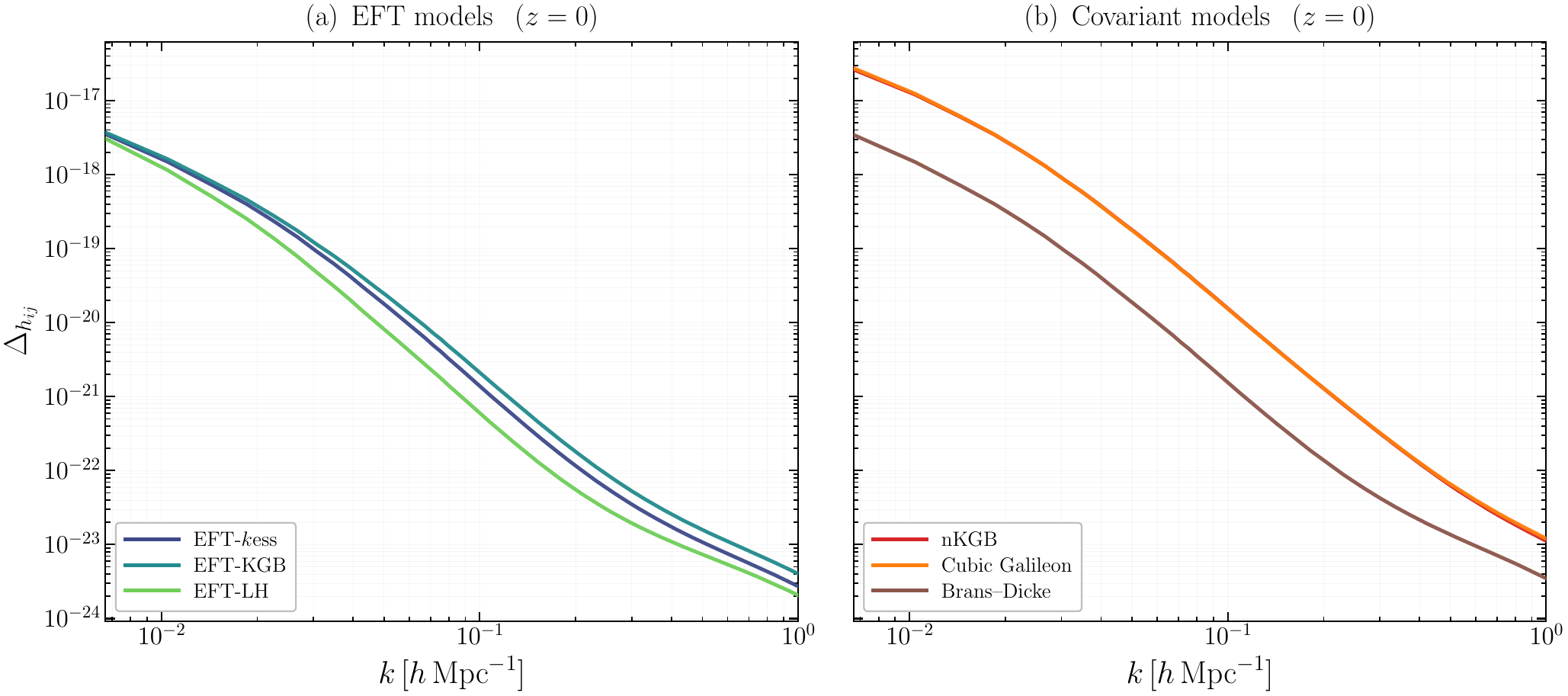}
    \caption{Dimensionless power spectra of the tensor metric perturbation $h_{ij}$ at $z=0$ for the EFT models (left) and the covariant models (right). Solid lines show the results from \mer.}
    \label{fig:hijPower}
\end{figure}

Figures~\ref{fig:BPower} and~\ref{fig:hijPower} show the vector and tensor metric perturbations generated during the nonlinear evolution. Both quantities are obtained from the spin-1 and spin-2 projections of the same trace-free spatial source. The linear Horndeski anisotropic stress is purely scalar and therefore does not contribute directly to either projection; instead, $B_i$ and $h_{ij}$ are sourced by the nonlinear matter anisotropic stress and by the quadratic metric contribution proportional to $\Phi_{,i}\Phi_{,j}$. Consequently, differences between the models partly reflect the different nonlinear matter and metric evolution already seen in the preceding spectra.

In addition, a running effective Planck mass affects both sectors through $\alpha_{\rm M}$. For the vector mode, which is evolved here using the parabolic spin-1 equation, the damping term becomes $(2+\alpha_{\rm M})\mathcal H B_i$. The effect is particularly clear when comparing EFT-LH with EFT-KGB: the two models share the same background evolution and the same $\alpha_{\rm K}$ and $\alpha_{\rm B}$, while only EFT-LH has $\alpha_{\rm M}\neq0$. The lower vector power in EFT-LH is therefore consistent with the additional $\alpha_{\rm M}$-dependent damping, together with the modifications that the evolving effective Planck mass induces in the nonlinear source terms. Similar behaviour occurs in the dynamical tensor equation, where $\alpha_{\rm M}$ modifies the friction term to $(2+\alpha_{\rm M})\mathcal H h'_{ij}$. The tensor results shown here use the elliptic treatment, which neglects time derivatives; their differences therefore arise through the source. Since $\alpha_{\rm T}=0$ for all models considered here, the tensor propagation speed remains unchanged. The lower vector and tensor amplitudes of Brans--Dicke are likewise consistent with its non-vanishing Planck-mass running together with its weaker nonlinear source, whereas nKGB and the Cubic Galileon, for which $\alpha_{\rm M}=0$, retain the largest amplitudes among the covariant models.

\subsection{Snapshot analysis}
\label{sect:snapshots}

In this section, we study the evolution of the dark-energy perturbations directly in real space. For this purpose, we perform additional simulations with $512^3$ grid points in a box of side length $L_{\rm box}=100\,h^{-1}{\rm Mpc}$. We focus on two cases, the EFT-$k$ess and EFT-LH models, using the same model parameters as listed in Table~\ref{tab:simulation_models}. The smaller volume provides a higher spatial resolution and allows the development of nonlinear structure to be followed more clearly. The snapshots are used to illustrate how an effectively nonlinear dark-energy distribution emerges in \mer even though the scalar-field perturbation equations themselves are implemented only at linear order. In \mer, the scalar perturbations evolve in the presence of the nonlinear matter and metric sources generated by structure formation, allowing matter inhomogeneities to imprint nonlinear features on $\delta_{\rm DE}$. By contrast, \hiclass evolves the same model entirely within linear perturbation theory.

\begin{figure}[!htbp]
    \centering
    \includegraphics[width=0.85\textwidth]{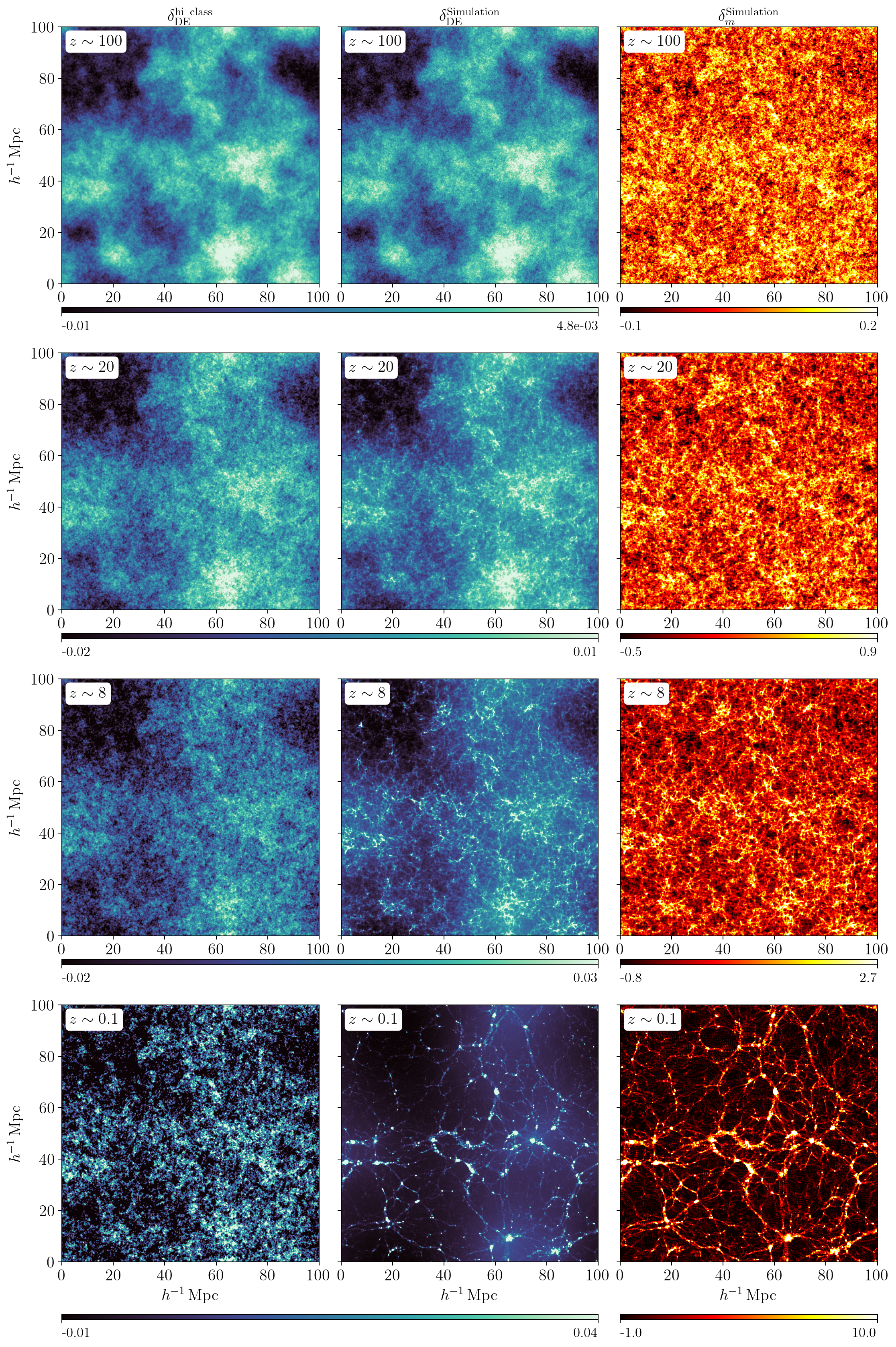}
    \caption{Real-space evolution of the density contrasts in the EFT-LH model with the parameters given in Table~\ref{tab:simulation_models}. The left and middle columns show the dark-energy density contrast obtained from \hiclass and \mer, respectively, while the right column shows the nonlinear matter density contrast in the simulation. For visualisation, the colour limits are restricted to the 1st--99th percentile range.}

    \label{fig:EFTLHSnapshots}
\end{figure}

\begin{figure}[!htbp]
    \centering
    \includegraphics[width=0.85\textwidth]{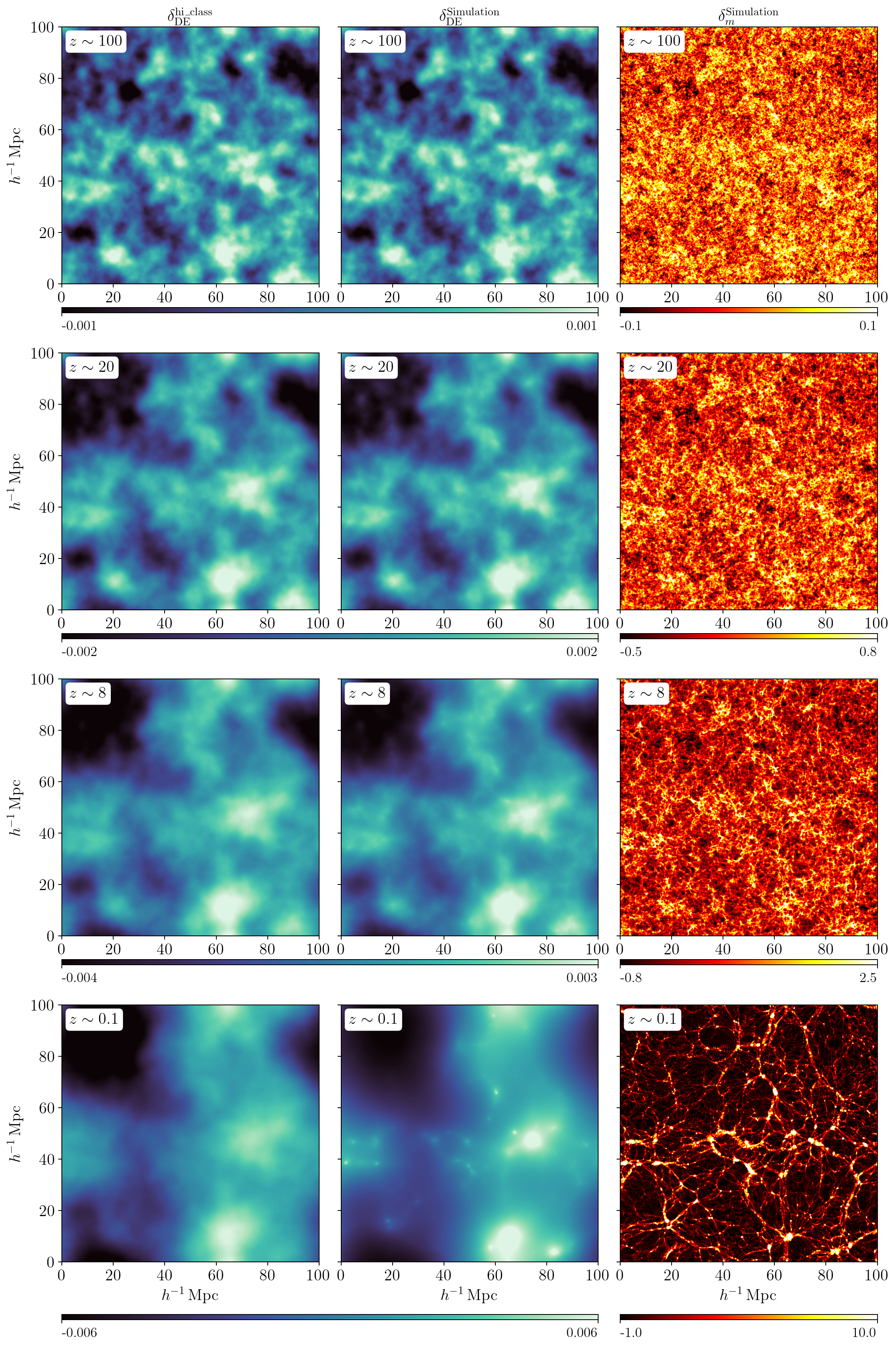}
    \caption{Same as Fig.~\ref{fig:EFTLHSnapshots}, but for the EFT-$k$ess model with the parameters given in Table~\ref{tab:simulation_models}.}
    \label{fig:kessSnapshots}
\end{figure}

Figure~\ref{fig:EFTLHSnapshots} shows this evolution for the EFT-LH model at four different redshifts. At the initial redshift, $z\simeq100$, the \hiclass and \mer dark-energy fields are nearly indistinguishable, as expected from the common linear initial conditions. As structure formation proceeds, differences between the two gradually develop. By $z\simeq20$ and more clearly by $z\simeq8$, the dark-energy field evolved in \mer begins to acquire spatial features correlated with the emerging matter distribution, while the \hiclass realisation retains the morphology predicted by linear evolution. At $z\simeq0.1$, the difference becomes particularly pronounced. The non-vanishing braiding parameter $\alpha_{\rm B}$ contributes to a strong response of the dark-energy perturbations to the nonlinear matter distribution, causing the \mer field to develop a filamentary structure that closely traces the cosmic web. In contrast, the corresponding \hiclass field remains comparatively smooth. This behaviour is consistent with our previous KGB analysis, where kinetic braiding was likewise found to drive enhanced dark-energy clustering \cite{Nouri-Zonoz:2025cul}. The non-vanishing $\alpha_{\rm M}$ and the associated evolution of the effective Planck mass $M_*^2$ also affect the dark-energy distribution. As discussed above, this occurs both through the explicit dependence of $\delta\rho_\phi$ on $M_*^2$ in Eq. \eqref{eq:deltarhoDE} and indirectly through the modified evolution of the scalar and metric perturbations. For the EFT-LH model considered here, its time evolution therefore modifies the overall amplitude of $\delta_{\rm DE}$, while also affecting the gradient-dependent contributions to $\delta\rho_\phi$. This effect is subdominant to the characteristic clustering induced by braiding, but further modifies the dark-energy field at late times.

Figure~\ref{fig:kessSnapshots} shows the corresponding evolution for the EFT-$k$ess model. As in the EFT-LH case, the \hiclass and \mer dark-energy fields agree closely at high redshift, when the evolution remains in the linear regime. At later times, differences between the two also begin to develop as the nonlinear matter and metric fields in \mer feed into the scalar perturbations. However, the departure from the purely linear \hiclass evolution is considerably weaker than in the EFT-LH model. In particular, the pronounced filamentary dark-energy structure seen in Fig.~\ref{fig:EFTLHSnapshots} at late times is much less prominent in the EFT-$k$ess case. This is consistent with the absence of kinetic braiding, $\alpha_{\rm B}=0$, for which the scalar sector responds less strongly to the nonlinear matter distribution.

The comparison between the two models is also informative at the linear level. The \hiclass snapshots in Figures~\ref{fig:EFTLHSnapshots} and \ref{fig:kessSnapshots} show that the EFT-LH and EFT-$k$ess models develop different dark-energy distributions even before the nonlinear response present in \mer is taken into account, reflecting their different linear dynamics. The additional separation between \hiclass and \mer at late times is nevertheless much stronger for the EFT-LH model. This comparison makes it possible to distinguish differences inherited from the linear dynamics of the two models from the additional structure generated through the coupling of the scalar perturbations to the nonlinear matter and metric fields in \mer.

For the models and parameters considered here, the dark-energy density perturbations remain small in both examples, even as the matter distribution becomes nonlinear. This supports the linear treatment of the scalar sector adopted in \mer, while allowing the dark-energy perturbations to respond to nonlinear matter clustering.\par \section{Discussion and conclusions}
\label{sec:discussion}

In this work, we have developed \mer as a relativistic $N$-body framework for Horndeski gravity within \gev. The main step beyond our previous implementations is the transition from specific subclasses of scalar--tensor theories to a common formulation based on the general linear Horndeski $\alpha$-basis. Together with the interface to \hiclass, this provides a common numerical pipeline in which the cosmological model can be specified either phenomenologically through the $\alpha$ functions or through an underlying covariant action, while the subsequent $N$-body evolution is performed within the same relativistic framework. The recovery of the previously implemented limits and the agreement with \hiclass in the linear regime provide the primary validation of this construction.

The present implementation should be viewed as an intermediate step between linear Horndeski calculations and a  nonlinear scalar--tensor simulation. Matter and the weak-field metric are evolved nonlinearly, and these nonlinear quantities enter the scalar-field evolution and its effective stress--energy tensor. The scalar sector can therefore respond dynamically to nonlinear structure formation and feed back on the gravitational field, rather than being imposed as an external linear realisation. At the same time, the intrinsic scalar dynamics are still described by the linear Horndeski perturbation equations. The nonlinear features that develop in the dark-energy distribution consequently arise from its coupling to the nonlinear matter and metric environment, rather than from nonlinear self-interactions of the Horndeski degree of freedom itself.

This distinction is also relevant for the interpretation of the $\alpha$-basis used by \mer. The functions $\alpha_{\rm K}$, $\alpha_{\rm B}$, $\alpha_{\rm M}$ and $\alpha_{\rm T}$ provide an efficient and largely model-independent characterisation of Horndeski gravity at the level of linear cosmological perturbations. Beyond linear order, however, the nonlinear scalar dynamics depend on higher-order operators and interactions that are not fixed by these four functions alone \citep{Frusciante:2019puu,Cusin:2017mzw}. Extending \mer beyond its present approximation will therefore require a corresponding nonlinear description of the theory, either by retaining the relevant higher-order operators in an extended effective formulation or by deriving the nonlinear equations directly from a specified covariant Horndeski action. The latter route is especially direct for the covariant models already supported by the code.

A major motivation for this extension is the possibility of studying screening within the same relativistic $N$-body framework. In scalar--tensor theories with sufficiently strong derivative interactions, nonlinearities in the scalar equation can suppress the response of the additional degree of freedom in regions where scalar gradients and derivative self-interactions become large, typically in and around nonlinear overdensities. Such behaviour cannot be reproduced by the present linear scalar equations, irrespective of how nonlinear the matter distribution becomes. Incorporating the nonlinear scalar terms would therefore allow \mer to follow the transition from the cosmological regime, where modifications to gravity may remain appreciable, to environments in which the scalar-mediated modification is dynamically suppressed. 
This would provide a framework for investigating screening mechanisms in relativistic cosmological simulations, including their dependence on the local environment and their impact on nonlinear structure formation.

Implementing this regime is not only a theoretical extension but also a substantially different numerical problem. The nonlinear scalar equation must be evolved together with the nonlinear contributions of the scalar field to the gravitational sources, while remaining consistent with the weak-field expansion used by \gev. 
In regions where nonlinear scalar interactions become dominant, the equations may develop strong spatial gradients and increasingly complex dynamics, requiring numerical methods capable of accurately resolving the transition between screened and unscreened regimes. Nonlinear scalar dynamics can also develop instabilities, as observed in \kev simulations at small sound speed and investigated analytically in \cite{Hassani:2022xyb, Hassani:2021tdd, Eckmann:2022wtd}. A nonlinear extension of \mer would also allow us to investigate whether similar behaviour occurs in broader classes of Horndeski models.
A nonlinear implementation will therefore require both an extension of the scalar solver and dedicated tests of its stability and convergence. The ability to recover the unscreened linear regime from the nonlinear solver will provide an important consistency test against the present version of \mer.

The framework is also naturally suited to extending the relativistic observables previously studied for KGB to a substantially broader class of scalar--tensor theories. Since \mer evolves the metric perturbations directly, its output can be combined with the light-cone infrastructure of \gev to construct observables sensitive to the gravitational field along the line of sight. Our previous KGB analysis showed that such observables can retain nonlinear signatures that are not captured by linear perturbation theory alone. Applying the same methodology to general Horndeski models will make it possible to connect modifications of the scalar and metric sectors to weak lensing, the ISW--RS effect, gravitational redshift, and other relativistic probes without relying on the matter power spectrum alone.

The present version of \mer therefore establishes the numerical link between the general linear Horndeski $\alpha$-basis implemented in \hiclass and relativistic nonlinear structure formation in \gev. The next step is to extend this framework to the nonlinear scalar sector, allowing scalar self-interactions, gravitational backreaction, and possible screening dynamics to be treated self-consistently alongside nonlinear matter within a single relativistic $N$-body framework.

\section*{Acknowledgements}
Ahmad Nouri-Zonoz, Julian Adamek and Martin Kunz acknowledge
funding from the Swiss National Science Foundation. Julian Adamek acknowledges additional support from the Dr.\ Tomalla Foundation for Gravity Research. Farbod Hassani acknowledges funding from the Research Council of Norway (Young Talent Grant, Project No. 345334). All simulations presented in this work were performed using
resources provided by Sigma2---the National Infrastructure
for High-Performance Computing and Data Storage in Norway
(project No.~NN12015K). Disclaimer: Co-funded by the European Union. Views and opinions expressed are, however, those of the author(s) only and do not necessarily reflect those of the European Union or the European Research Executive Agency. Neither the European Union nor the granting authority can be held responsible for them.

\section*{Code availability and computational performance}
The numerical results presented in this work were obtained using the CPU and GPU implementations of the same \mer code, which yield consistent numerical results but differ in computational performance. The power-spectrum analysis was carried out using the CPU version of \mer, whereas the simulation snapshots were obtained with the GPU version. Both versions of the code are publicly available at: \url{https://github.com/anourizo}. The GPU implementation of \mer is based on the GPU-accelerated framework of \gev~2.0, for which a reduction in computational cost by a factor of approximately $4.5$, measured in node-hours, was reported relative to the earlier CPU implementation \cite{Adamek:2026vof}. A detailed performance and scaling comparison of the CPU and GPU implementations of \mer, including the computational overhead of the additional scalar-field evolution, is left for future work.

\appendix
\section{Convergence tests}
\label{app:convergence}

In this appendix, we assess numerical convergence at the level of both the individual power spectra and their ratios, using the EFT-LH model as a representative case. We first compare the matter and dark-energy density power spectra and their ratio, and then examine the EFT-LH to EFT-$k$ess power-spectrum ratios used in the main analysis. We repeat the simulations with $N_{\rm grid}=N_{\rm pcl}=1536^3$, $2048^3$, and $3072^3$, keeping the box size fixed at $L_{\rm box}=4000\,h^{-1}{\rm Mpc}$. The corresponding Nyquist wavenumbers are $k_{\rm Ny}\simeq1.21$, $1.61$, and $2.41\,h\,{\rm Mpc}^{-1}$, respectively. All simulations use the same cosmological and model parameters as in the main analysis.

\begin{figure}[!htbp]
    \centering
    \includegraphics[width=0.9\textwidth]{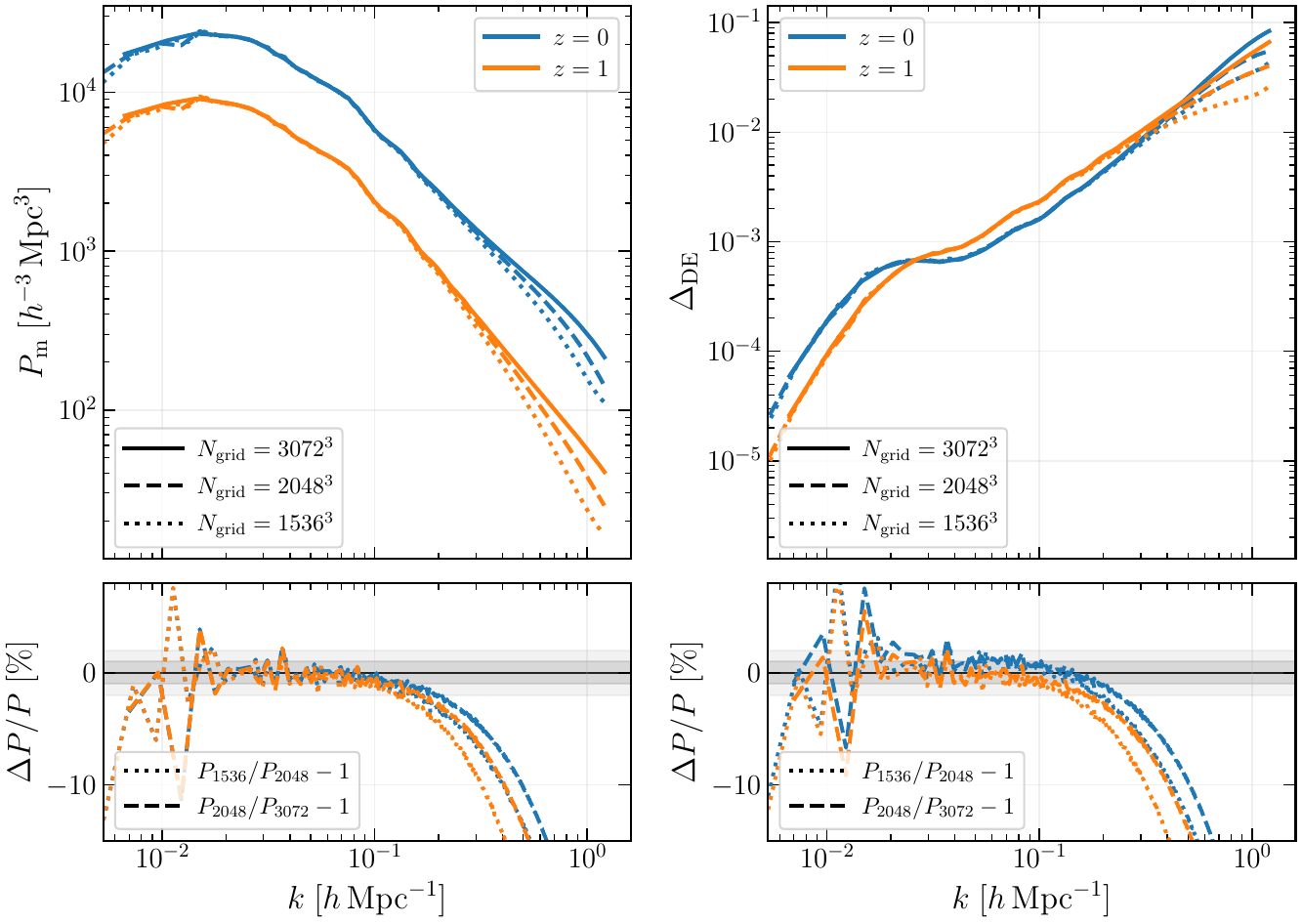}
    \caption{ Resolution convergence for the EFT-LH model in a
    $L_{\rm box}=4000\,h^{-1}{\rm Mpc}$ box.
    The upper panels show the matter (left) and dark-energy density (right) power spectra for $N_{\rm grid} = N_{\rm pcl}=1536^3$, $2048^3$, and $3072^3$ at $z=0$ and $z=1$. The lower panels show the relative differences between successive resolutions. The shaded regions indicate $1\%$ and $2\%$ deviations.}
    \label{fig:convtest}
\end{figure}

Figure~\ref{fig:convtest} shows the individual power spectra at $z=0$ and $z=1$, together with the relative differences between successive resolutions, $P_{1536}/P_{2048}-1$ and $P_{2048}/P_{3072}-1$. Here and below, the resolution subscript denotes the number of grid points along each direction. The spectra agree well on large scales, while finite-resolution effects become more important towards larger wavenumbers. The departure shifts to smaller physical scales as the resolution is increased, with $P_{2048}/P_{3072}-1$ remaining closer to zero than $P_{1536}/P_{2048}-1$ over the resolved range.

\begin{figure}[!htbp]
    \centering
    \includegraphics[width=1\textwidth]{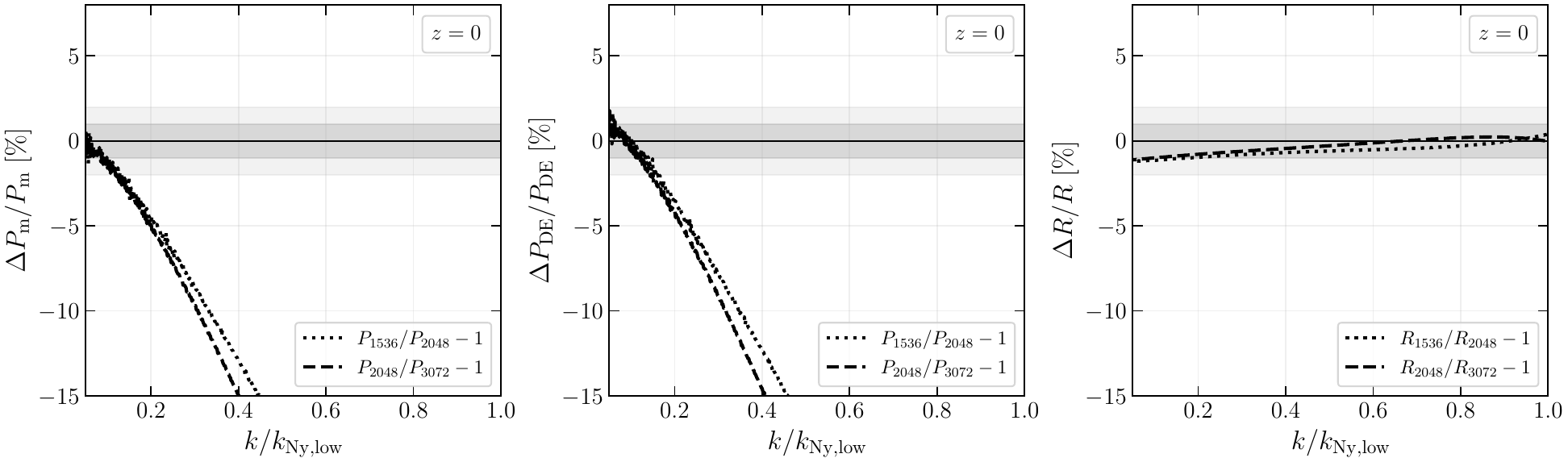}

\caption{Relative differences between successive resolutions at $z=0$, shown as a function of $k/k_{\rm Ny,low}$ for the matter power spectrum (left), dark-energy density power spectrum (middle), and their ratio $R(k)=P_{\mathrm{m}}(k)/P_{\rm DE}(k)$ (right). Here, $k_{\rm Ny,low}$ denotes the Nyquist wavenumber of the lower-resolution simulation in each pair.}

    \label{fig:convkny}
\end{figure}

To test whether this departure follows the grid scale, we rescale the wavenumber by the Nyquist wavenumber of the lower-resolution simulation in each pair, $q=k/k_{\rm Ny,low}$. Figure~\ref{fig:convkny} shows the resulting relative differences at $z=0$ for the matter and dark-energy density power spectra, together with those of their ratio $R(k)\equiv P_{\mathrm{m}}(k)/P_{\rm DE}(k)$. The two resolution pairs show a similar scale dependence in $q$ for both power spectra, suggesting that the discrepancy begins at approximately the same fraction of the Nyquist wavenumber.

As a quantitative reference, the $2\%$ deviation occurs at $q\simeq0.12$--$0.13$ for the matter power spectrum and at $q\simeq0.15$--$0.16$ for the dark-energy density power spectrum. If this scaling continues to the production resolution, $N_{\rm grid}=3072^3$, its Nyquist wavenumber gives the estimated $2\%$ resolution scales
\begin{equation}
k_{\rm 2\%}^{m}\simeq0.29\,h\,{\rm Mpc}^{-1}\,,
\qquad
k_{\rm 2\%}^{\rm DE}\simeq0.36\,h\,{\rm Mpc}^{-1}\,.
\end{equation}

These extrapolated scales provide an indication of resolution sensitivity rather than a direct measurement of the absolute numerical error of the production runs. In contrast to the individual spectra, the ratio $R(k)$ shows much smaller differences between resolutions, with $\Delta R/R$ remaining within approximately $1\%$ over most of the plotted range. This indicates a substantial cancellation of resolution effects in this ratio.

\begin{figure}[!htbp]
    \centering
    \includegraphics[width=1\textwidth]{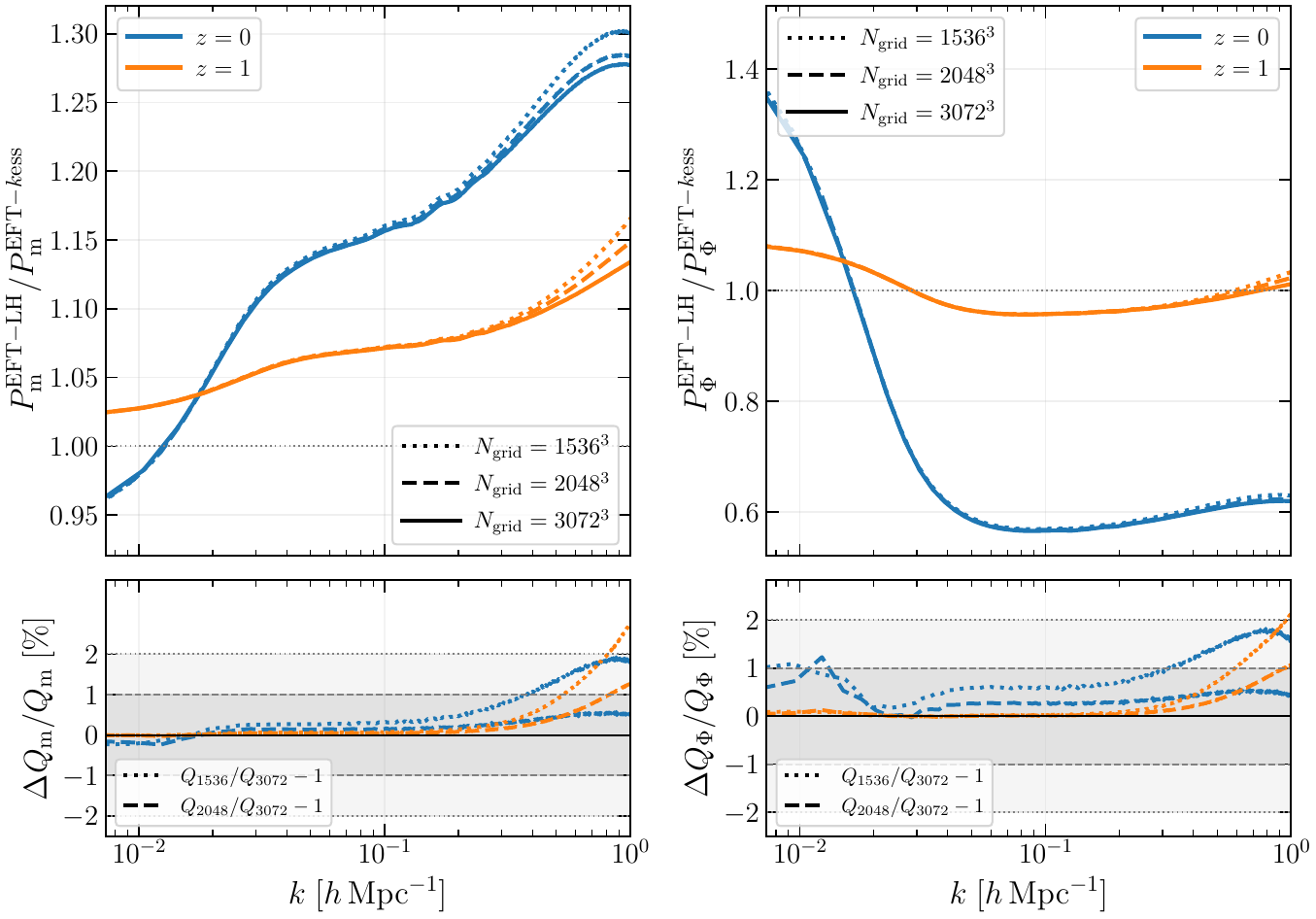}
    \caption{Ratios of the EFT-LH to EFT-$k$ess power spectra for matter (left) and the spatial metric potential $\Phi$ (right), shown at $z=0$ and $z=1$ for simulations with $N_{\rm grid}=N_{\rm pcl}=1536^3$, $2048^3$, and $3072^3$. The lower panels show the relative differences with respect to the $N_{\rm grid}=3072^3$ simulation, with shaded bands indicating $1\%$ and $2\%$ deviations.}
    \label{fig:ratioConv}
\end{figure}

We next test whether a similar cancellation occurs in the model comparisons presented in the main text. Figure~\ref{fig:ratioConv} shows $Q_{X,N}\equiv P_{X,N}^{\mathrm{EFT\text{-}LH}}/P_{X,N}^{\mathrm{EFT\text{-}}k\mathrm{ess}}$, with $X\in\{\mathrm{m},\Phi\}$, at all three resolutions. Taking $N_{\rm grid}=3072^3$ as the reference, we define $\Delta Q_X/Q_X\equiv Q_{X,N}/Q_{X,3072}-1$, with $N\in\{1536,2048\}$. For $N_{\rm grid}=2048^3$, these differences remain within approximately $2\%$ over the plotted range at $z=0$ and $z=1$. The $N_{\rm grid}=1536^3$ simulation shows larger differences on small scales, exceeding $2\%$ at $z=1$. The reduced resolution sensitivity of these model ratios supports a more robust interpretation of the relative model signatures on smaller scales than is possible from the individual spectra alone.

\newpage
 \bibliographystyle{JHEP}
\bibliography{biblio.bib}
\end{document}